\documentclass{aa}
\usepackage{txfonts}
\usepackage{amsmath}
\usepackage{color}
\usepackage{xcolor}
\usepackage[normalem]{ulem}
\usepackage{multicol}
\usepackage[colorlinks=true,citecolor=blue,linkcolor=blue,urlcolor=cyan]{hyperref}
\usepackage{textcomp}
\usepackage{subcaption}
\usepackage{graphicx}
\usepackage{float}
\usepackage{placeins}
\usepackage{multirow}
\usepackage{changepage}
\usepackage{colortbl}
\usepackage{amsfonts}
\usepackage{booktabs}
\usepackage{siunitx}
\usepackage{amssymb} 
\usepackage{enumitem}
\usepackage{multirow}
\usepackage{tabularx}
\usepackage{booktabs}
\usepackage{gensymb}

\author{
J.M.G.H.J. de Jong\inst{\ref{inst:Leiden}} 
\and R.J. van Weeren\inst{\ref{inst:Leiden}}
\and F. Sweijen\inst{\ref{inst:Durham},\ref{inst:Astron},\ref{inst:Leiden}}
\and J.B.R. Oonk\inst{\ref{inst:surf},\ref{inst:Leiden},\ref{inst:Astron}}
\and T.W. Shimwell\inst{\ref{inst:Astron},\ref{inst:Leiden}}
\and A.R. Offringa\inst{\ref{inst:Astron},\ref{inst:Groningen}}
\and L.K. Morabito\inst{\ref{inst:Durham},\ref{inst:Durham2}}
\and H.J.A. R\"ottgering\inst{\ref{inst:Leiden}}
\and R. Kondapally\inst{\ref{inst:edin}}
\and E.L. Escott\inst{\ref{inst:Durham}}
\and P.N. Best\inst{\ref{inst:edin}}
\and M. Bondi\inst{\ref{inst:INAF}}
\and H. Ye\inst{\ref{inst:Leiden},\ref{inst:cambridge}}
\and J.W. Petley\inst{\ref{inst:Durham}}
}

\institute{
Leiden Observatory, Leiden University, PO Box 9513, 2300 RA Leiden, The Netherlands\relax\label{inst:Leiden}
\and
Centre for Extragalactic Astronomy, Department of Physics, Durham University, Durham, DH1 3LE, UK\relax\label{inst:Durham} 
\and 
ASTRON, The Netherlands Institute for Radio Astronomy, Postbus 2, 7990 AA Dwingeloo, The Netherlands\relax\label{inst:Astron} 
\and
SURF/SURFsara, Science Park 140, 1098 XG Amsterdam, The Netherlands\relax\label{inst:surf}
\and
Kapteyn Astronomical Institute, P.O. Box 800, 9700 AV Groningen, Netherlands\relax\label{inst:Groningen}
\and 
Institute for Computational Cosmology, Department of Physics, Durham University, South Road, Durham DH1 3LE, UK\relax\label{inst:Durham2}
\and
Institute for Astronomy, University of Edinburgh, Royal Observatory, Blackford Hill, Edinburgh, EH9 3HJ, UK \relax\label{inst:edin}
\and
INAF - Istituto di Radioastronomia, Via Gobetti 101, 40129, Bologna, Italy\relax\label{inst:INAF} 
\and
Cavendish Astrophysics, University of Cambridge, Cambridge, UK\relax\label{inst:cambridge}
}

\begin{document}

\title{Into the depths: Unveiling ELAIS-N1 with LOFAR's deepest sub-arcsecond wide-field images}

\date{Received XXX 2024 / Accepted XXX 2024}

\begin{abstract} 
    {We present the deepest wide-field 115-166~MHz image at sub-arcsecond resolution spanning an area of 2.5\degree$\times$2.5\degree~centred at the ELAIS-N1 deep field. To achieve this, we improved the direction-independent (DI) and direction-dependent (DD) calibrations for the International LOw Frequency ARray (LOFAR) Telescope. This enhancement enabled us to efficiently process 32~hrs of data from four different 8-hr observations using the high-band antennas (HBAs) of all 52 stations, covering baselines up to approximately 2,000~km across Europe. The DI calibration was improved by using an accurate sky model and refining the series of calibration steps on the in-field calibrator, while the DD calibration was improved by adopting a more automated approach for selecting the DD calibrators and inspecting the self-calibration on these sources. For our brightest calibrators, we also added an additional round of self-calibration for the Dutch core and remote stations in order to refine the solutions for shorter baselines. To complement our highest resolution at 0.3\arcsec, we also made intermediate resolution wide-field images at 0.6\arcsec~and 1.2\arcsec. Our resulting wide-field images achieve a central noise level of 14~$\mu$Jy~beam$^{-1}$ at 0.3\arcsec, doubling the depth and uncovering four times more objects than the Lockman Hole deep field image at comparable resolution but with only 8~hrs of data. Compared to LOFAR imaging without the international stations, we note that due to the increased collecting area and the absence of confusion noise, we reached a point-source sensitivity comparable to a 500-hr ELAIS-N1 6\arcsec~image with 16 times less observing time. Importantly, we have found that the computing costs for the same amount of data are almost halved (to about 139,000 CPU~hrs per 8~hrs of data) compared to previous efforts, though they remain high. Our work underscores the value and feasibility of exploiting all Dutch and international LOFAR stations to make deep wide-field images at sub-arcsecond resolution.}
    \maketitle
\end{abstract}

\section{Introduction}

The International LOw Frequency ARray (LOFAR) Telescope (ILT) is an interferometer uniquely designed to measure low frequency radio waves between 10 and 80~MHz with the low-band antennas (LBAs) and between 110 and 240~MHz with the high-band antennas (HBAs) \citep{haarlem2013}. With its baselines extending up to \textasciitilde2,000~km coupled with a degree-scale field of view, it can thus produce wide-field images at sub-arcsecond resolution. Nevertheless, reducing data from all 38 Dutch and 14 international stations of LOFAR for wide-field imaging is nontrivial, as it requires a carefully optimised calibration strategy to correct for various corrupting effects on the data and extensive computing facilities to handle the substantial data volumes and perform the final imaging \citep{sweijen2022, ye2023}.

In recent years, significant efforts have been devoted to automatically calibrating and imaging observations with the Dutch HBA stations located in the Netherlands. This has led to the LOFAR Two-metre Sky Survey \citep[LoTSS;][]{shimwell2017, shimwell2019, shimwell2022, williams2019} and the LoTSS-Deep Fields \citep{kondapally2021, duncan2021, tasse2021, sabater2021, best2023, bondi2024}, which have provided wide-field images of the northern sky at 144~MHz and 6\arcsec~resolution. 
Despite the fact that these works discovered many new radio sources at the lowest frequencies, approximately 90\% of these sources at 6\arcsec~remain unresolved at 144~MHz. This becomes an issue when, for instance, one aims to examine the detailed dynamics of bright radio-loud active galactic nuclei (RLAGN) \citep[e.g.][]{mahatma2023}, derive source size distributions at the smallest angular scales (e.g. Sweijen et al., in prep.), separate radio emission from (radio-quiet) AGN and star formation \citep[e.g.][]{morabito2022a}, or when the 6\arcsec~resolution limit introduces selection effects in the study of the cosmic evolution of resolved RLAGN \citep[e.g.][]{dejong2024}. This, among other scientific objectives, underscores the scientific value of the sub-arcsecond wide-field imaging capabilities of LOFAR. 

Calibrating data at low frequencies is challenging due to direction-dependent effects (DDEs), which are variations of data corruption across the field of view. At low frequencies, most DDEs are posed by the ionosphere, resulting in propagation delays of radio waves \citep{intema2009, smirnov2011, vanweeren2016, tasse2018}. Correcting these factors inadequately results in image fidelity issues due to calibration artefacts that extend from arcsecond up to arcminute scales. These effects are destructive for the quality of the high-resolution images if not properly corrected. Incorporating data from all international LOFAR stations during calibration makes the data reduction more complicated, as extra phase delays are induced by the fact that the international stations have independent clocks \citep{morabito2022}. Moreover, the fact that the availability of bright sources reduces towards higher resolutions complicates the calibration strategy, as this heavily relies on (self-)calibration of sources with a high S/N on all baselines. On top of this is the substantial volume of visibility data that needs to be processed. A typical LOFAR observation with an integration time of 8~hrs and a time and frequency resolution of 1~sec and 12.21~kHz is in the order of 16~TB, which can be reduced to 4~TB with Dysco compression \citep{offringa2016}. To process LOFAR data, it is therefore essential to have access to machines with enough computational power and with extensive storage capacities.

Early pioneering works have demonstrated how to utilize all HBA antennas from LOFAR's international stations to produce images at sub-arcsecond resolutions \citep[e.g.][]{varenius2015, varenius2016, ramirez2018, harris2019}. Subsequent efforts by \cite{morabito2022} standardised and partly automated the calibration and imaging process with the international stations, which resulted in a first version of the LOFAR Very Long Baseline Interferometry (VLBI) calibration workflow.\footnote{\url{https://github.com/LOFAR-VLBI}} The value of their work is directly evident through the large number of studies that have already utilised their workflow \citep[e.g.][]{sweijen2022a, sweijen2023, bonnassieux2022, timmerman2022, timmerman2022b, harwood2022, pranav2022, morabito2022a, mahatma2023, venkattu2023}. During the same time, \cite{sweijen2022} extended this strategy to perform wide-field imaging and produced with 8~hrs of LOFAR data from the Lockman Hole the first 6.6 deg$^{2}$ wide-field image at a resolution of 0.30\arcsec$\times$0.38\arcsec~at 144~MHz with a sensitivity down to 25$~\mu$Jy~beam$^{-1}$. This image, produced with a computational cost of 250,000 CPU~hrs, captured in one snapshot 2,483 high-resolution sources, each with peak intensities five times greater than their local RMS. These types of wide-field images contain approximately 10 billion pixels, which makes imaging the most dominant part of the total computational costs. Another recent study by \cite{ye2023} adopted a similar calibration approach to \cite{sweijen2022} but with the aim to make an intermediate resolution wide-field image of the ELAIS-N1 deep field at 1.2\arcsec$\times$2\arcsec. This resolution serves as a scientifically valuable intermediary that improves the 6\arcsec~resolution from LoTSS and recovers extended emission that is lost at the finer 0.3\arcsec~resolution, such as from low-excitation radio galaxies (LERGs) \citep{ye2023}. Since imaging represents the main computational bottleneck for the complete data processing pipeline, \cite{ye2023} achieved a total computing time speedup of nearly a factor five compared to \cite{sweijen2022}.

Even though one of the primary objectives of achieving higher resolutions is to resolve more sources, high-resolution images are less suitable for studying low surface brightness structures, as these are more likely to be resolved out. For instance, \cite{sweijen2022} showed that only 40\% of the sources that are detected and unresolved at 6\arcsec~are detected in their corresponding 8-hr radio map at 0.3\arcsec, of which 11\% of these are resolved at 0.3\arcsec. The number of 6\arcsec~counterparts at 1.2\arcsec~doubles, as shown for ELAIS-N1 by \cite{ye2023}. Hence, in order to recover more resolved sources at higher resolutions, it is essential to enable the production of deeper images through the use of multiple 8-hour observations of the same field and to get more information out of the data by making images at intermediate resolutions (between 0.3\arcsec~and 6\arcsec) as well. This approach is further supported by the fact that confusion noise limits the sensitivity obtainable by deep wide-field imaging using only the Dutch stations, as was demonstrated by \cite{sabater2021} in their imaging of ELAIS-N1 with 163.7~hrs of LOFAR observations. They reached a best noise level of approximately \textasciitilde17$~\mu\text{Jy}~\text{beam}^{-1}$, which is expected to be achievable with about ten times less LOFAR observing time when including both the Dutch and international stations.

We aim in this paper to produce the deepest wide-field images currently available at sub-arcsecond resolution at 140~MHz (115-166~MHz) by jointly calibrating four LOFAR observations, totalling 32~hrs, of the ELAIS-N1 deep field. Building upon the work from \cite{morabito2022}, \cite{sweijen2022}, and \cite{ye2023}, we refined the direction-independent (DI) calibration steps and improved the direction-dependent (DD) calibrator selection. This enabled us to obtain the final merged calibration solutions for Dutch and international LOFAR stations, which are required for facet-based imaging at (sub-)arcsecond resolutions. Utilising the calibrated data, we produced wide-field images at 0.3\arcsec, 0.6\arcsec, and 1.2\arcsec~resolution. This allowed us to make source catalogues at different resolutions and thereby analyse source detections across different resolutions and sensitivities.\footnote{All data products are published at \url{https://lofar-surveys.org/hd-en1.html}.}

In Section \ref{sec:datadescription}, we discuss the details of our four LOFAR observations, which leads into a detailed discussion of the calibration process in Section \ref{sec:calibration}. Following calibration, we address the imaging process in Section \ref{sec:imaging} and then detail the creation of the associated source catalogues in Section \ref{sec:catalogue}. Our discussion extends to evaluating the quality of our image and catalogue outputs in Section \ref{sec:discussion}, and we finish with our conclusions in Section \ref{sec:conclusion}.

\section{Data description}\label{sec:datadescription}

\begin{table*}[htbp]
\caption{Metadata from the four ELAIS-N1 observations used in this paper.}
\centering
\begin{tabular}{lllll} \toprule
    \textbf{Observation ID} & \textbf{\textit{L686962}} & \textbf{\textit{L769393}} & \textbf{\textit{L798074}} & \textbf{\textit{L816272}} \\ \midrule
     \textbf{Project} & LT10\_012 & LT10\_012 & LT14\_003 & LT14\_003 \\
    \textbf{Calibrator} & 3C\,295 (L686958) & 3C\,295 (L769389) & 3C\,295 (L798082) & 3C\,295 (L816280) \\
   \textbf{Observation date} & 26-11-2018 & 24-05-2020 & 14-11-2020 & 13-5-2021 \\
   \textbf{Pointing centres} & 16:11:00, +54.57.00 & 16:11:00, +54.57.00 & 16:11:00 +55.00.00 & 16:11:00 +55.00.00 \\
   \textbf{Integration time} & 8~hrs & 8~hrs & 8~hrs & 8~hrs \\
   \textbf{Frequency range} & 120-166~MHz & 120-166~MHz & 115-164~MHz & 115-164~MHz \\
   \textbf{Stations (International)} & 51 (13) & 51 (13) & 50 (12) & 52 (14) \\
   \bottomrule
\end{tabular}
\label{table:data}
\end{table*}

The area covered by ELAIS-N1 has been studied in the optical \citep[e.g.][]{mcmahon2001, aihara2018}, infrared \citep[e.g.][]{lawrence2007, mauduit2012}, ultraviolet \citep[e.g.][]{martin2005}, X-rays \citep[e.g.][]{manners2003}, and radio \citep[e.g.][]{ciliegi1999, sirothia2009, croft2013, ocran2020}. This extensive multi-wavelength coverage has made ELAIS-N1 an invaluable field for extra-galactic science and it was therefore selected as one of the LOFAR Deep Fields \citep{sabater2021, kondapally2021, best2023}.

In order to make the deepest sub-arcsecond resolution wide-field radio map of this field with LOFAR, we selected four 8-hrs LOFAR observations of ELAIS-N1 by examining calibration solutions of calibrator sources 3C~295 or 3C~48 that were already observed for 10~minutes before or after 16 different available ELAIS-N1 observations stored in the LOFAR long-term archive (LTA). This calibration step is an important part of the entire calibration (as discussed in Section \ref{sec:initdutchcalibration}) and provides a computationally cheap way to assess the ionospheric conditions and the quality of the data (as discussed in Section \ref{sec:initdutchcalibration}). Although we could select more than these 4 observations, it is important to stress that the compute costs for calibrating and imaging data at sub-arcsecond resolutions are expensive and limit us to selecting more than 4 observations (as highlighted by \cite{sweijen2022} and by us in Section \ref{sec:computingcosts}). Our selected observations are part of two different observing projects (\texttt{LT10\_012} and \texttt{LT14\_003}, PI: P.N. Best) and were retrieved from the LTA.\footnote{\url{lta.lofar.eu}} 

\begin{figure}[htbp]
 \centering
    \includegraphics[width=0.96\linewidth]{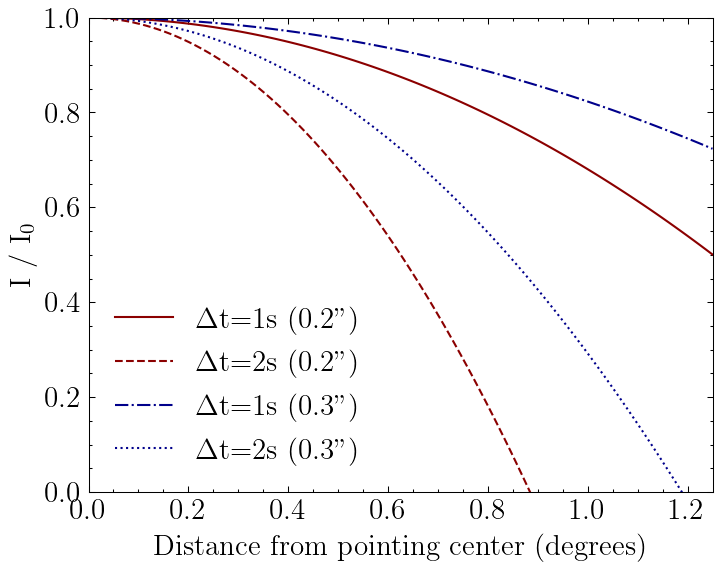}
  \caption{Intensity $I$ as a function of distance from the pointing centre due to a combination of bandwidth and time smearing over the original intensity at the pointing centre $I_{0}$. We used the smearing formulas according to \cite{bridle1999} with a central frequency of 140~MHz and a bandwidth of 12.21~kHz. This plot includes the smearing from the longest baseline (1980~km) between the LOFAR stations in Birr (Ireland) and Łazy (Poland), reaching a resolution of about 0.2\arcsec~(red). We also added $I/I_{0}$ for our target resolution of 0.3\arcsec~(blue), corresponding to a baseline of 1470~km. This figure shows both the smearing for the 1~sec and 2~sec pre-averaged datasets.}
\label{fig:theoretical_smearing}
\end{figure}

We provide a description of our selected observations in Table \ref{table:data}. All four observations have 3C\,295 as the primary calibrator. The pointing centres of two observations are 0.03\degree~offset from the other two, which necessitates a phase-shift correction to a common right ascension (RA) of 16:11:00 and declination (DEC) of +54.57.00 to enable combined imaging (see Section \ref{sec:imaging}). Prior to the storage of the observations L798074 and L816272 on the LTA, their data was averaged from a time resolution of 1~sec to 2~sec. The pre-averaging leads to additional time smearing effects on the data. This was unfortunately only noticed after doing most of the calibration discussed in Section \ref{sec:calibration} and thus we kept the data in our final images. Whilst the time smearing cannot be completely mitigated (see Figure \ref{fig:theoretical_smearing}), we reduce the impact during calibration by flagging the baselines that are most severely affected (see Section \ref{sec:ddcal}). 

\begin{figure*}[ht]
\begin{subfigure}{.5\textwidth}
\includegraphics[width=1\linewidth]{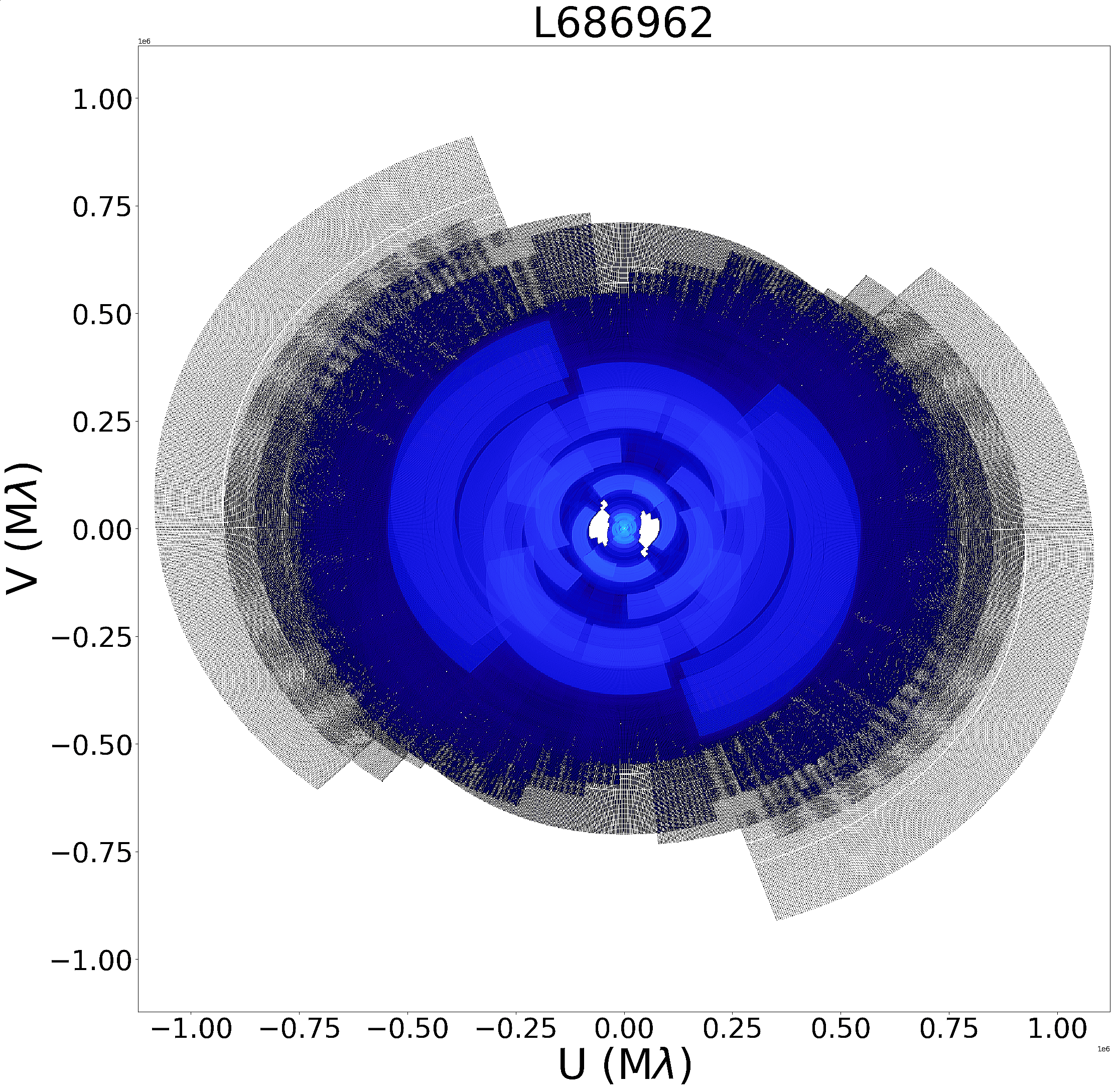}
\end{subfigure}
\begin{subfigure}{.5\textwidth}
\includegraphics[width=1\linewidth]{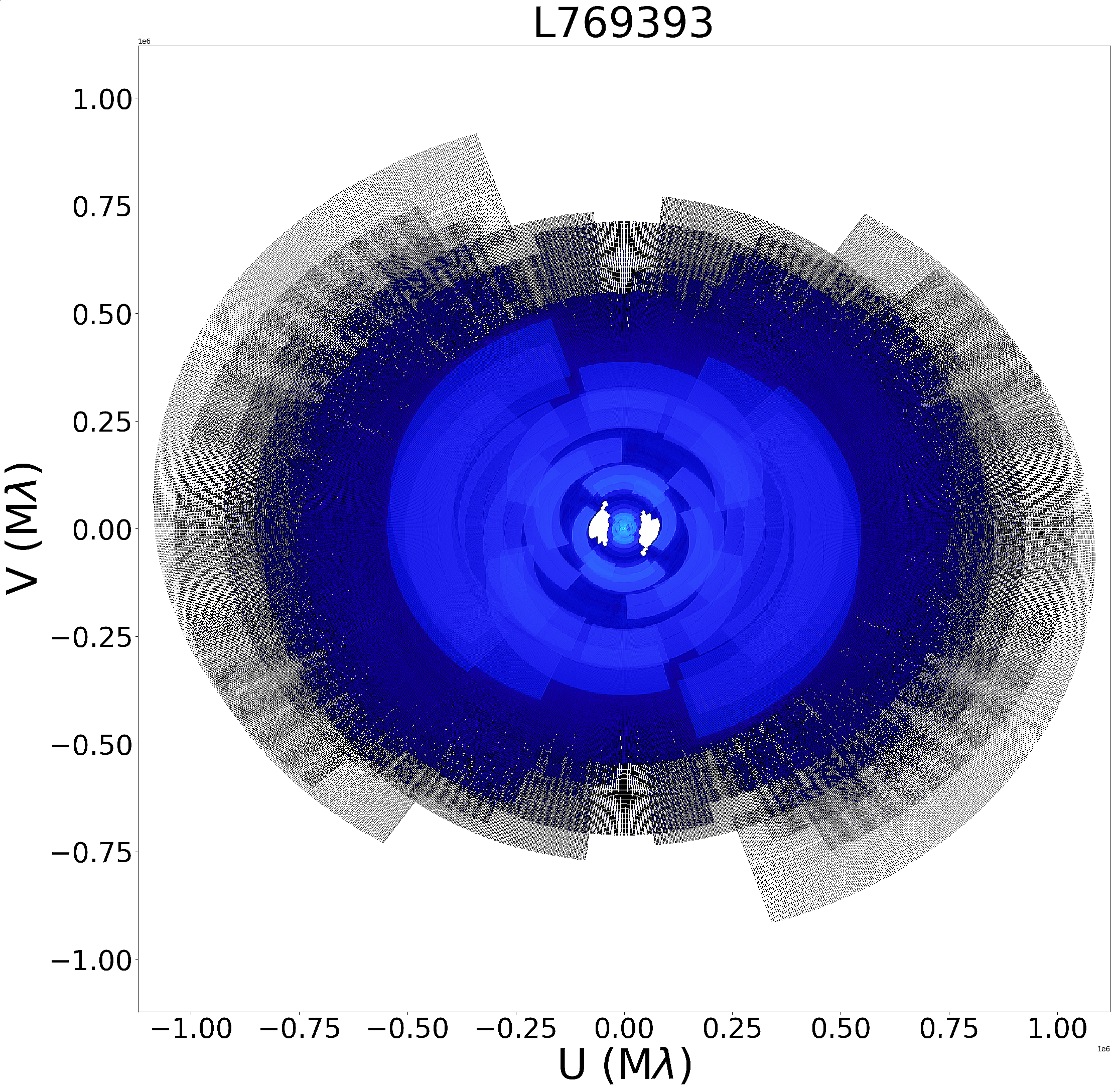}
\end{subfigure}
\begin{subfigure}{.5\textwidth}
\includegraphics[width=1\linewidth]{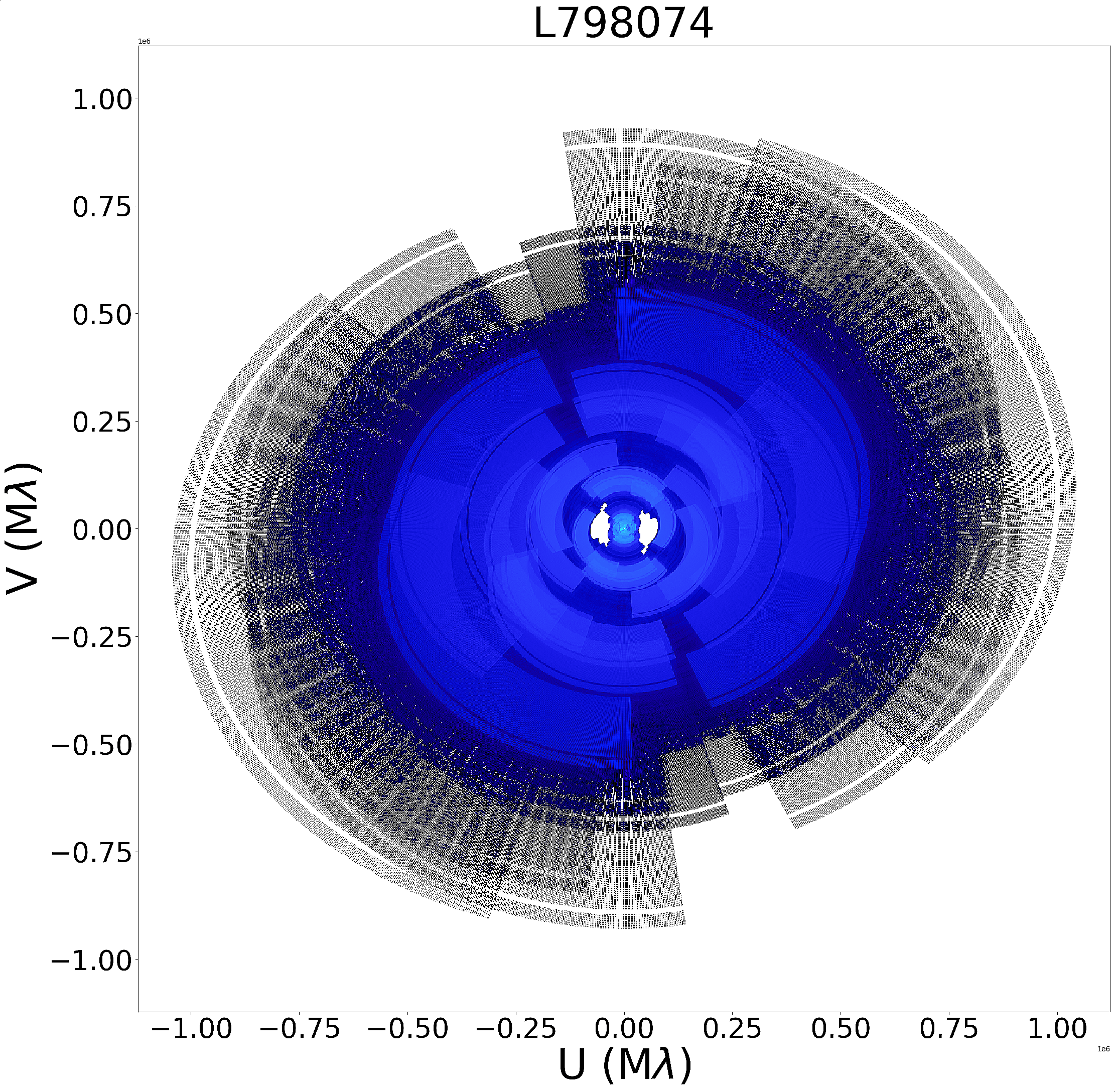}
\end{subfigure}
\begin{subfigure}{.5\textwidth}
\includegraphics[width=1\linewidth]{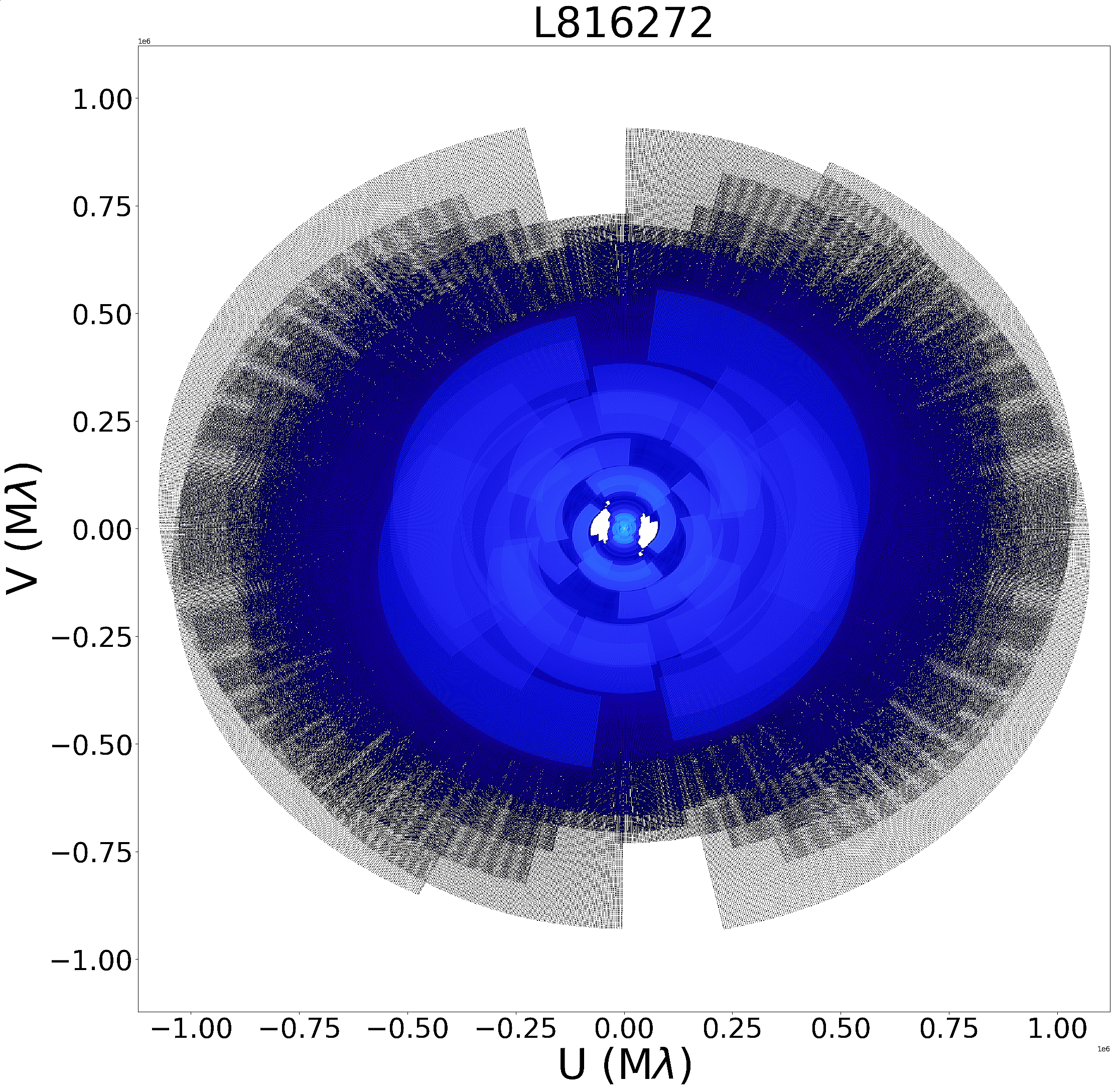}
\end{subfigure}
\caption{\textit{uv}-coverage of all four LOFAR observations utilised in this paper. These define the shape of the dirty beam. The $uv$-coverages in this plot include flagging and are plotted with conjugate $uv$ points. They also include the full frequency bandwidth, which produces the radial extent. These figures are made with the Python library \texttt{shadems}.}
\label{fig:uvcoverage}
\end{figure*}

The observations we have used have variations in the stations used. The observation with ID L816272 has the largest number of stations (52), as it includes also the Latvian station that only recently became operational \citep{latvia2020}. This adds more baselines longer than \textasciitilde1700 km with the stations located in Ireland and France (see Table \ref{table:baselines}). Observation IDs L686962 and L769393 have the same stations as L816272 but without the Latvian station. Observation L798074 includes the Latvian station but misses one German and the Polish station in Łazy, leading to the absence of the longest LOFAR baseline (see Table \ref{table:baselines}). The different combinations of LOFAR stations result in different $uv$-coverages, as displayed in Figure \ref{fig:uvcoverage}. The $uv$ sampling gaps between 80 and 180~km are due to the sparsity of LOFAR stations between the Dutch remote and German stations.

\begin{table*}[ht]
\centering
\begin{tabular}{c||c|c|c|c|c}
 & & & & & \\
& \textbf{Łazy} & \textbf{Birr} & \textbf{Bałdy} & \textbf{Nan\c{c}ay} & \textbf{Borówiec} \\
& (PL611HBA) & (IE613HBA) & (PL612HBA) & (FR606HBA) & (PL610HBA) \\
\hline \hline
 & & & & & \\
\textbf{Irbene} & X & 1930~km & X & 1735~km & X \\ 
(LV614HBA) & & & & & \\
\hline
 & & & & & \\
\textbf{Birr} & 1980~km & X & 1880~km & X & 1679~km \\
(IE613HBA) & & & & & \\
\end{tabular}
\caption{Five longest baselines between LOFAR stations according to the Euclidean distance of the Earth-centered coordinates of these stations. The station IDs are given between brackets.}
\label{table:baselines}
\end{table*}

\section{Calibration}\label{sec:calibration}

Our calibration strategy of all our observations builds upon the procedures described in \cite{morabito2022} and \cite{sweijen2022}, where we further refined parts of their calibration strategy. \cite{sweijen2022} averaged their data to a time resolution of 2~sec. Given that half of our data is averaged to 2~sec while the remainder is at a 1-second resolution, it follows that our data volume is about 6 times larger than the data from \cite{sweijen2022}. This introduces additional challenges regarding storage and computational demands, leading to our decision to utilize a high-throughput compute cluster named Spider for our full data processing.\footnote{\url{https://doc.spider.surfsara.nl}} Spider enables us to run many of our jobs embarrassingly parallel, which reduces the wall-clock time of our full data processing.

In the following subsections, we discuss the calibration workflow starting with downloading the data to arriving at the final DI and DD corrected solutions necessary for imaging. We will also highlight the improvements we have made compared to previous work.

\subsection{Initial Dutch calibration}\label{sec:initdutchcalibration}

\begin{figure*}[htbp]
 \sidecaption
    \includegraphics[width=12cm]{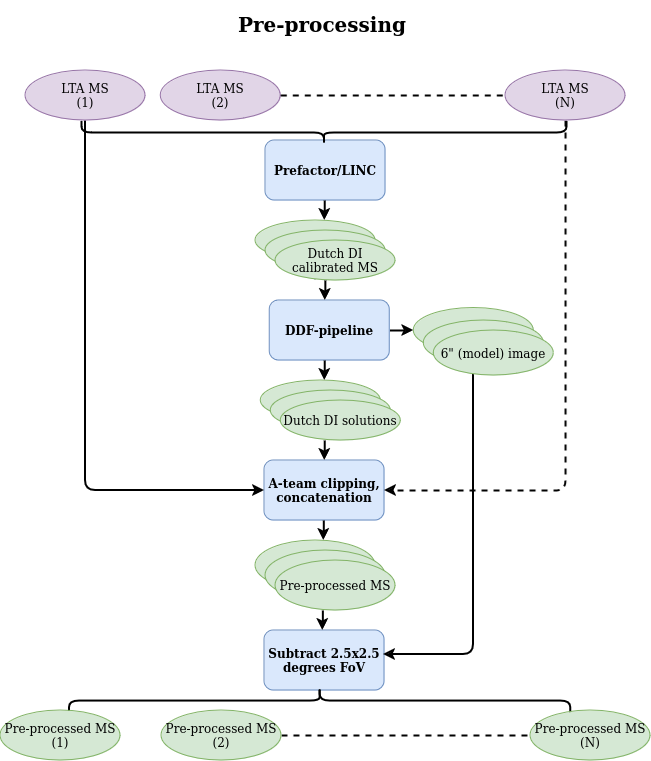}
  \caption{Workflow corresponding to the calibration steps explained in Section \ref{sec:initdutchcalibration} for the general case with \textit{N} observations.  The workflow starts with the $uv$-data pulled from the LTA and ends with pre-calibrated $uv$-data, ready for calibrating the international stations for \textit{N} different observations of the same field. Purple ovals are input data, blue boxes are operations on the data, and green ovals are output data. Stacked ovals imply that there are output products for each observation. Dashed lines indicate the presence of numerous observations that can run for this workflow in parallel. For a description of the calibration operations, we refer to Table \ref{table:calibration_operations}.}
\label{fig:pipeline0}
\end{figure*}

The first calibration steps focus on calibrating the $uv$-data of the Dutch stations. This follows the standard procedure, similar to LoTSS \citep{shimwell2017, shimwell2019, shimwell2022}, but with the goal to pre-process our data up to the stage where we
can start with the calibration of the international stations \citep{morabito2022, sweijen2022}. The steps, as described in this section, are summarized in the workflow from Figure \ref{fig:pipeline0}.

After downloading the data from the LTA, we ran the standard data reduction pipeline from \texttt{Prefactor}\footnote{Predecessor of the LOFAR Initial Calibration (LINC) pipeline.} on our four separate datasets \citep{vanweeren2016, gasperin2019}. This pipeline starts with a calibration of our primary calibrator 3C\,295. \texttt{Prefactor} corrects for all stations the phase differences between XX and YY polarisations, and derives constant clock offsets between the stations, and the bandpass. The pipeline proceeds with the target pipeline where the goal is to correct the Dutch stations of the target data of ELAIS-N1 for DI effects. This starts by transferring the calibrator solutions to the target data and removing the international stations to reduce the data volume. The pipeline involves flagging bad data and problematic stations, finding Faraday corrections with \texttt{RMextract} \citep{mevius2018}, and phase calibration against a sky model from the TIFR Giant Metrewave Radio Telescope (GMRT) Sky Survey \citep[TGSS,][]{intema2017}. This procedure results in the first DI solutions for all Dutch stations. We utilised the solution inspection plots from \texttt{LoSoTo}\footnote{\url{https://github.com/revoltek/losoto/}} to conduct a first assessment of the quality of our observations and to notify whether there were substantial parts of the data flagged or entire stations removed.

Using the output from the \texttt{Prefactor} target pipeline, we also ran the \texttt{DDF}-pipeline\footnote{\url{https://github.com/mhardcastle/ddf-pipeline}} to obtain DI and DD corrections, DDE-corrected images, and DDE-corrected models for the Dutch core and remote stations \citep{shimwell2019, tasse2021}. This pipeline uses \texttt{KillMS}\footnote{\url{https://github.com/saopicc/killMS}} \citep{tasse2014, tasse2014b, smirnov2015} to derive phase and amplitude corrections, which are applied during imaging with \texttt{DDFacet}\footnote{\url{https://github.com/saopicc/DDFacet}} \citep{tasse2018}. The resulting 6\arcsec~wide-field images of each of our 8-hrs observations were used to assess the quality of the corrections on the data from the Dutch stations and to gauge the calibratability of the ionosphere during each of our selected observations.

Following the strategy from \cite{morabito2022}, we prepare our data for international DI calibration by transferring the Dutch DI calibration solutions, predicting and flagging (part of) the response of bright off-axis so-called `A-team' sources (Cassiopeia A, Cygnus A, Taurus A, and Virgo A), and perform concatenation of datasets into subbands of 1.95~MHz. Unlike when processing just Dutch stations, we do not perform any averaging. Given that the full width at half maximum (FWHM) is smaller at sub-arcsecond resolutions, due to the size difference between Dutch and international LOFAR stations, we adopt a narrower field of view compared to the 6\arcsec~resolution. We made therefore use of the \texttt{DDF}-pipeline models and solutions to subtract sources outside a box of 2.5~$\times$~2.5 deg$^{2}$ centred on the pointing centre. This box size sets the field of view of our final image products. The subtraction step suppresses artefacts induced by sources outside this field of view. After performing these steps, we have a final total data volume of \textasciitilde12~TB after compression \citep{offringa2016}. This total consists of 4~TB for the two datasets with 1-sec resolution and 2~TB for the two datasets with 2-sec resolution. The data is now prepared for calibration with international stations using DI.

\subsection{Direction independent calibration of full array}\label{sec:infieldcal}

\begin{figure*}
 \sidecaption
    \includegraphics[width=12cm]{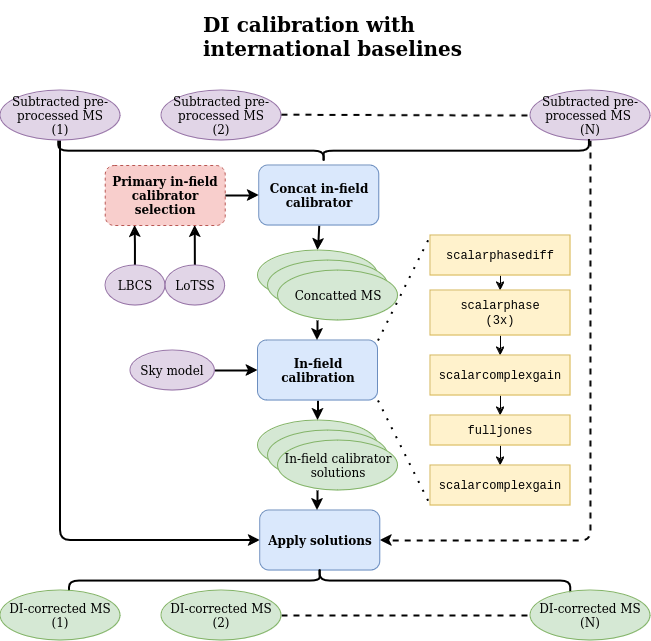}
  \caption{Workflow corresponding to the calibration steps explained in Section \ref{sec:infieldcal} for the general case with \textit{N} observations. The workflow starts with pre-processed $uv$-data and ends with DI-calibrated $uv$-data for \textit{N} different observations of the same field. These steps follow after the workflow in Figure \ref{fig:pipeline0}. Purple ovals are input data, blue boxes are operations on the data, red boxes are data filters, yellow boxes are calibration steps, and green ovals are output data. Stacked ovals imply that there are output products for each observation. Dashed lines indicate the presence of numerous observations that can run in parallel. For a description of the calibration operations we refer to Table \ref{table:calibration_operations}.}
\label{fig:pipeline1}
\end{figure*}

After obtaining the pre-calibrated data using existing pipelines, as described in the previous subsection, we proceed with the initial DI calibration of the international stations. This step is challenging as there are fewer suitable calibrators available with enough S/N compared to observations at lower resolutions \citep{morabito2022, jackson2022}. Finding the best fitting calibration strategy remains partly empirical and therefore needs additional attention. In Figure \ref{fig:pipeline1}, we illustrate the workflow starting with the pre-calibrated data, as detailed in Section \ref{sec:initdutchcalibration}, and concluding with the final DI calibrated data, discussed in this subsection.

\subsubsection{Direction independent calibrator selection}\label{sec:infieldselection}

An important step in the DI calibration is the selection of a suitable primary in-field calibrator. Not every bright source is a good primary in-field calibrator \citep{jackson2016, jackson2022}. Proxies for good DI calibrators are: 
\begin{itemize}[leftmargin=*,labelsep=5mm]
\item[$\square$] \textbf{S/N}: The source must exhibit high S/N on the longest baselines, ensuring sufficient signal to calibrate the phases and amplitudes of the data from the international stations. The source should ideally be one of the brightest within the field of view, having a peak intensity of at least \textasciitilde25~mJy~beam$^{-1}$ at 0.3\arcsec.
\item[$\square$] \textbf{Position}: The primary in-field calibrator needs to be well within the FHWM of the international station beam to avoid too much attenuation due to the primary beam. Therefore, it is desirable to have a source located within 1\degree~of the pointing centre.
\item[$\square$] \textbf{Polarisation (optionally)}: If the information is available, it is an advantage to select an unpolarised in-field calibrator, as this allows for polarisation calibration on the in-field calibrator; we explain this further in Section \ref{sec:dical_strategy}.
\end{itemize}
 Identifying the best in-field calibrator is essential as bad amplitude or phase corrections will be largely irreversible. This is due to the higher time and frequency resolution that we use when calibrating the primary in-field calibrator compared to the time and frequency resolution used when correcting DDEs, as we later discuss in Section \ref{sec:ddcal}.

Fortunately, we already knew from the calibration of ELAIS-N1 by \cite{ye2023} which source satisfied the in-field calibrator selection criteria above. For their selection, they used the Long-Baseline Calibrator Survey \citep[LBCS,][]{jackson2022}, and selected the Seyfert 2 galaxy identified by ICRF\,J160607.6+552135 \citep{charlot2020, sexton2022}. This source, with a compact flux density of \textasciitilde0.28~Jy at 140~MHz, is located about 0.8\degree~away from the ELAIS-N1 pointing centre. Moreover, there is no evidence to suggest that this source is polarized \citep{ruiz2021, callingham2023}.

\subsubsection{Sky model}\label{sec:skymodel}

\begin{figure}
  \includegraphics[width=1\linewidth]{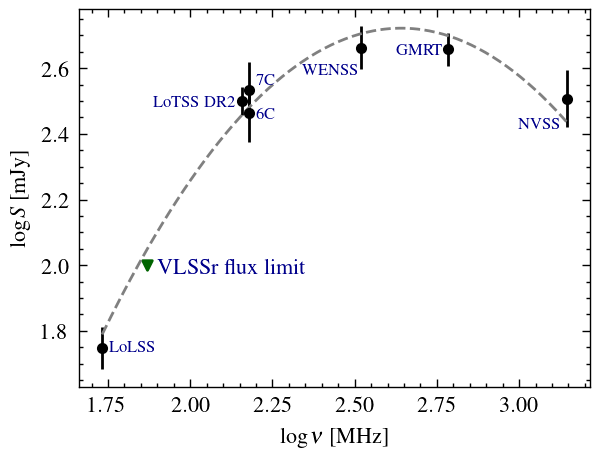}
  \caption{Fit of the radio spectrum of our primary in-field calibrator using data from WENSS at 330~MHz, the GMRT at 610~MHz, NVSS at 1.4~GHz, LoTSS DR2 at 144~MHz, the 6C and 7C surveys at 151~MHz, and the LoLSS image at 54~MHz constructed by us. We added the green downward triangle from the VLSSr flux density limit to illustrate the accuracy of the spectrum at lower frequencies.}
\label{fig:delayfit}
\end{figure}

For the calibration of the primary in-field calibrator, we constructed a point source sky model, as our source does not show any structure at sub-arcsecond scale. In order to determine the spectral index for our calibrator, we used the measured flux densities from observations by the NRAO VLA Sky Survey at 1.4~GHz \citep[NVSS;][]{condon1998}, the GMRT at 610~MHz \citep{garn2008}, the Westerbork Northern Sky Survey at 325~MHz \citep[WENSS;][]{rengelink1997}, LoTSS DR2 at 144~MHz \citep{shimwell2022}, and the 6C and 7C survey at 151~MHz \citep{vollmer2010}. We found the flux density to turn over between WENSS and LOFAR HBA frequencies, which led us to decide to better characterise the spectrum by processing an LBA observation at 54~MHz from the ELAIS-N1 field using the \texttt{LiLF}\footnote{\url{https://github.com/revoltek/LiLF}} calibration pipeline \citep{degasperin2018, degasperin2019, degasperin2020}. We have also added the flux density limit from the Very Large Array Low-frequency Sky Survey Redux \citep[VLSSr;][]{lane2014}. This supports the accuracy of our fitted spectrum. With the flux densities and frequencies, we fitted a second-order logarithmic polynomial
$$\log{S(\nu)} = \log{S_{0}} + c_{0}\log\left(\frac{\nu}{\nu_{0}}\right) + c_{1}\log\left(\frac{\nu}{\nu_{0}}\right)^{2},$$
where $S$ is the flux density as a function of frequency $\nu$. Using $\nu_{0}$ as the reference frequency at 141~MHz, we found $\log{S_{0}}=2.45$, $c_{0}=1.11$ and $c_{1}=-1.13$. This gave the fit shown in Figure \ref{fig:delayfit}. With these results, we obtained a flux density at 141~MHz of 0.28 Jy and the following spectral index as a function of frequency:
$$\alpha=\frac{\delta \log{S(\nu)}}{\delta \log{\nu}}=-2.26\log\left({\frac{\nu}{\nu_{0}}}\right) + 1.11.$$
Based on \cite{charlot2020}, we also have the coordinates of our in-field calibrator with a positional precision for the RA and DEC of dRA=1.88~mas and dDEC=1.83~mas respectively.
We used this information together with the source spectral index as input for our point source sky model. Our sky model ensures an accurate astrometry and flux density scale (as we later demonstrate in Sections \ref{sec:astrometry} and \ref{sec:fluxscale}).

\subsubsection{In-field calibration}\label{sec:dical_strategy}

Before performing any calibration on our selected in-field calibrator, we first phase-shifted the visibilities to the position of our calibrator source, after which we averaged the data down to 488~KHz and 32~sec, which decreases the data volume by a factor 640 or 1280 (depending on the original 2~sec or 1~sec data resolutions). This was followed by a primary beam correction at the position of the in-field calibrator. Applying the primary beam after averaging helps reduce computational time and is justified by the fact that the beam only varies very slowly as a function of time and frequency.
We express the operations to prepare our data for calibration mathematically by using the radio interferometry measurement equation~\citep[RIME;][]{hamaker1996, smirnov2011a, smirnov2011} as follows:
\begin{equation*}
    V_{\text{IF}}=\mathbf{B}_{\text{IF}} \langle \mathbf{P}_{\text{IF}}V_{\text{pre}}\mathbf{P}^{H}_{\text{IF}}\rangle \mathbf{B}^{H}_{\text{IF}},
\end{equation*}
where $V_{\text{pre}}$ are the visibilities with pre-applied Dutch solutions and the subtracted 6\arcsec~model outside our 2.5\degree$\times$2.5\degree~field of view. $\mathbf{B_{\text{IF}}}$ is the beam correction in the direction of the in-field calibrator, $\mathbf{P_{\text{IF}}}$ is the phase-shift to the position of the in-field calibrator, $V_{\text{IF}}$ the corresponding visibilities centred on the in-field calibrator as starting point of the calibration. The angular brackets represent the averaging over time and frequency, while $H$ denotes the conjugate transposed matrix.

\begin{figure*}
 \centering
    \includegraphics[width=0.95\linewidth]{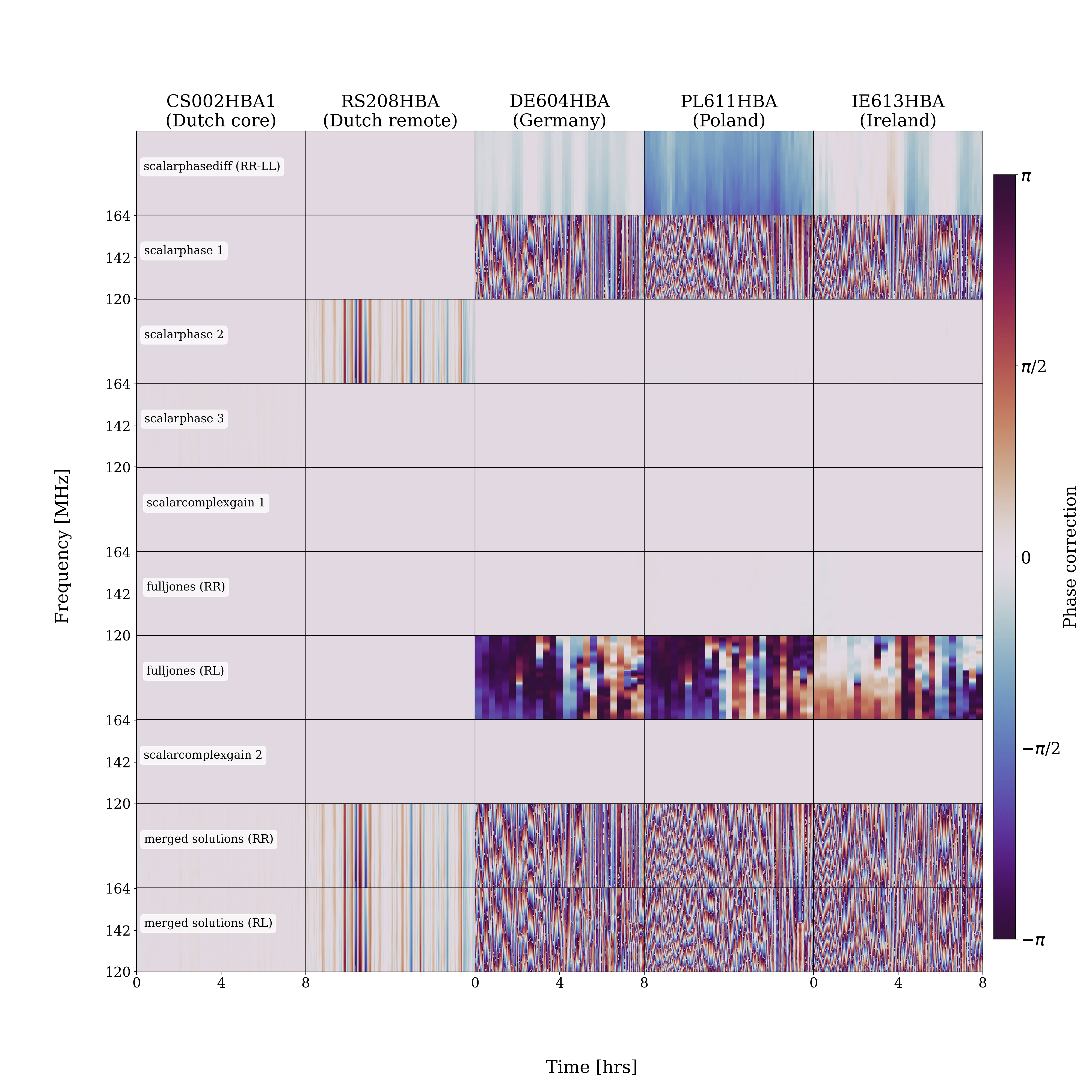}
  \caption{Phase calibration solution plots corresponding to the different calibration steps (\textit{rows}) and different stations, given by their station IDs (\textit{columns}), for calibrating the primary in-field calibrator. These solutions are relative to the CS001HBA0 Dutch core station. For the full-Jones corrections, we only show the RR and RL solutions. The solutions on the last two rows show how these solutions are combined into a final merged solutions. It is important to note that the \texttt{scalarcomplexgain} and \texttt{fulljones} corrections have small phase corrections for RR (and LL) polarisations, due to the fact that these are already corrected in the previous steps. However, the same \texttt{scalarcomplexgain} steps do correct significantly for amplitudes (see Figure \ref{fig:delay_amplitude_solutions}) and the \texttt{fulljones} step for the RL (and LR) polarisations.}
\label{fig:delay_phase_solutions}
\end{figure*}

\begin{figure*}[htbp]
 \centering
    \includegraphics[width=0.95\linewidth]{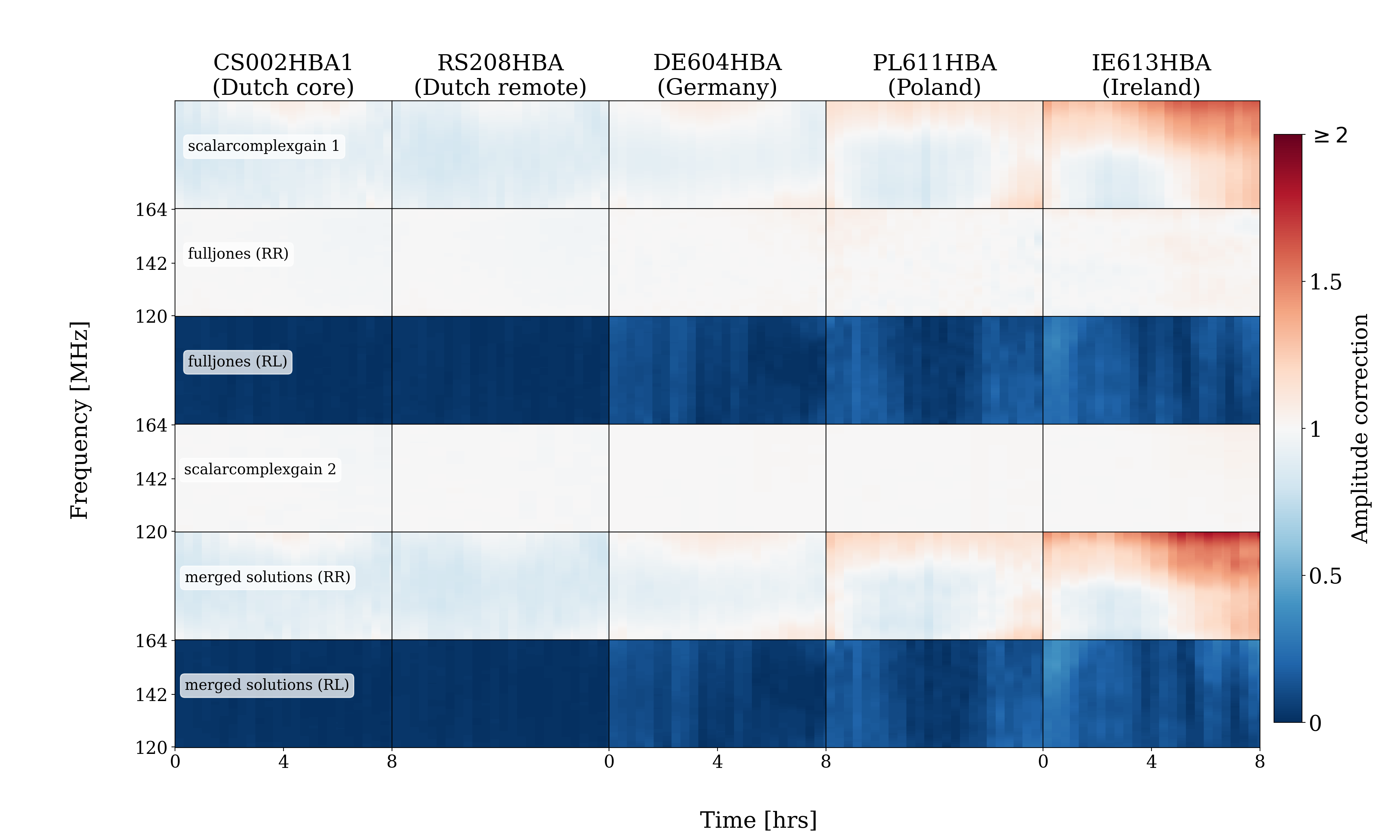}
  \caption{Amplitude calibration solution plots corresponding to the different calibration steps (\textit{rows}) and different stations, given by their station IDs (\textit{columns}), for calibrating the primary in-field calibrator. For the full-Jones corrections we only show the RR and RL solutions. The solutions on the last two rows show how these solutions are combined into a final merged solutions.}
\label{fig:delay_amplitude_solutions}
\end{figure*}

For the DI calibration of our in-field calibrator, we used \texttt{facetselfcal},\footnote{\url{https://github.com/rvweeren/lofar_facet_selfcal}} which utilizes the Default Preprocessing Pipeline (\texttt{DP3}, \citet{dp3, dijkema2023}) and \texttt{WSClean} \citep{wsclean} to perform (self-)calibration on a source. This calibration algorithm allowed us to derive the best phase and amplitude solutions on station level through minimisation of the difference between our sky model and the input visibilities.
\texttt{facetselfcal} uses an ``iterative-perturbative'' approach, where after each calibration step the model is adjusted with the new solutions before going to the next step. This procedure gives us full flexibility to incorporate our own calibration strategy to correct for different effects on different time, frequency, and antenna selections, on our data. The calibration steps used in this paper are described in Table \ref{table:calibration_operations}. To allow ourselves to tailor the magnitude of phase corrections as a function of time and frequency for different subsets of antennas, we split below the \texttt{scalarphase} calibration up into three separate iterations (\texttt{scalarphase} I, II, and III), where in each iteration we reset the solutions for a set of antennas to phase 0 and amplitude 1 values after running the calibration operation. After experimenting with various solution intervals, smoothness constraints, and calibration steps from \texttt{facetselfcal}, we found that the strategy described below performed best on our in-field calibrator source. This strategy is illustrated by a selection of solution plots from different LOFAR stations in Figures \ref{fig:delay_phase_solutions} and \ref{fig:delay_amplitude_solutions}.

\begin{table}
\caption{Description of the calibration operations used in this paper.}
\centering
\begin{tabularx}{\columnwidth}{lX}
\toprule
\textbf{Operation name} & \textbf{Description} \\
\midrule
\texttt{scalarphase} & Solving for phase errors as a function of time and frequency, independent of polarisation. \\ 
\midrule 
\texttt{scalarcomplexgain} & Solving for phase and amplitude errors as a function of time and frequency, independent of polarisation. \\ 
\midrule 
\texttt{fulljones} & Solving for phase and amplitude errors as a function of time and frequency for both diagonal and cross-hand polarisations simultaneously. \\ 
\midrule 
\texttt{scalarphasediff} & Derive a diagonal phase correction to eliminate any phase difference present between circular RR and LL (or linear XX and YY) polarisations as a function of time and frequency. \\
\bottomrule 
\end{tabularx}
\tablefoot{The names originate from the operation names used by \texttt{DP3} \citep{dp3, dijkema2023} and \texttt{facetselfcal} \citep{vanweeren2021}.}
\label{table:calibration_operations}
\end{table}

\begin{enumerate}[label=\textbf{\arabic*.}, leftmargin=*, labelsep=2mm, itemsep=2mm, font=\sffamily]
    \item \textbf{\texttt{scalarphasediff}}: Our in-field calibrator is unpolarised \citep{tremblay2016, ruiz2021, callingham2023}. The absence of a signal in Stokes~V polarisation enables us to employ \texttt{scalarphasediff} calibration in circular polarisation basis to correct for differential Faraday rotation after converting our data polarisation basis from linear to circular. We constrain the Dutch stations for this step to have the same solutions, as the effect of differential Faraday rotation is negligible on shorter baselines. We found a suitable solution interval for this step to be 8~min and the frequency to be best constrained by a smoothness kernel of 10~MHz. The varying calibration solutions for the international stations are illustrated in the first row of Figure \ref{fig:delay_phase_solutions}.
    \item \textbf{\texttt{scalarphase I}}: After having corrected the RR and LL polarisation phase difference, we derive polarisation-independent corrections for phase errors with the \texttt{scalarphase} solve. In the first \texttt{scalarphase} iteration, we solve for `fast' phase variations for the international stations by taking a solution interval of 32~sec and a small frequency smoothness kernel of 1.25~MHz. These are the smallest solution interval and frequency smoothness constraints, as we expect the largest phase variability across the longest baselines. The reset option setting the phase solutions to 0 and amplitude solutions to 1 for the Dutch core and remote stations results in only solutions for the international stations. On the second row in Figure \ref{fig:delay_phase_solutions} we see how the solutions corresponding to this step are wrapping fast from $-\pi$ to $\pi$ radians for the international stations.
    \item \textbf{\texttt{scalarphase II}}: In the second \texttt{scalarphase} iteration, we solve again for `fast' phase changes with a solution interval of 32~sec. However, we now include the Dutch remote stations by only resetting the solutions for the Dutch core stations to phase solutions equal to 0 and amplitude solutions equal to 1 after the solve. Compared to the previous \texttt{scalarphase} solve, we found a larger frequency constraint with a smoothness kernel of 10~MHz to work best. The solutions are most significant for remote stations because the phases for the international stations are already corrected, as is illustrated on the third row in Figure \ref{fig:delay_phase_solutions}.
    \item \textbf{\texttt{scalarphase III}}: In the third \texttt{scalarphase} iteration we solve for `slow' phase changes for all stations, including the Dutch core stations, by taking a solution interval of 20~min and without using a reset of solutions. With a smoothness kernel of 20~MHz we use a larger frequency constraint compared to the other two \texttt{scalarphase} iterations. The Dutch core stations observe a similar ionosphere and were already corrected for DI effects (see Section \ref{sec:initdutchcalibration}). This results in small corrections between these stations, as illustrated on the fourth row in Figure \ref{fig:delay_phase_solutions}. 
    \item \textbf{\texttt{scalarcomplexgain} I}: After correcting for phase errors, we also incorporate polarisation-independent phase and amplitude corrections by doing a `slow' \texttt{scalarcomplexgain} solve with a solution interval of 20~min. We constrained the frequency axis here by a smoothness kernel of 7.5~MHz. On the fifth row in Figure \ref{fig:delay_phase_solutions} we see that the phase corrections are negligible, due to the phase corrections from the previous iterations. The amplitude corrections on the first row of Figure \ref{fig:delay_amplitude_solutions} are most significant for the more distant international stations.
    \item \textbf{\texttt{fulljones}}: After having corrected for phases and amplitudes for the diagonal RR and LL polarisation directions, we also correct with a full-Jones correction for leakage in the RL and LR cross-hands. As we have already applied full-Jones DI corrections for the Dutch stations (see Section \ref{sec:initdutchcalibration}), we expect the leakage of Dutch stations to be similar. Hence, we constrained these stations to have the same value. This also boosts the calibration signal at these stations. We opt for solution intervals of 20~min and constrain the frequencies with a smoothness frequency kernel of 5~MHz. In Figures \ref{fig:delay_phase_solutions} and \ref{fig:delay_amplitude_solutions}, we find the most significant corrections for the off-axis polarisations of the international stations.
    \item \textbf{\texttt{scalarcomplexgain} II}: Finally, we performed an additional final round of scalar corrections, by using a slow \texttt{scalarcomplexgain} solve set to a solution interval of 40~min and a frequency smoothness constraint of 7.5~MHz. This step serves as a final verification to ensure the stability of the solutions. In Figures \ref{fig:delay_phase_solutions} (eighth row) and \ref{fig:delay_amplitude_solutions} (fourth row), we see that the corrections are minor compared to the solutions from the previous calibration steps. This confirms the reliability of the iterative calibration up to the full-Jones calibration.
\end{enumerate}

Throughout the calibration we ignored baselines with a length smaller than 20,000 times the wavelength ($\lambda$), by setting a constraint on the \texttt{uvmin} parameter. This is to prevent possible issues related to having an incomplete sky model. This \texttt{uvmin} corresponds to a largest angular scale (LAS) of \textasciitilde10\arcsec~at 140~MHz. Each calibration step returned an \texttt{h5parm} solution table. We merged all solutions derived for each of the four observations to obtain four final solution tables with phase and amplitude corrections. If we let $\mathbf{G_{\text{DI}}}$ represent the final solutions after merging all solutions, we express the RIME equation~to obtain the final DI calibrated visibilities on the $V_{\text{pre}}$ visibilities with pre-applied Dutch solutions as
\begin{equation}\label{eq:DI}
    V_{\text{DI}} = \mathbf{B}_{\text{IF}}^{-1}\mathbf{G}_{\text{DI}} \mathbf{B}_{\text{IF}} V_{\text{pre}}\mathbf{B}^{H}_{\text{IF}} \mathbf{G}^{H}_{\text{DI}}\mathbf{B}^{H-1}_{\text{IF}},
\end{equation}
where $\mathbf{B}_{\text{IF}}^{-1}$ is the inverse beam correction from the centre of the in-field calibrator back to the pointing centre of the ELAIS-N1 observation.


It is important to stress that the order of merging solutions is essential, as our \texttt{scalarphasediff} and \texttt{fulljones} corrections do not commute. This implies that we need to merge the solutions in the order of the steps we have iteratively solved for. Similarly, the order of applying the beam corrections ($\mathbf{B_{\text{IF}}}$ and $\mathbf{B_{\text{IF}}^{-1}}$) and the full-Jones solutions from ($\mathbf{G_{\text{DI}}}$) in Equation~\ref{eq:DI} are not commutative either, due to the fact that they are typically not simultaneously diagonalisable Jones matrices \citep{smirnov2011a, smirnov2011}.


\subsection{Direction-dependant calibration of full array}\label{sec:ddcal}

\begin{figure*}[!ht]
 \sidecaption
    \includegraphics[width=12cm]{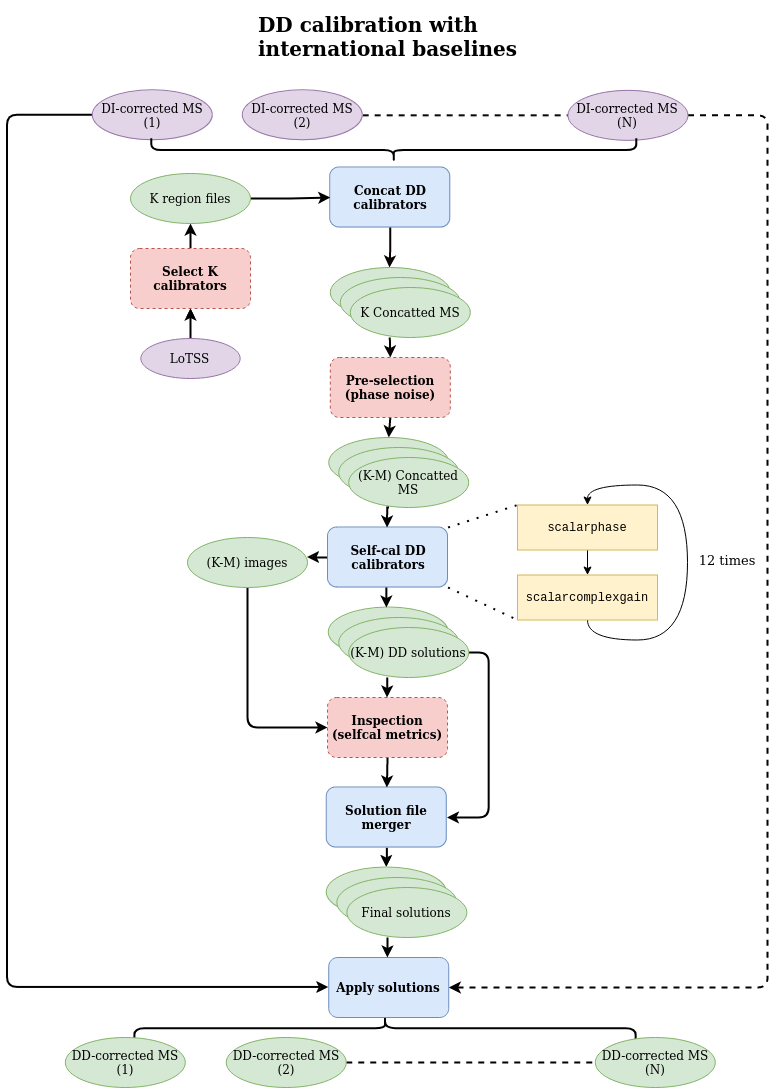}
  \caption{Workflow corresponding to the calibration steps explained in Section \ref{sec:ddcal} for the general case with \textit{N} observations. The workflow starts with DI-corrected $uv$-data and ends with DD-corrected $uv$-data for \textit{N} different observations of the same field. These steps follow after the workflow in Figure \ref{fig:pipeline1}. In the first source selection, based on the brightest sources from the LoTSS catalogue, we select $K$ sources, after which $M$ of these are filtered out during the phase noise selection metric (see Section \ref{sec:ddsourceselection}). This leaves us with $(K-M)$ solutions for each of the $N$ observations. We note that the \texttt{scalarcomplexgain} is only optionally triggered in \texttt{facetselfcal} for brighter sources \citep[See][]{vanweeren2021}. Purple ovals are input data, blue boxes are operations on the data, red boxes are data filters, yellow boxes are calibration steps, and green ovals are output data. Stacked ovals imply that there are output products for each observation. Dashed lines indicate the presence of numerous observations that can run in parallel. For a description of the calibration operations we refer to Table \ref{table:calibration_operations}.}
\label{fig:pipeline2}
\end{figure*}

The ionosphere and errors in the beam model introduce DDEs that corrupt the `real' visibilities across the field of view. These are not corrected by the DI calibration, as they depend on the direction of the calibration. We therefore divided the sky area up into smaller facets by selecting and calibrating for bright compact secondary calibrators distributed across ELAIS-N1 \citep{vanweeren2016, williams2016}.
The main challenge in the selection is that from the best existing radio images we only have source information available at 6\arcsec~resolution, while we need to find compact calibrators that have enough S/N to calibrate at 0.3\arcsec~resolution. It is therefore vital, after the initial selection and performing self-calibration on the DD calibrators, to examine both the calibration solutions and the resulting images to ensure that we have selected good calibrators with good calibration solutions. The workflow discussed in this subsection is illustrated in the diagram in Figure \ref{fig:pipeline2} for the general case of $N$ observations.

\subsubsection{Direction-dependant calibrator selection}\label{sec:ddsourceselection}

To initiate the search for compact sources, we used the ELAIS-N1 deep-field catalogue constructed from a 6\arcsec~resolution LOFAR HBA map \citep{kondapally2021, sabater2021}. From this catalogue we selected a sample of 86~sources with peak intensities above 25~mJy~beam$^{-1}$ inside our 2.5\degree$\times$2.5\degree~field of view. To investigate whether these sources may be good calibrators, the sources were first all split off by phase-shifting the DI corrected visibilities from Equation~\ref{eq:DI}. We averaged the phase-shifted data down to 32~sec and 390.56~kHz, which decreased the data volume by a factor 512 or 1024 (depending on the original 2~sec or 1~sec data resolutions). The averaging also reduced the effects from other nearby sources, without introducing smearing effects in our calibrator data. The full procedure can be expressed as
\begin{eqnarray*}
    V_{\text{S},n}&=&\mathbf{B}_{\text{S},n}\langle\mathbf{P}_{\text{S},n}\mathbf{B}_{\text{IF}}^{-1}\mathbf{G}_{\text{DI}}\mathbf{B}_{\text{IF}} V_{\text{pre}} \mathbf{B}^{H}_{\text{IF}}\mathbf{G}^{H}_{\text{DI}}\mathbf{B}^{H-1}_{\text{IF}}\mathbf{P}^{H}_{\text{S},n}\rangle\mathbf{B}^{H}_{\text{S},n}\\
    &=&\mathbf{B}_{\text{S},n}\langle\mathbf{P}_{\text{S},n}\mathbf{V_{\text{DI}}}\mathbf{P}^{H}_{\text{S},n}\rangle\mathbf{B}^{H}_{\text{S},n},
\end{eqnarray*}
where $V_{\text{S},n}$ are the visibilities after applying the DI corrections, phase-shifting and beam corrections in the direction of source $n$, and where we substituted Equation~\ref{eq:DI} on the second line.

We expect a significant fraction of the selected sources to be resolved out at 0.3\arcsec~resolution. If we were to run self-calibration naively on all 86 candidates and visually examine the results, it would not only cost extensive manual inspection time but it would also be computationally demanding. We therefore came up with a computationally cheap but reliable metric to identify which of our 86~candidate sources have enough S/N at the longest baselines, as we outline below.

For the selection metric we use the fact that circularly polarised sources are very rare at low frequencies, as
\cite{callingham2023} found in their 20\arcsec~V-LoTSS survey at 144~MHz only 68 circularly polarised sources across 5634~deg$^{2}$. One of their detections appears within our field of view \citep{callingham2021} but is not in our list of candidate calibrators. Considering that \cite{callingham2023} reports a completeness above 1~mJy and all our ELAIS-N1 calibrators have flux densities exceeding 25~mJy, it is reasonable to assume that none of our 86 calibrators are strongly circularly polarized. This implies that differences between corrections on RR and LL polarisations of our calibrators are attributed to the amount of noise on the solutions (ignoring polarisation leakage variations and the small effect of DD differential Faraday rotation, see further below). Therefore, calibrators with a high S/N will have more similar phase corrections on both RR and LL polarisations, whereas diffuse sources with a low S/N will exhibit noisier phase corrections.

To quantify the differences in RR and LL polarisations we first run one round of \texttt{scalarphasediff} calibration with \texttt{facetselfcal} (see Table \ref{table:calibration_operations}). 
To make the assessment consistent when comparing different sources, we use a fixed solution interval of 10~min. Since we are interested in the amount of S/N at the longest baselines, we only consider the \texttt{scalarphasediff} solutions from the Dutch and German stations. From the obtained solutions we take the discrete difference along the frequency axis to account for small differential Faraday rotation, after which we use the circular standard deviation as a measure of the phase noise. The circular standard deviation serves as an alternative to the traditional standard deviation to account for phase wrapping \citep[e.g.][]{mardia1972}. The formula for the circular standard deviation is given by
$$\sigma_{c} = \sqrt{-2 \ln{R}},$$
where $R$ is the mean resultant length given by
$$R=\sqrt{\bar{C}^{2}+\bar{S}^{2}},$$
with the mean cosine angles
$$\bar{C}=\frac{1}{N}\sum_{i=1}^{N}\cos{\theta_{i}}$$
and the mean sine angles
$$\bar{S}=\frac{1}{N}\sum_{i=1}^{N}\sin{\theta_{i}}$$
for $N$ phase solutions ($\theta$).
To test this metric and find a rejection threshold for the standard deviation, we applied this first on 40 sources from one of our four observations. With this sample, we empirically found sources below a circular standard deviation of 2.3~rad to be sufficient for self-calibration with the international LOFAR stations. Upon rejecting sources above this threshold from our initial 86~sources, we were left with 30 candidate sources. One of the 30 sources was less than 0.1\degree~away from a neighbouring selected calibrator, which made us decide to pick the source with the lowest circular standard deviation value. This procedure narrowed our selection down to 29 candidates.

This phase noise metric takes \textasciitilde1~CPU~hr for each source, which includes averaging down the $uv$-data to calculate the circular standard deviation. By implementing this metric we reduced the number of self-calibration runs by about a factor of 3, compared to running self-calibration on all 86~original candidates. This decreases the total computing time in our case of using 4 observations by \textasciitilde18,000 CPU~hrs. While this is a small fraction of the total computing costs (see Section \ref{sec:computingcosts}), it does remove a large part of the visual inspection when doing this fully automated (see also Appendix \ref{sec:automation}).

\subsubsection{Self-calibration}\label{sec:selfcal}

\begin{figure*}[htbp]
 \centering
    \includegraphics[width=0.95\linewidth]{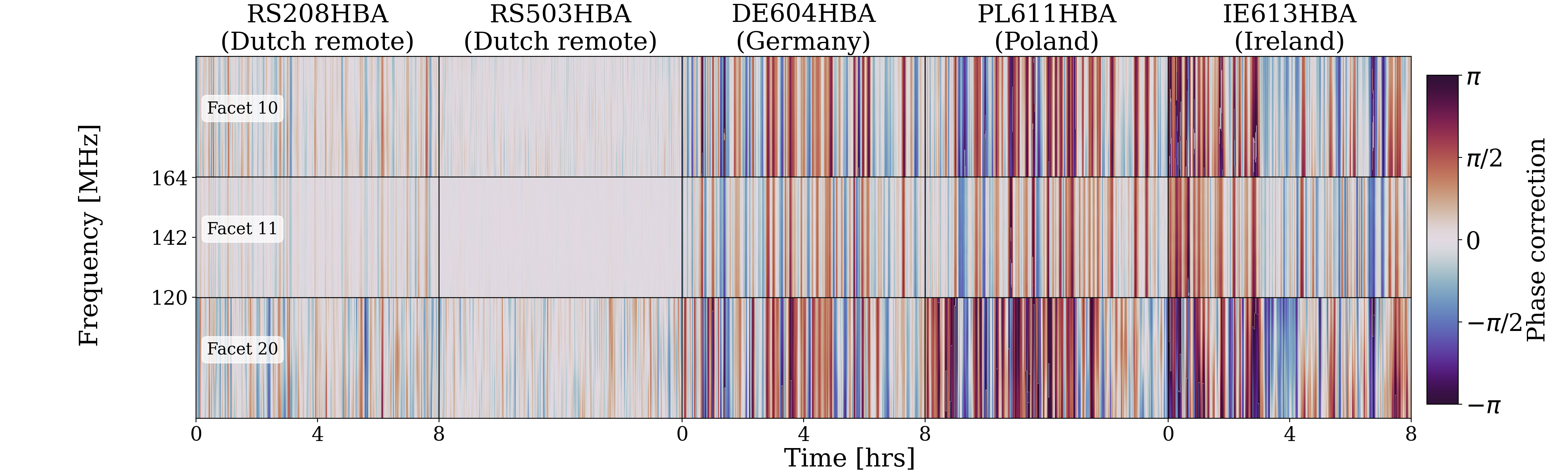}
  \caption{Merged phase calibration solution plots corresponding to the different facets (\textit{rows}) and different stations, given by their station IDs (\textit{columns}). These solutions are relative to the CS001HBA0 Dutch core station. The facets have the DD solutions from their corresponding calibrator, as depicted in Figure \ref{fig:facets}.}
\label{fig:dd_phase_solutions}
\end{figure*}

\begin{figure*}[htbp]
 \centering
    \includegraphics[width=0.95\linewidth]{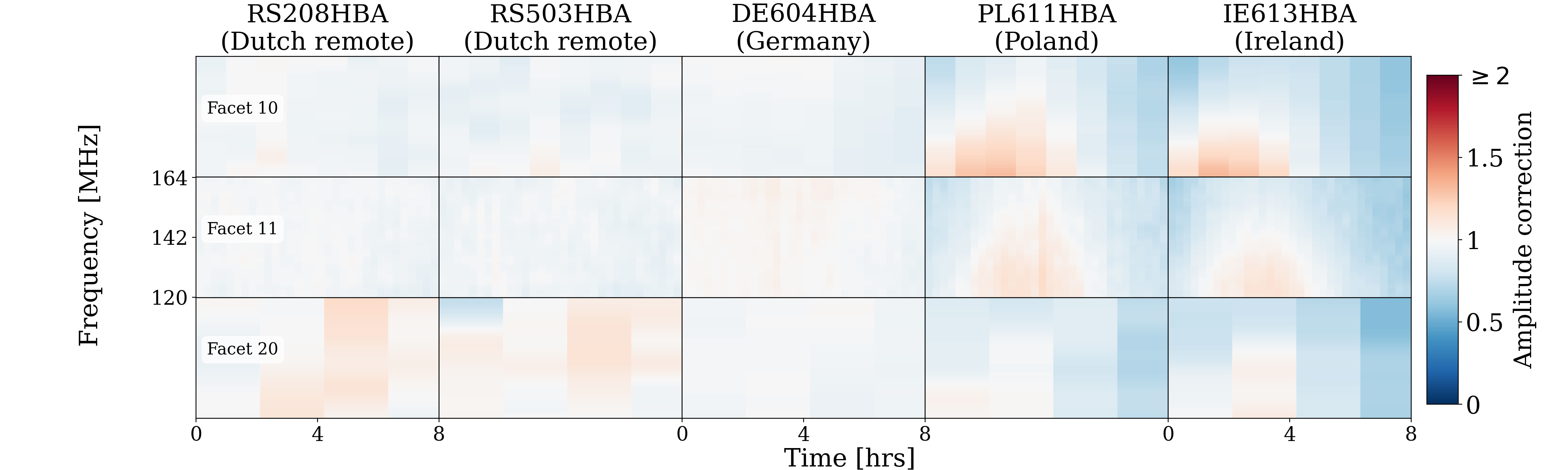}
  \caption{Merged amplitude calibration solution plots corresponding to the different facets (\textit{rows}) and different stations, given by their station IDs (\textit{columns}). The facets have the DD solutions from their corresponding calibrator, as depicted in Figure \ref{fig:facets}.}
\label{fig:dd_amplitude_solutions}
\end{figure*}

\begin{figure*}[!ht]
    \begin{minipage}{0.75\textwidth}
        \centering
        \includegraphics[width=0.72\textwidth]{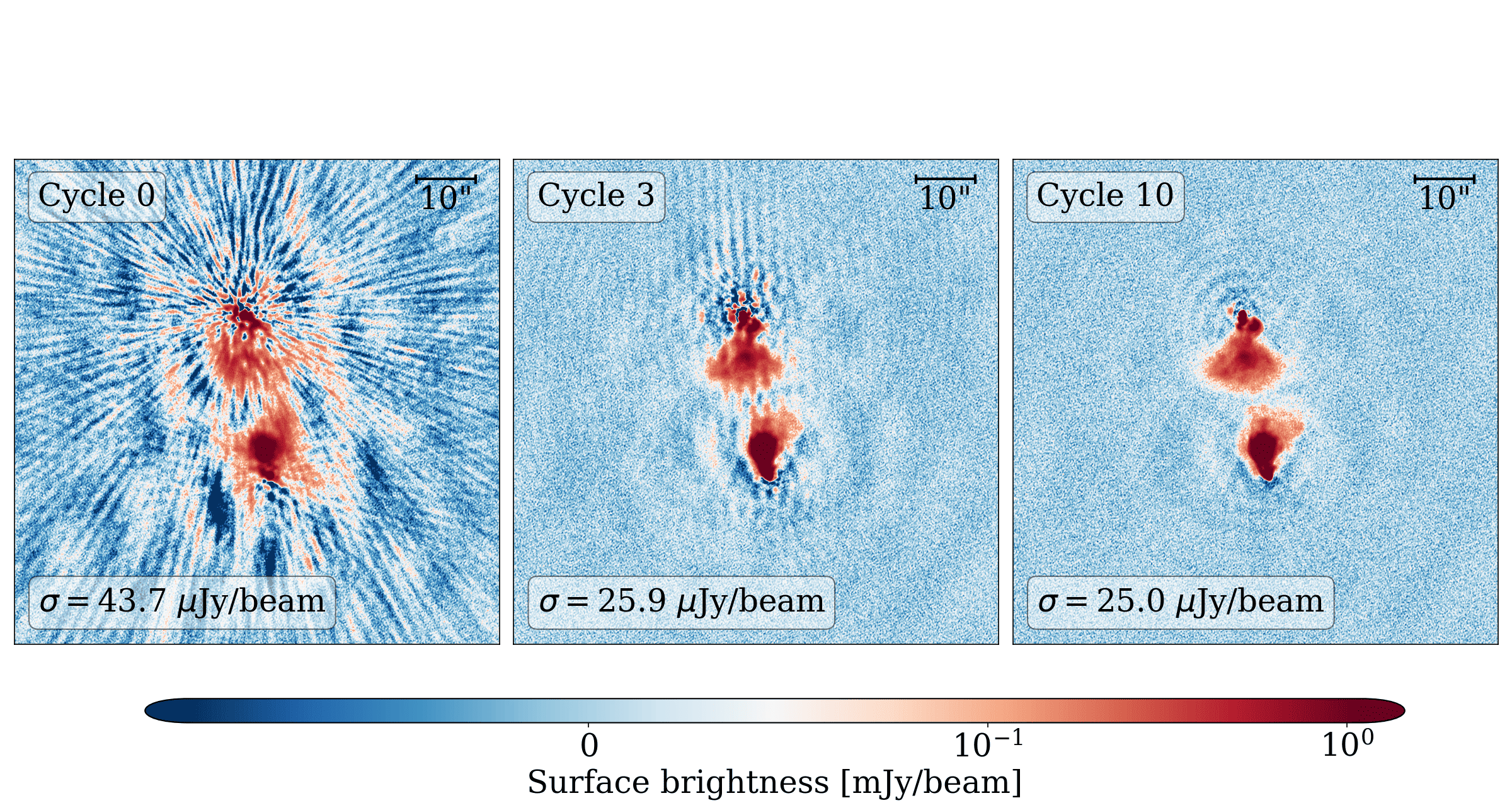}\\
        \includegraphics[width=0.72\textwidth]{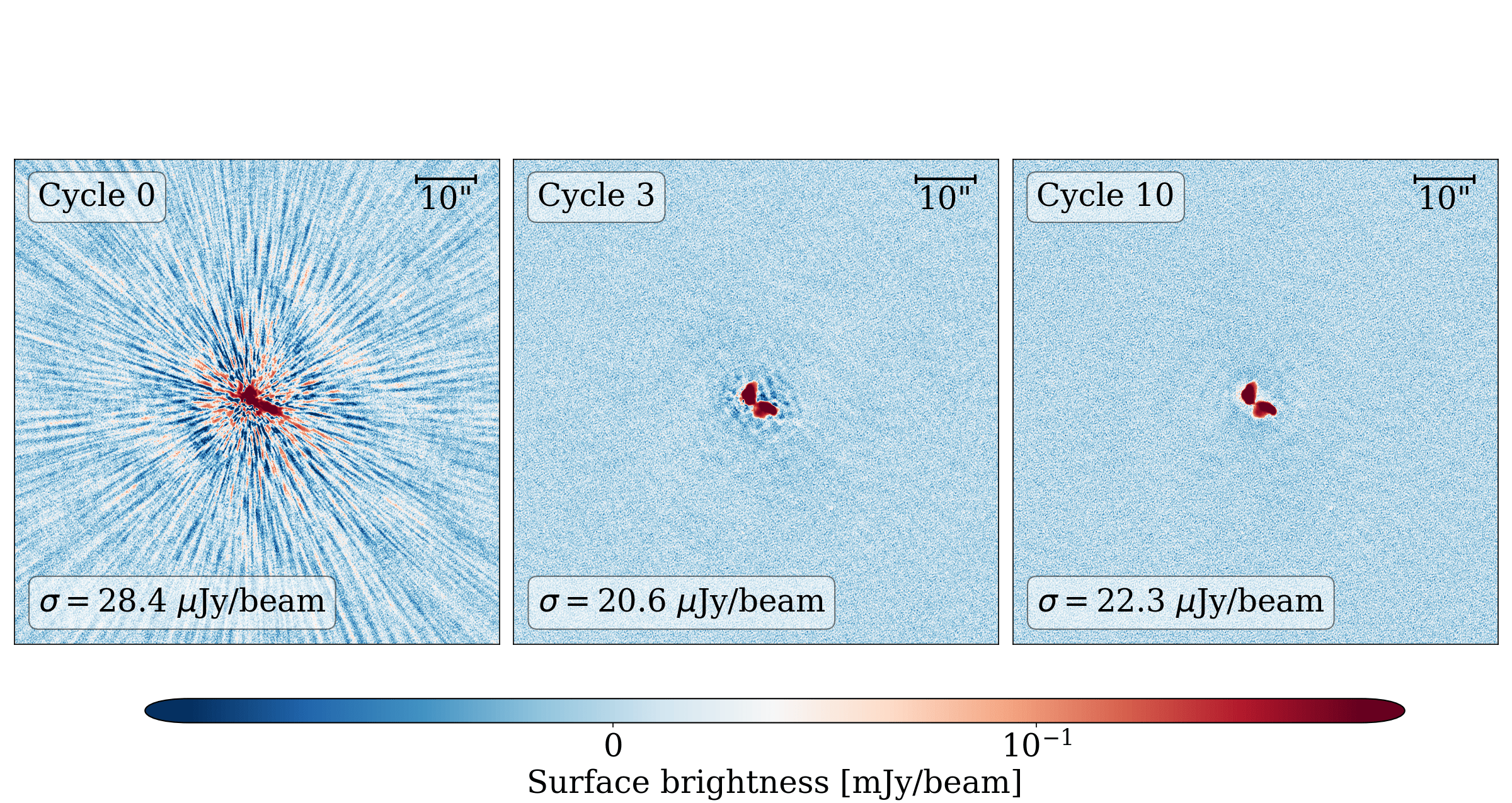}\\
        \includegraphics[width=0.72\textwidth]{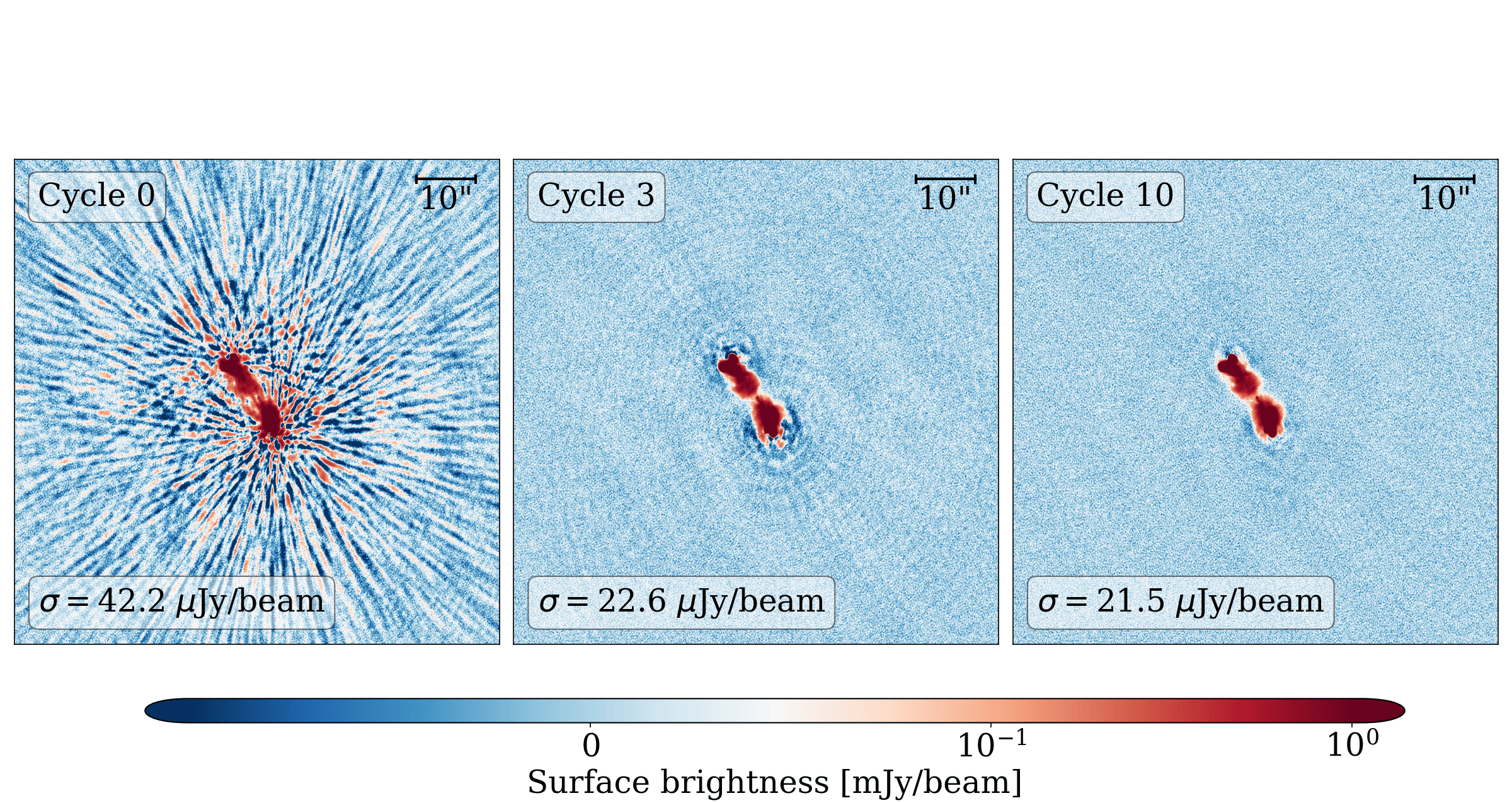}\\
        \includegraphics[width=0.72\textwidth]{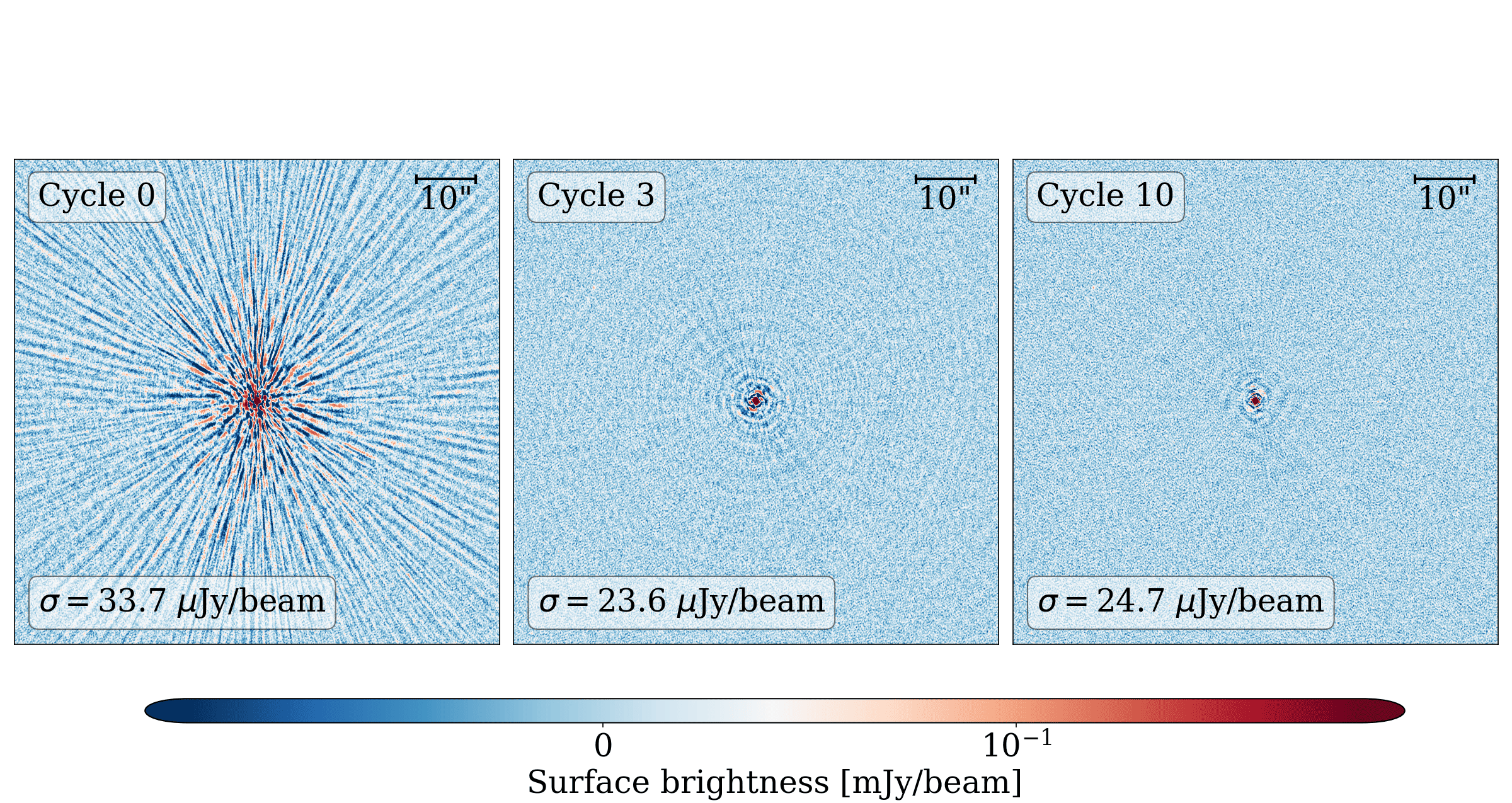}
    \end{minipage}\hfill
    \begin{minipage}{0.24\textwidth}
    \vspace{0.45\textheight}
        \caption{Four examples of self-calibration from our selected DD calibrators. Cycle 0 is the first image with only DI solutions applied from the in-field calibrator (see Section \ref{sec:infieldcal}). Cycle 3 corresponds to the self-calibration image after 3 rounds of \texttt{scalarphase} calibration. After this cycle, \texttt{scalarcomplexgain} calibration is added. This also calibrates for amplitude errors. Cycle 10 shows the result after the 10th self-calibration round. In some cases, the RMS noise (given by $\sigma$ in the figures) goes slightly up when comparing cycles 3 and 10. This is due to the introduction of amplitude corrections, which can cause slight increases or decreases in the overall local RMS values when for instance larger corrections for stations closer to the Dutch core are obtained. The angular size scale is indicated in the right top corner.}
        \label{fig:selfcals}
    \end{minipage}
\end{figure*}

For the remaining sources, we carried out up to 12 rounds of self-calibration by employing the \texttt{auto} option in \texttt{facetselfcal}. This calibration step is essential to calibrate for the ionospheric differences across the field of view. The number of cycles was set based on experience \citep{sweijen2022, ye2023}, as it has consistently been shown to achieve convergence for good calibrators. The \texttt{auto} setting automatically adjusts the solution intervals, and frequency smoothness constraints, among other parameters, based on the available flux density from the source, as discussed in \cite{vanweeren2021}. As we also have two observations that were pre-averaged by a factor two in time (see Section \ref{sec:datadescription}), we utilised an additional setting, \texttt{flagtimesmeared}, that flags visibilities where the amplitude reduction due to time smearing is more than a factor of two. This is especially important at the longest baselines and for calibrators more distant from the pointing centre, as these suffer most from smearing (as we see in Figure \ref{fig:theoretical_smearing} and later in Section \ref{sec:smearing}). Before the first calibration, we let \texttt{facetselfcal} also apply a phase-up of the Dutch core stations into a superstation in the centre of the Dutch array. This suppresses the signal from nearby sources on shorter baselines. To further tune the calibration for structures on small angular scales, we apply the same \textit{uv}-cut at 20,000$\lambda$ as used for the DI calibration of the in-field calibrator (corresponding to a LAS of \textasciitilde10\arcsec). The phase-up also reduces the data volume from each measurement by about 80\% \citep{morabito2022}, which therefore speeds up the self-calibration significantly. The \texttt{auto} setting performs calibration during the first four rounds for phases by applying a \texttt{scalarphase} solve, while the subsequent eight rounds might, depending on the available S/N from the calibrator, apply \texttt{scalarcomplexgain} calibration to find amplitude corrections as well \citep{vanweeren2021}. To make sure that the flux density scale, after applying amplitude corrections, is not drifting, we normalize the global amplitude corrections over all antennas and our four observations to one.

The final phase and amplitude solutions for a selection of stations from three selected DD calibrators are given in Figures \ref{fig:dd_phase_solutions} and \ref{fig:dd_amplitude_solutions}. This shows that we allow phases to have shorter solution intervals because we anticipate these to vary more rapidly over time than amplitudes. From the amplitude solutions we also see the solution interval size differences, which is because \texttt{facetselfcal} ensures that there is enough S/N on the longest baselines. To illustrate the self-calibration image quality, we show in Figure \ref{fig:selfcals} four examples of sources with self-calibration cycles 0 (no correction), 3 (phase correction only), and 10 (phase and amplitude corrections). These demonstrate how the phase and amplitude corrections have improved the image fidelity.

\subsubsection{Direction-dependant calibration inspection}\label{sec:ddsourceinspection}

\begin{figure*}[htbp]
 \centering
    \includegraphics[width=0.49\linewidth]{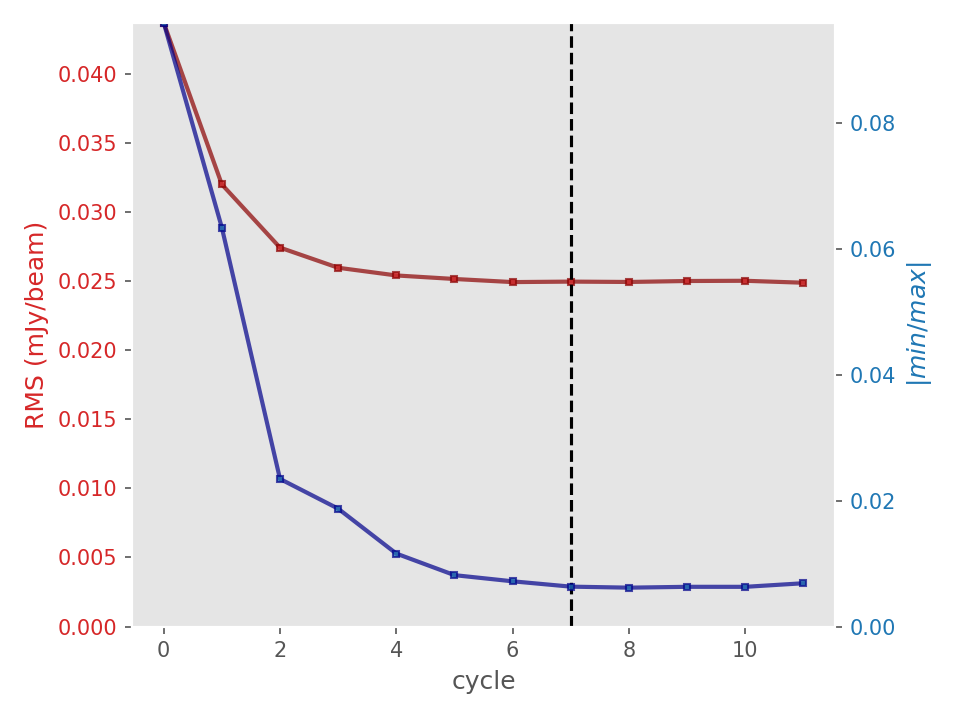}
    \includegraphics[width=0.49\linewidth]{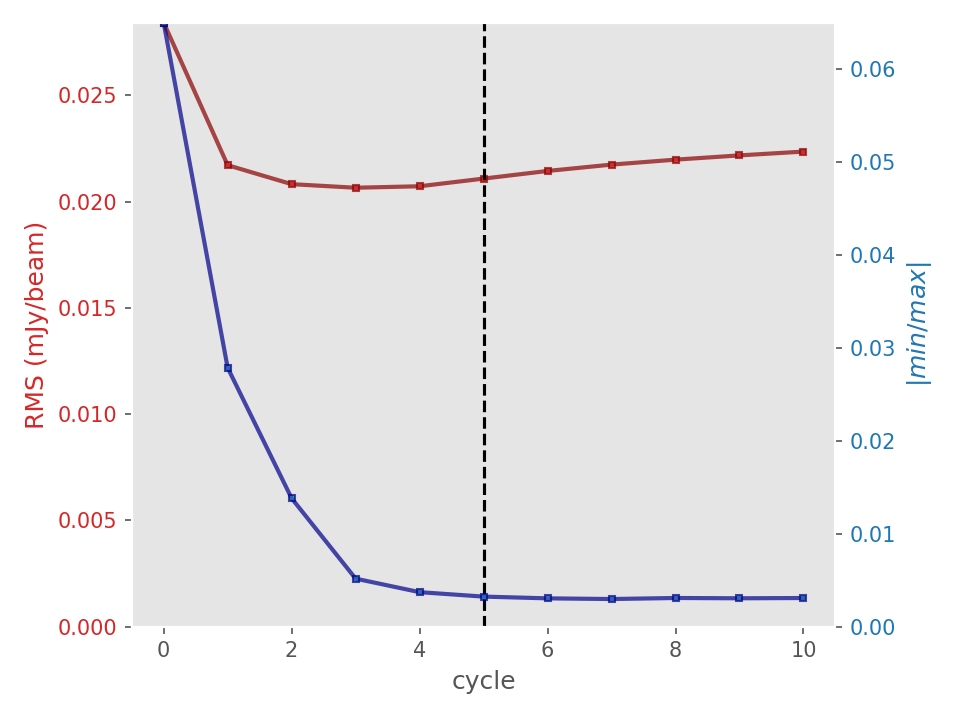}
  \caption{Self-calibration image stability for two different sources. The red line displays the progression of the RMS over self-calibration cycles, while the blue line represents the dynamic range (absolute min/max pixel) over self-calibration cycles. The black dashed line is the best calibration cycle according to a combined assessment of the solution and image stability. \textit{Left panel}: This example corresponds to the self-calibration cycles of the source in the first row of Figure \ref{fig:selfcals}. \textit{Right panel}: This example corresponds to the self-calibration cycles of the source in the second row of Figure \ref{fig:selfcals}.}
\label{fig:imagestability}
\end{figure*}

\begin{figure*}[htbp]
 \centering
 \includegraphics[width=0.49\linewidth]{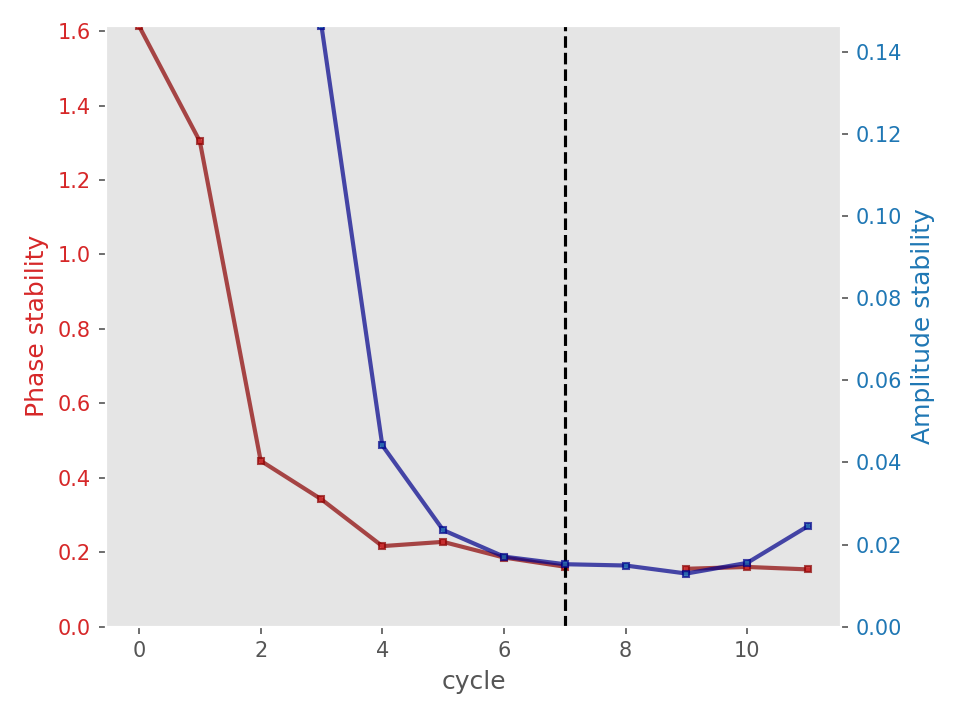}
    \includegraphics[width=0.49\linewidth]{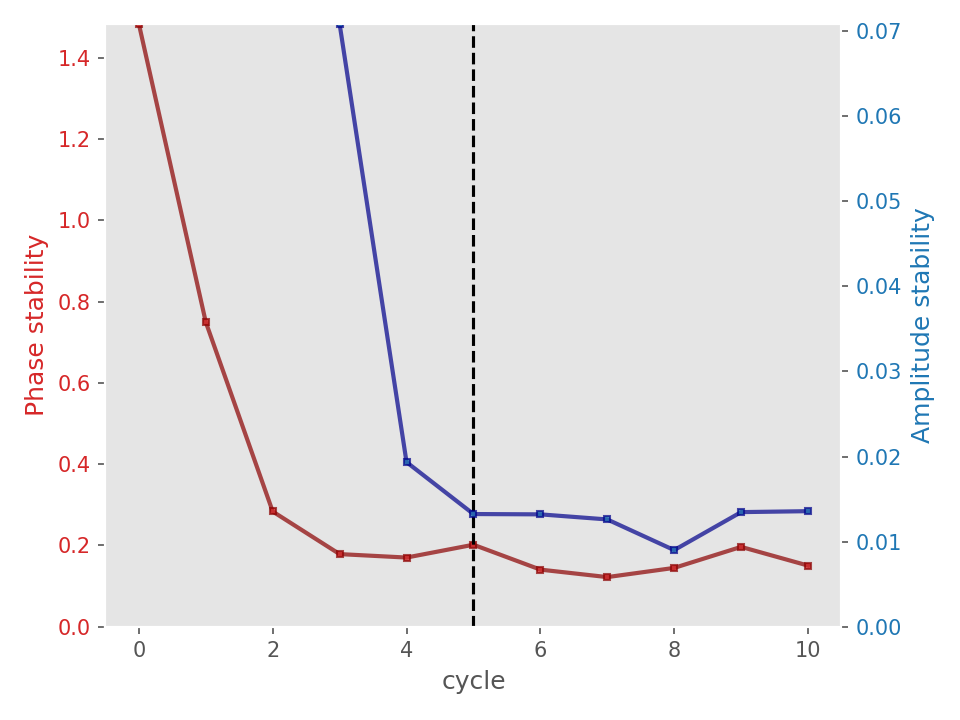}
  \caption{Self-calibration solution stability for two different sources (corresponding to the sources from Figure \ref{fig:imagestability}). The red line displays the circular standard deviation of the phase solution difference between the current and previous self-calibration cycle for each time and frequency value for each station. The blue line gives the standard deviation of the amplitude ratio of each time and frequency value for each station between the current and previous self-calibration cycle. We note that amplitude solves are only optionally triggered from cycle 3 onward, when the S/N is deemed sufficient by metrics from \texttt{facetselfcal}. The black dashed line corresponds to the selected calibration cycle, based on a combined assessment of the solution and image stability.}
\label{fig:solutionsstability}
\end{figure*}

Although self-calibration is well-established \citep[e.g.][]{cornwell1999} and \texttt{facetselfcal} has proven to be reliable to calibrate our best candidate DD calibrators \citep[e.g.][]{vanweeren2021, dejong2022, sweijen2022, ye2023}, it is essential to perform a final quality control on the self-calibration output products, as was done by \cite{ye2023}. This ensures our phase noise selection metric discussed in Section \ref{sec:ddsourceselection} did not include any false-positives and the calibration algorithm from \texttt{facetselfcal} performed as expected.

We inspect for each self-calibration cycle of each source the following characteristics:
\begin{itemize}[leftmargin=*,labelsep=5mm]
  \item[$\square$] \textbf{RMS noise}: We expect for self-calibration improvements on calibrators with compact emission the RMS to be significantly lower compared to the image that is only calibrated with DI solutions (cycle 0 in Figure \ref{fig:selfcals}). In the left panel of Figure \ref{fig:imagestability} we have an example of a stable improvement of the RMS. The source corresponding to the curve in the right panel has an RMS that is going up after cycle 4, which is due to the effects of the \texttt{scalarcomplexgain} calibration lifting the amplitude values. Although the increase is minor, this example demonstrates that relying exclusively on the image RMS to assess the self-calibration quality is insufficient.
  \item[$\square$] \textbf{Dynamic range}: Since the RMS does not fully convey the quality of the image, we also evaluate the dynamic range of the self-calibration images. We define the dynamic range in the figures as the absolute value of the minimum pixel value divided by the maximum pixel value. We expect for image improvements the most negative pixels to get closer to 0, which improves our measure of the dynamic range. Both cases in Figure \ref{fig:imagestability} corresponding to two of our DD calibrator sources show dynamic range improvements.
  \item[$\square$] \textbf{Solution stability}: Another important metric is the stability of the solutions over self-calibration cycle, as we expect the solutions to converge over self-calibration cycles. To assess this, we subtract the phase solutions for each time and frequency solution value between two consecutive self-calibration cycles and take the circular standard deviation as a measure of the stability. Similarly, to examine the behaviour of the amplitude solutions, we calculate the standard deviation of the ratio between solutions from two consecutive self-calibration cycles. Both measures should for stable self-calibration converge to small values, depending only on the solution noise. This converging behaviour is illustrated in Figure \ref{fig:solutionsstability} for the same sources as in Figure \ref{fig:imagestability}.
\end{itemize}
Using these metrics, we can quickly assess both the quality of self-calibration and the best self-calibration cycle (see Appendix \ref{sec:automation} for automatic approaches). This aligns with the ad-hoc calibrator selection criteria implemented by \citet{ye2023}. Among our 29 self-calibrated sources we did not find any diverging behaviour. These results also reassured us that the phase noise selection metric, discussed in Section \ref{sec:ddsourceselection}, did not select false-positive candidates.

During testing of our selection metrics, we did also run self-calibration on some of the sources that were above our phase noise selection threshold (see Section \ref{sec:ddsourceselection}). Although sources with scores close to our selection threshold did slightly improve after self-calibration, were the corrections too small to sufficiently improve the DDEs within a facet. In Appendix \ref{appendix:selfcalibrationissues} we discuss two examples with strong divergent calibration behaviour. These illustrate the effectiveness of performing an additional inspection of the image and solution quality.

\subsubsection{Facet layout}

After selecting the best self-calibration cycles of each source, we merge all solutions from each direction together into one multi-direction solution file for each of our observations, stored in \texttt{HDF5} format \citep{hdf5}. We also add phase and amplitude solutions of 0 and 1, respectively, from our primary in-field calibrator, as we did not have to apply self-calibration on this source after doing the thorough calibration, which includes DD calibration, as discussed in Section \ref{sec:dical_strategy}. The positions of the 30 calibrator sources (1 primary in-field calibrator source and the 29 DD calibrator sources) determine our facet layout through a Voronoi tessellation. This assigns each point in our field of view to the solutions of their nearest calibrator source. Across each facet, we assume that calibration solutions are constant \citep{schwab1984, vanweeren2016}. 

\begin{figure*}[htbp]
 \sidecaption
    \includegraphics[width=12cm]{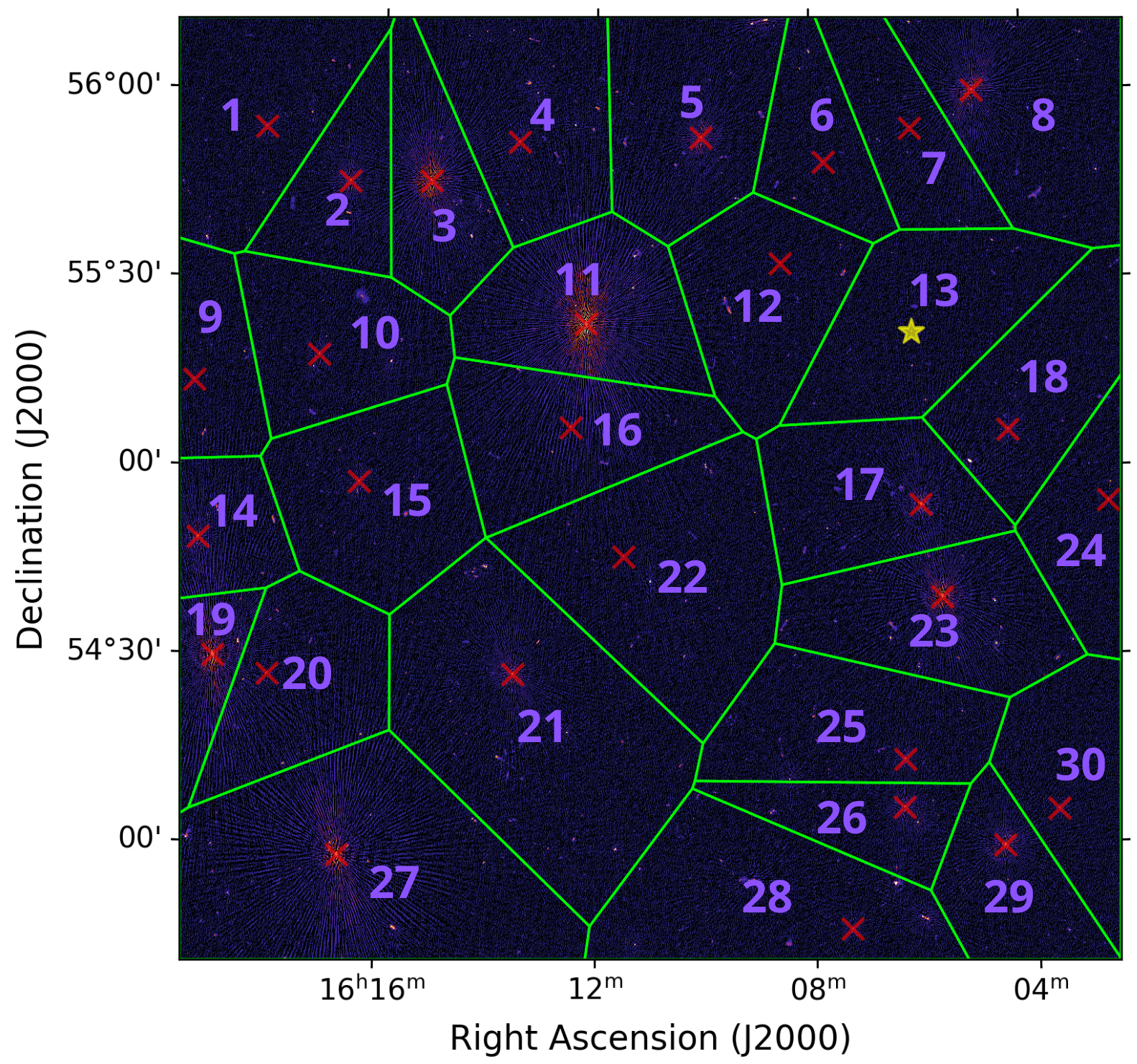}
  \caption{Final facet-layout on top of the 1.2\arcsec~DI image, created by applying DI corrections on data from one of our observations. The yellow star indicates the position of the primary in-field calibrator, the red crosses correspond to the DD calibrators, and the green boundaries show the Voronoi tessellation corresponding to these calibrators. The numbers are used for reference throughout this paper.}
\label{fig:facets}
\end{figure*}

In Figure \ref{fig:facets} we show an 1.2\arcsec~DI image from one of our observations, which we produced with \texttt{WSClean}\footnote{\url{https://gitlab.com/aroffringa/wsclean}} \citep{wsclean}, after applying the solutions from our DI calibrator. On top of this image, we projected the Voronoi tessellation corresponding to our 30 calibrators. The figure illustrates the successful correction of DDEs near the DI calibrator (indicated by the yellow star) and highlights how strong the DDEs are around our selected calibrator. Imaging including our final DD solutions is discussed in Section \ref{sec:imaging}.

\subsubsection{Refining Dutch calibration}\label{sec:extradutch}

\begin{figure*}[!ht]
    \sidecaption
    \includegraphics[width=12cm]{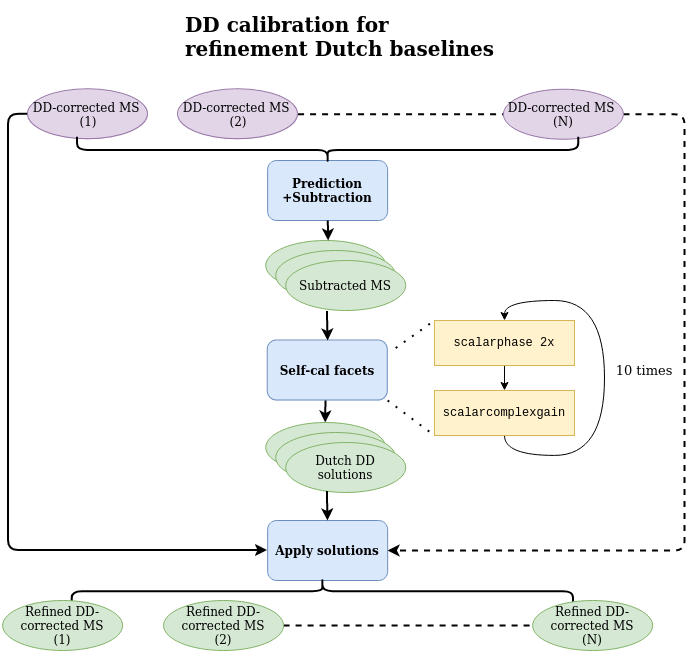}
        \caption{Workflow corresponding to the calibration steps explained in Section \ref{sec:extradutch} for the general case with \textit{N} observations. The workflow starts with the DD-corrected $uv$-data and ends with the refined DD-corrected $uv$-data for \textit{N} different observations of the same field. These steps follow after the workflow in Figure \ref{fig:pipeline2}. The prediction and subtraction steps are in more detail explained in Section \ref{sec:imaging_method}. Purple ovals are input data, blue boxes are operations on the data, red boxes are data filters, yellow boxes are calibration steps, and green ovals are output data. Stacked ovals imply that there are output products for each observation. Dashed lines indicate the presence of numerous observations that can run in parallel. For a description of the calibration operations we refer to Table \ref{table:calibration_operations}.}
        \label{fig:pipeline3}
\end{figure*}

\begin{figure*}[htbp]
 \centering
    \includegraphics[width=0.95\linewidth]{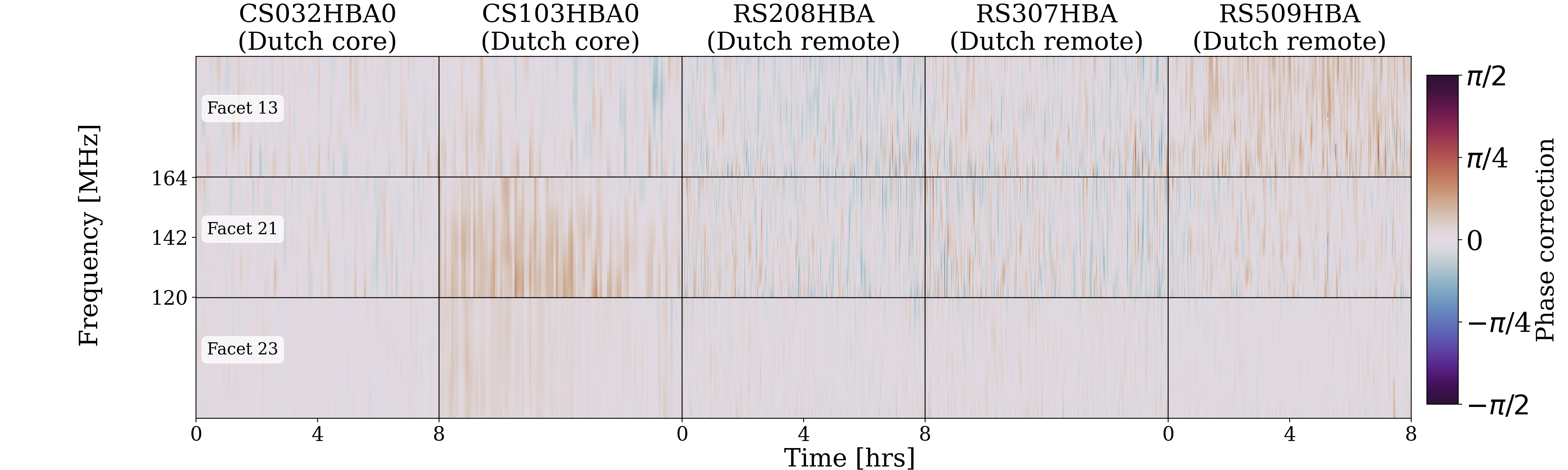}
  \caption{Merged phase calibration solution plots corresponding to the different facets (\textit{rows}) and different stations, given by their station IDs (\textit{columns}). These solutions are relative to the CS001HBA0 Dutch core station. The facets have the DD solutions from their corresponding calibrator, as depicted in Figure \ref{fig:facets}.}
\label{fig:dutch_phase_solutions}
\end{figure*}

\begin{figure*}[htbp]
 \centering
    \includegraphics[width=0.95\linewidth]{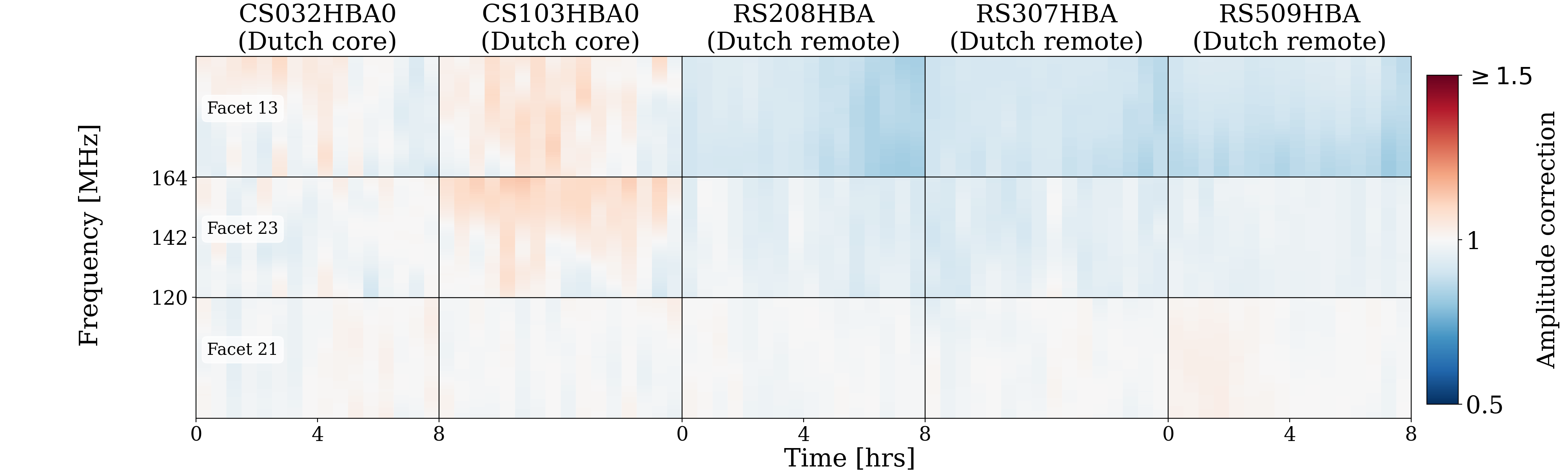}
  \caption{Merged amplitude calibration solution plots corresponding to the different facets (\textit{rows}) and different stations, given by their station IDs (\textit{columns}). The facets have the DD solutions from their corresponding calibrator, as depicted in Figure \ref{fig:facets}.}
\label{fig:dutch_amplitude_solutions}
\end{figure*}

\begin{figure*}[htbp]
 \centering
     \includegraphics[width=0.8\linewidth]{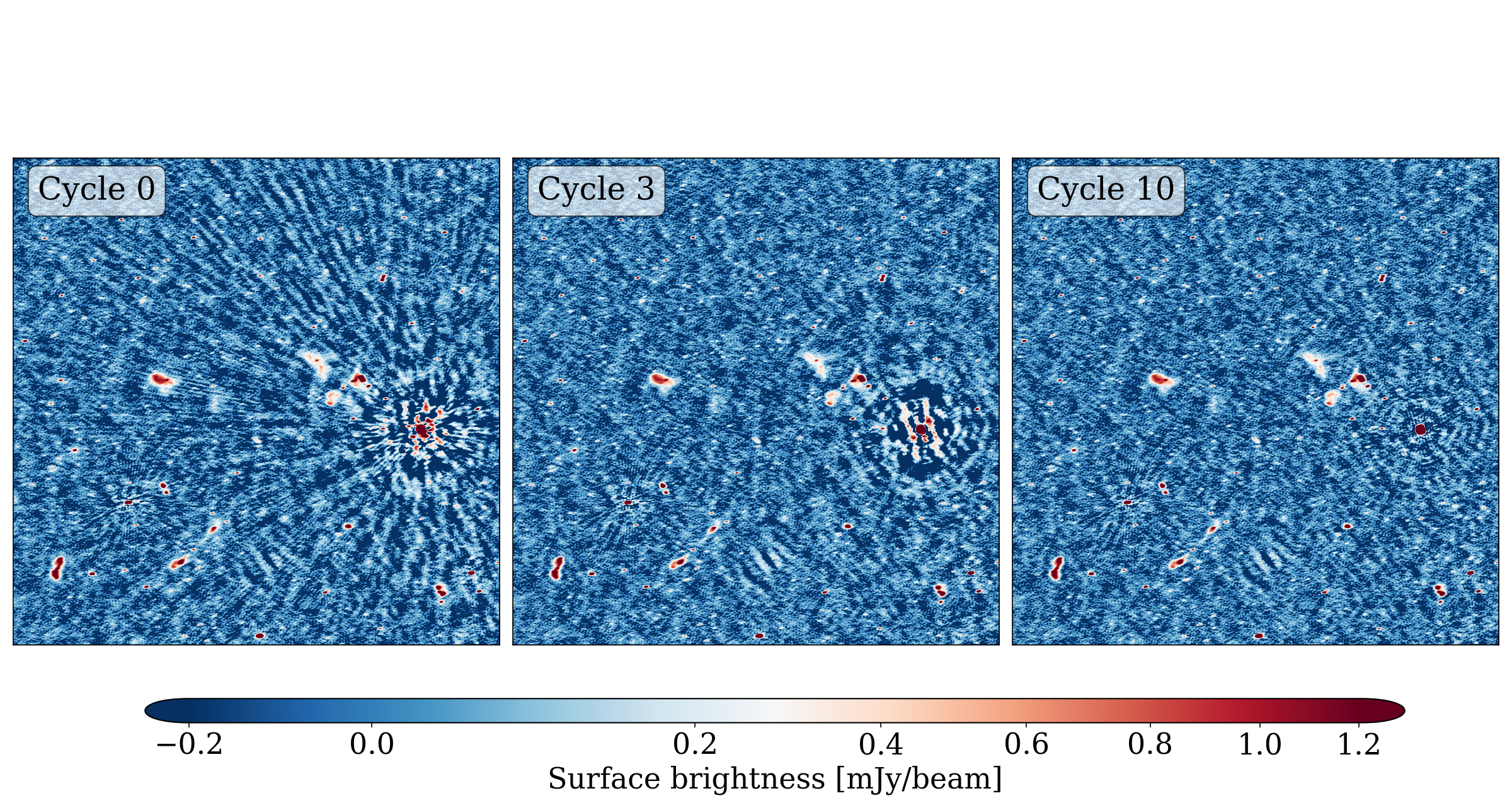}
  \caption{Example of the image improvements after applying extra self-calibration with only Dutch core and remote stations at 6\arcsec~on facet 17 (see Figure \ref{fig:facets}) in a 0.4\degree$\times$0.4\degree~cutout image. Cycle 0 has only DI solutions applied, cycle 3 had the first round of \texttt{scalarcomplexgain} solve where both phases and amplitudes are calibrated, while cycle 10 shows the result after the final round of self-calibration.}
\label{fig:6asec_selfcal}
\end{figure*}

\begin{figure*}[htbp]
 \centering
     \includegraphics[width=0.85\linewidth]{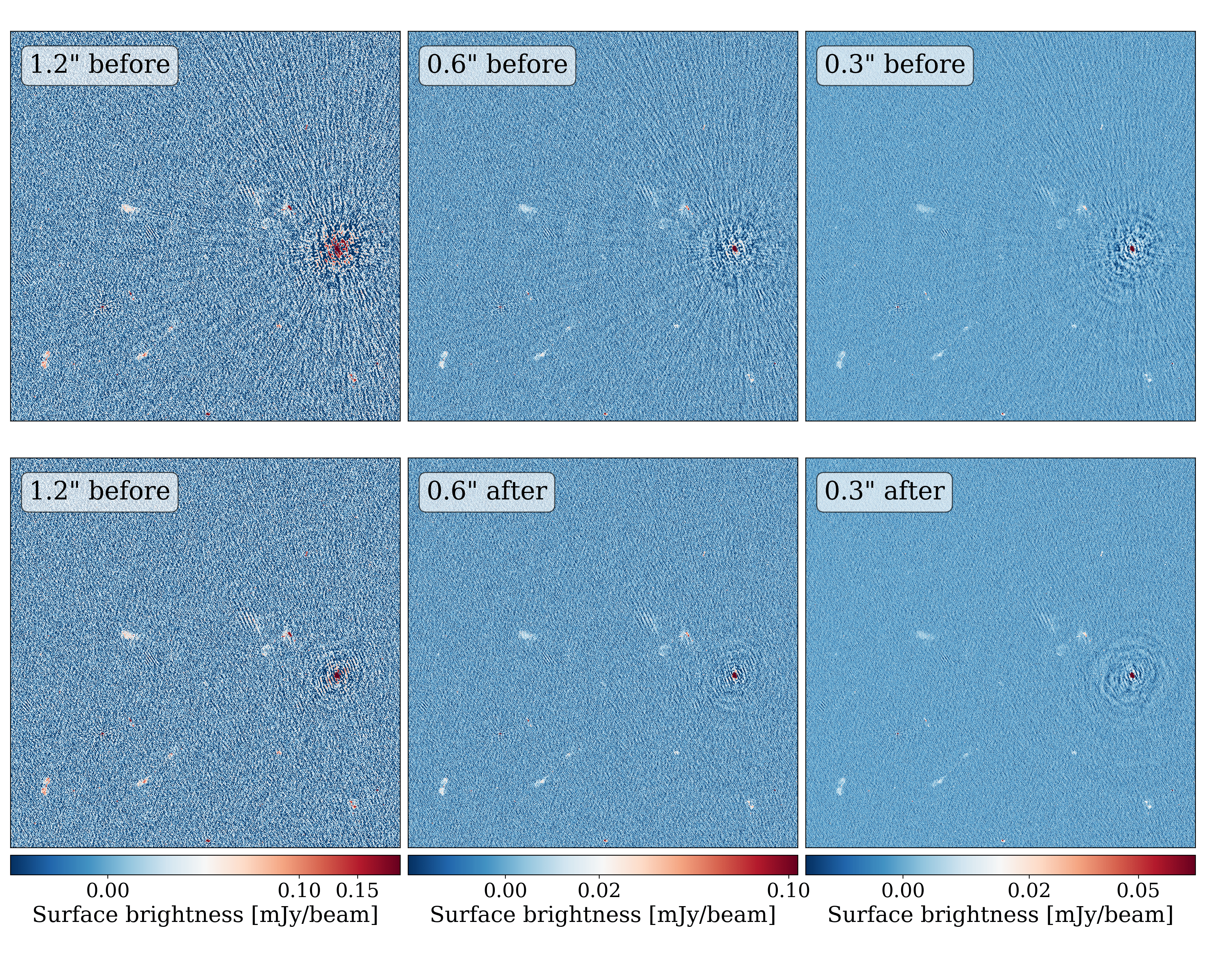}
  \caption{Image quality changes across four resolutions after applying extra self-calibration with only Dutch core and remote stations on the calibrator from facet 17 (see Figure \ref{fig:facets}) in a 0.4\degree$\times$0.4\degree~cutout image. The top row displays part of the facet images before adding the extra Dutch solutions and the bottom row displays the results after adding the extra Dutch solutions.}
\label{fig:compare_res_improvements}
\end{figure*}

Our calibration strategy is specifically designed to enhance the calibration solutions for international stations, by incorporating the phase-up of the Dutch core stations and excluding baselines shorter than 20,000$\lambda$ (see Section \ref{sec:selfcal}). However, after applying DD calibration corrections, we observed strong artefacts from 18 of our 30 calibrators when we created facet images at 6\arcsec. This arises from bright sources outside the facet boundaries that affect the calibration of shorter baselines during DD self-calibration. Adding subtraction of sources near our calibrators before DD calibration, using for instance the 6\arcsec~models, would have been too expensive, as this brings an additional cost of \textasciitilde800,000 CPU~hrs. However, as we are using subtraction around our facets before imaging (see Section \ref{sec:imaging_method}), we could still further refine the calibration solutions for the shorter baselines, as we outline below. The workflow for refining the solutions corresponding to the Dutch LOFAR stations, discussed in this subsection, is also presented in the diagram in Figure \ref{fig:pipeline3} for the general case of $N$ observations.

To suppress calibration issues introduced by bright sources beyond the facet boundaries, we subtracted sources from our visibility data that correspond to sources outside our facets, using image models at 1.2\arcsec. These model images were produced for each of our four observations with the merged DD solutions discussed in Section \ref{sec:selfcal}. As the subtraction of sources is part of our imaging procedure, we refer to Section \ref{sec:imaging_method} for further discussion about this process. After subtraction, we conducted extra rounds of self-calibration on the entire facet with only the Dutch core and remote stations, using a \textit{uv}-cut of 750$\lambda$ (corresponding to a LAS of \textasciitilde275\arcsec). Since we calibrate at 6\arcsec~and our facets are smaller than the entire wide-field image, we are allowed to average our data to 20~sec and 244~kHz and apply on top of this additional averaging based on the facet size. With the resulting $uv$-data sets we found 10 rounds of self-calibration to be sufficient to reach convergence. After experimenting with different settings, we found for each self-calibration cycle the following steps to work best (see Table \ref{table:calibration_operations} for the operation description):
\begin{enumerate}[label=\textbf{\arabic*.}, leftmargin=*, labelsep=2mm, itemsep=2mm, font=\sffamily]
    \item \textbf{\texttt{scalarphase} I}: We start by solving for `fast' phase variations for the Dutch remote stations, by applying \texttt{scalarphase} corrections with a solution interval of 1~min and a frequency smoothness kernel of 10~MHz for the Dutch remote stations. This is because the remote stations have faster phase variations.
    \item \textbf{\texttt{scalarphase} II}: To solve for the slower varying phases for the Dutch core stations, we then employed \texttt{scalarphase} corrections with solution intervals of 5~min and a larger frequency smoothness kernel of 20~MHz.
    \item \textbf{\texttt{scalarcomplexgain}}: While the first two self-calibration cycles only have \texttt{scalarphase} corrections, we introduce in the third self-calibration also \texttt{scalarcomplexgain} calibration to correct for scalar amplitude effects as well. This step solves with a solution interval of 30~min and a frequency smoothness kernel of 15~MHz.
\end{enumerate}
The final merged solutions from three different facets for different Dutch stations across the Netherlands are shown in Figures \ref{fig:dutch_phase_solutions} and \ref{fig:dutch_amplitude_solutions}. Figure \ref{fig:6asec_selfcal} demonstrates for 1 facet the significant image improvements at 6\arcsec~resolutions. We merged the resulting Dutch core and remote solutions back into our full merged solutions for all stations that we obtained after DD calibration (see Section \ref{sec:selfcal}). To compare how these new solutions improve the image quality at 3 different resolutions, we show for the same facet in Figure \ref{fig:compare_res_improvements} the image quality.

This additional step was implemented after we already completed the full imaging of all facets at all resolutions (see Section \ref{sec:imaging}). Since the imaging procedure is computationally expensive and increases with the image size and the size of our $uv$-data, we opted to only use these solution refinements for the imaging at 1.2\arcsec~resolution and just for the 5 facets that we found visually to be most affected by the Dutch solution issues at 0.6\arcsec~and 0.3\arcsec~resolutions (such as the facet from Figure \ref{fig:compare_res_improvements}). This is also motivated by the fact that the Dutch core and remote stations have the most short baselines, which implies that the calibration issues for those stations reduce towards higher resolutions.

\section{Wide-field imaging}\label{sec:imaging}

Following the completion of the calibration procedures and collecting our merged phase and amplitude solutions, we performed wide-field imaging to obtain our final image products. In this section, we discuss the imaging method, show parts of our imaging results, and discuss the computational costs of this step in the process compared to calibration.

\subsection{Method}\label{sec:imaging_method}

\begin{figure*}[!ht]
 \centering
    \includegraphics[width=0.9\linewidth]{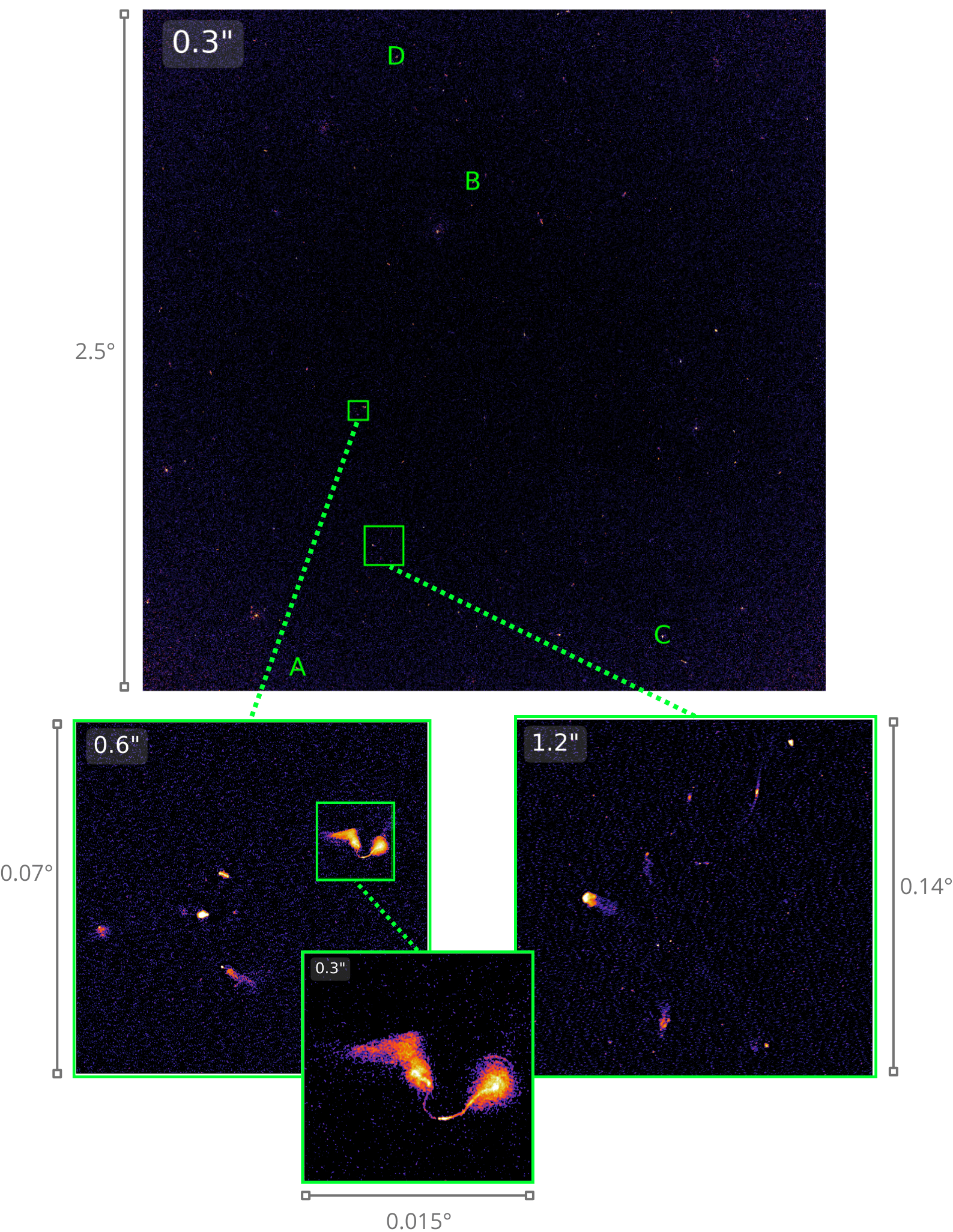}
  \caption{Our final 0.3\arcsec~wide-field image centred on RA=242.75\degree~and DEC=54.95\degree~with cutouts at 0.3\arcsec, 0.6\arcsec, and 0.6\arcsec~from selected areas. The letters correspond to the selection of sources in Figure \ref{fig:compare_res}.}
\label{fig:widefield}
\end{figure*}

\begin{figure*}[htpb]
 \centering
    \includegraphics[width=0.95\linewidth]{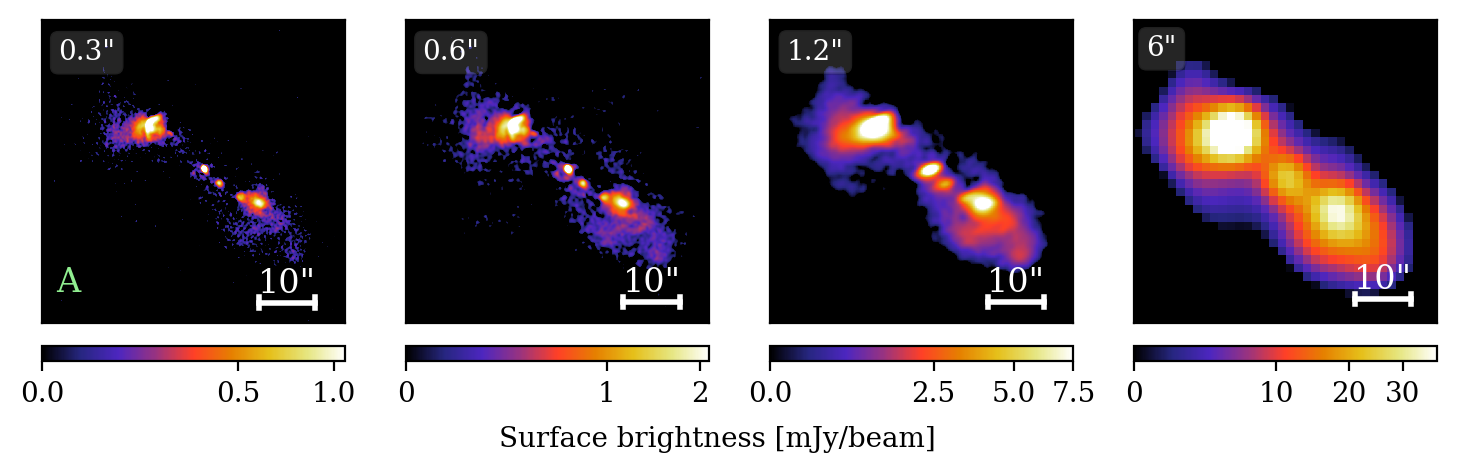}
    \includegraphics[width=0.95\linewidth]{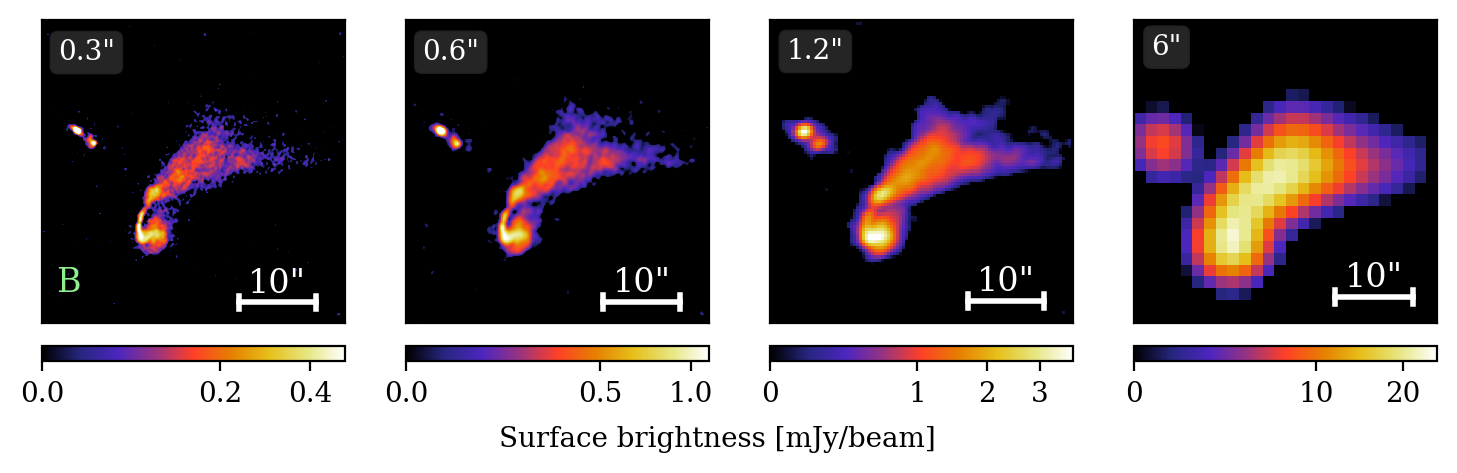}
    \includegraphics[width=0.95\linewidth]{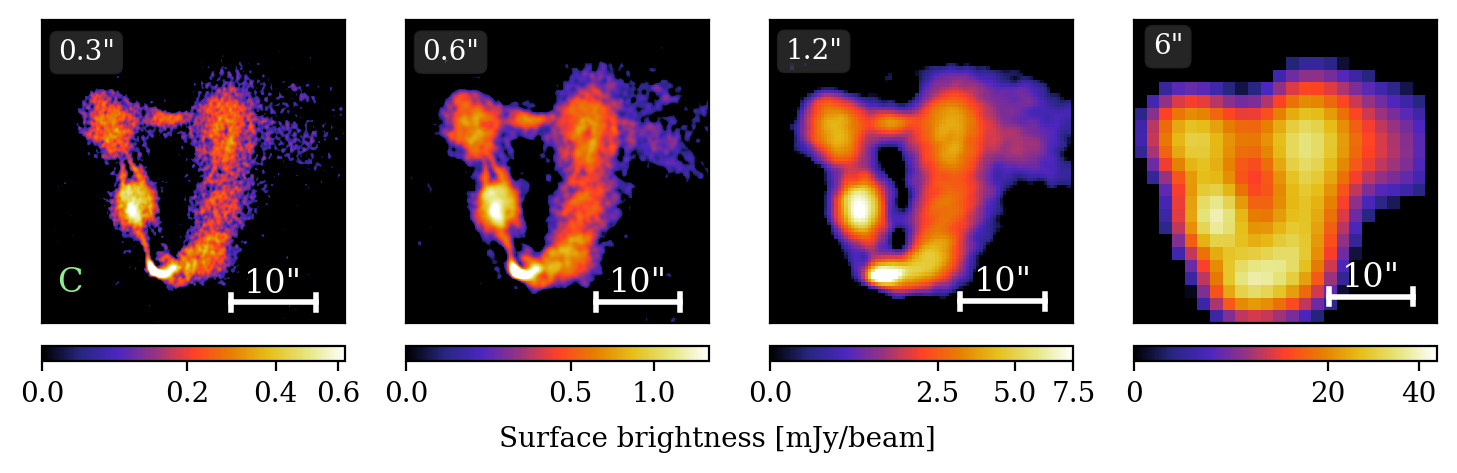}
    \includegraphics[width=0.95\linewidth]{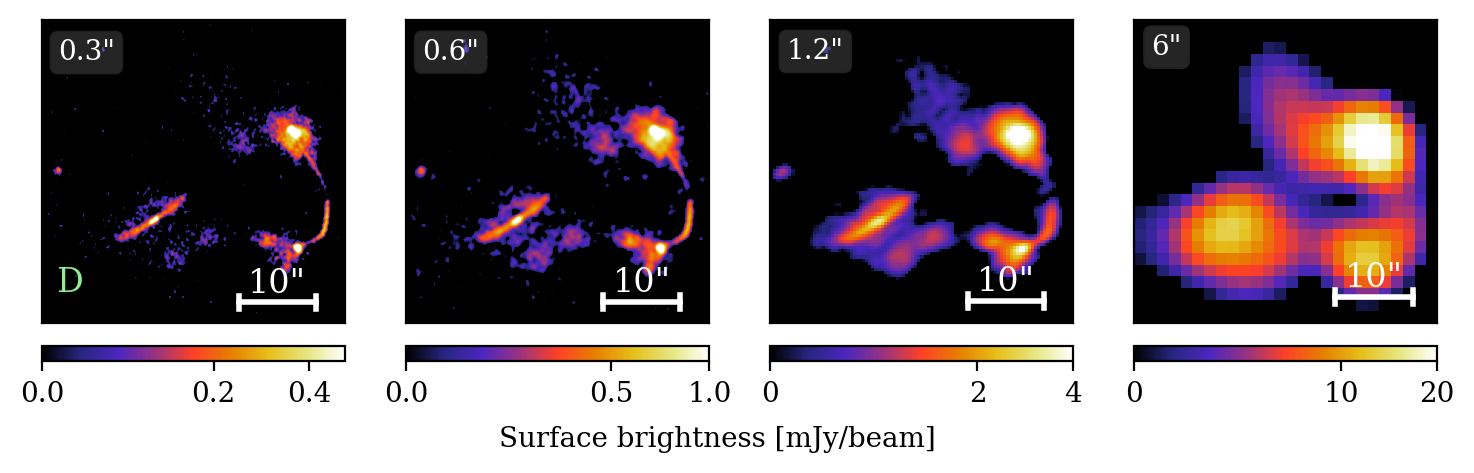}
  \caption{Different radio galaxies (\textit{rows}) across resolution (\textit{columns}) for a selection of cutouts in the wide-field images produced in this paper. The green letters correspond to the letters depicted in the wide-field image from Figure \ref{fig:widefield}. The 0.3\arcsec, 0.6\arcsec, and 1.2\arcsec~images are produced by us, the 6\arcsec~images are from the wide-field image of ELAIS-N1 produced by Shimwell et al. (in prep.). The angular size scale is indicated in the right lower corner.}
\label{fig:compare_res}
\end{figure*}

\begin{table*}
\caption{Resolutions, RMS noises, and source densities of each individual facet.}
\centering
\begin{tabular}{|c| c c c | c c c | c c c |}
\arrayrulecolor{gray}\toprule
 & \multicolumn{3}{c}{{\large\textbf{0.3\arcsec}}} & \multicolumn{3}{c}{{\large\textbf{0.6\arcsec}}} & \multicolumn{3}{c}{{\large\textbf{1.2\arcsec}}} \\ \midrule
{\small \textcolor{darkgray}{\textbf{Facet}}} & {\small \textcolor{darkgray}{\textbf{Resolution}}} & {\small \textcolor{darkgray}{\textbf{RMS}}} & \textcolor{black}{\textbf{$\rho$}} & {\small \textcolor{darkgray}{\textbf{Resolution}}} & {\small \textcolor{darkgray}{\textbf{RMS}}} & \textcolor{black}{\textbf{$\rho$}} & {\small \textcolor{darkgray}{\textbf{Resolution}}} & {\small \textcolor{darkgray}{\textbf{RMS}}} & \textcolor{black}{\textbf{$\rho$}}  \\
& {\small (arcsec$^{2}$)} & {\small ($\mu$Jy~beam$^{-1}$)} & {\small (degree$^{-2}$)} & {\small (arcsec$^{2}$)} & {\small ($\mu$Jy~beam$^{-1}$)} & {\small (degree$^{-2}$)} & {\small (arcsec$^{2}$)} & {\small ($\mu$Jy~beam$^{-1}$)} & {\small (degree$^{-2}$)}\\ [0.15cm] \midrule
\textbf{1} & 0.36$\times$0.45 & 29 & 668 & 0.61$\times$0.67 & 37 & 797 & 1.25$\times$2.74 & 60 & 668 \\ 
\textbf{2} &  0.35$\times$0.42 & 21  &  1278 & 0.58$\times$0.63 & 29 & 1388 & 1.09$\times$2.03 & 56 & 943 \\ 
\textbf{3} & 0.34$\times$0.40 & 20 & 1824 & 0.56$\times$0.61 & 27 & 1824 & 1.05$\times$1.71 & 50 & 1450 \\ 
\textbf{4} & 0.34$\times$0.41 & 20 & 1491 & 0.57$\times$0.62 & 28 & 1336 & 1.06$\times$1.81 & 49 & 1052 \\ 
\textbf{5} & 0.34$\times$0.41 & 19 & 1594 & 0.57$\times$0.62 & 26 & 1403 & 1.05$\times$1.78 & 51 & 989 \\ 
\textbf{6} & 0.34$\times$0.41 & 19 & 1596 & 0.57$\times$0.62 & 27 & 1500 & 1.06$\times$1.87 & 47 & 1093 \\ 
\textbf{7} & 0.35$\times$0.42 & 22 & 928 & 0.58$\times$0.63 & 30 & 877 & 1.06$\times$1.97 & 54 & 577 \\ 
\textbf{8} & 0.36$\times$0.43 & 28 & 718 & 0.62$\times$0.65 & 36 & 778 & 1.15$\times$2.51 & 62 & 692 \\ 
\textbf{9} & 0.35$\times$0.43 & 22 & 875 & 0.58$\times$0.64 & 30 & 967 & 1.08$\times$2.11 & 55 & 611 \\ 
\textbf{10} & 0.34$\times$0.41 & 18 & 1615 & 0.56$\times$0.61 & 25 & 1456 & 1.04$\times$1.68 & 47 & 1023 \\ 
\textbf{11} & 0.33$\times$0.39 & 15 & 2118 & 0.55$\times$0.59 & 22 & 1764 & 1.02$\times$1.45 & 48 & 900 \\ 
\textbf{12} & 0.33$\times$0.39 & 16 & 2281 & 0.55$\times$0.60 & 22 & 1811 & 1.02$\times$1.52 & 41 & 1057 \\ 
\textbf{13} & 0.34$\times$0.40  & 17 & 1843 & 0.56$\times$0.61 & 25 & 1740 & 1.03$\times$1.63 & 44 & 1214 \\ 
\textbf{14} & 0.34$\times$0.42 & 21 & 1435 & 0.57$\times$0.62 & 28 & 1373 & 1.05$\times$1.78 & 56 & 1071 \\ 
\textbf{15} & 0.34$\times$0.40 & 16 & 1860 & 0.55$\times$0.60 & 22 & 1571 & 1.02$\times$1.53 & 41 & 1062 \\ 
\textbf{16} & 0.33$\times$0.38 & 14 & 2071 & 0.54$\times$0.59 & 21 & 1678 & 1.00$\times$1.40 & 40 & 912 \\ 
\textbf{17} & 0.33$\times$0.39 & 15 & 1983 & 0.55$\times$0.59 & 21 & 1721 & 1.01$\times$1.48 & 39 & 1132 \\ 
\textbf{18} & 0.34$\times$0.41 & 20 & 1269 & 0.57$\times$0.62 & 27 & 1218 & 1.04$\times$1.78 & 49 & 860 \\ 
\textbf{19} & 0.35$\times$0.43 & 25 & 1115 & 0.59$\times$0.64 & 33 & 1158 & 1.09$\times$2.14 & 58 & 974 \\ 
\textbf{20} & 0.35$\times$0.42 & 21 & 1203 & 0.58$\times$0.63 & 28 & 1239 & 1.06$\times$1.92 & 49 & 848 \\ 
\textbf{21} & 0.34$\times$0.40 & 17 & 1942 & 0.55$\times$0.60 & 24 & 1733 & 1.02$\times$1.54 & 43 & 1152 \\ 
\textbf{22} & 0.33$\times$0.38 & 15 & 2315 & 0.55$\times$0.59 & 21 & 1966 & 1.00$\times$1.41 & 41 & 1145 \\ 
\textbf{23} & 0.33$\times$0.40 & 17 & 1826 & 0.56$\times$0.61 & 24 & 1756 & 1.03$\times$1.60 & 44 & 1178 \\ 
\textbf{24} & 0.35$\times$0.42 & 21 & 863 & 0.58$\times$0.63 & 30 & 1052 & 1.05$\times$1.94 & 52 & 821 \\ 
\textbf{25} & 0.34$\times$0.40 & 19 & 1573 & 0.56$\times$0.61 & 26 & 1433 & 1.03$\times$1.63 & 51 & 948 \\ 
\textbf{26} & 0.35$\times$0.42 & 20 & 1365 & 0.57$\times$0.63 & 27 & 1455 & 1.07$\times$2.01 & 54 & 997 \\ 
\textbf{27} & 0.35$\times$0.43 & 28 & 712 & 0.60$\times$0.64 & 37 & 785 & 1.12$\times$2.32 & 63 & 630 \\ 
\textbf{28} & 0.35$\times$0.42 & 24 & 1044 & 0.58$\times$0.63 & 32 & 1121 & 1.09$\times$2.11 & 58 & 742 \\ 
\textbf{29} & 0.36$\times$0.45 & 28 & 679 & 0.56$\times$0.61 & 38 & 658 & 1.28$\times$2.82 & 71 & 679 \\ 
\textbf{30} & 0.35$\times$0.44 & 28 & 734 & 0.60$\times$0.66 & 36 & 824 & 1.14$\times$2.50 & 61 & 657 \\ 
\bottomrule
\end{tabular}
\tablefoot{The facet numbers correspond to the numbers in Figure \ref{fig:facets}. The source density ($\rho$) is based on the catalogues after cleaning our source detections, as discussed in Section \ref{sec:catalogue}.}
\label{table:facetinformation}
\end{table*}

We employed \texttt{WSClean} Version 3.3 to produce the wide-field images. This imager has, using the \texttt{wgridder} module \citep{arras2021, ye2022}, a facet-based imaging mode that enables wide-field imaging with solutions for different facets. While this option has proven to be fast and reliable for making large wide-field images corrected for DD effects \citep[e.g.][]{dejong2022, ye2023}, with the large data volumes the computational demands for 0.3\arcsec-imaging are so high, at this resolution, it would take over four~months to make at this resolution images up to $10^{10}$ pixels directly with four observations (see Section \ref{sec:computingcosts}).
For an 1.2\arcsec~resolution wide-field image, it only takes up to four days for an 8~hr dataset. So, we decided to only utilize the facet-based imaging for making wide-field images at 1.2\arcsec~resolution for each of the four individual observations, as this gives us an image to assess the quality of our fully calibrated data across the entire field of view for each observation. Additionally, the model images from \texttt{WSClean} were essential for imaging our facets separately. This involved predicting and subtracting data outside each facet, a process we describe in detail below.

By making images of our facets separately, we allowed ourselves to average the visibilities without introducing smearing effects. 
The averaging factors in both time and frequency are determined by the facet size and vary from 3 to 7. This speeds up the imaging, compared to the original 1~sec (or 2~sec) and 12.21~kHz resolutions of the datasets before averaging. We note that the datasets that were originally averaged to 2~sec will have smaller time averaging factors. To remove emission outside a facet, we first derived model visibilities corresponding to the sky outside each facet by utilising the model images from the 1.2\arcsec~resolution radio maps corresponding to each observation. To achieve this, we masked the facet in the model image and used this masked model image to predict visibilities with applied DD solutions in \texttt{WSClean}. This yields the model data visibilities that we then subtracted from the original DI corrected visibilities. As this process can be done in parallel over frequency blocks, we did the prediction and subtraction for smaller frequency sub-band for each of our observations, which helped us reduce the processing wall-clock time by a factor 16 (see Section \ref{sec:computingcosts}). We note that this does not reduce the total CPU~time. After phase-shifting the subtracted data to the centre of the facet, applying the solutions from the DD calibrator, and accounting for the beam at the facet's centre, we averaged the data for each observation before proceeding with the final imaging using \texttt{WSClean}. 

We imaged each facet with all observations together, using a Briggs weighting of $-1.5$ \citep{briggs1995}, a minimum \textit{uv}-value of 80$\lambda$ (corresponding to a LAS of \textasciitilde43\arcmin), pixel sizes of 0.1\arcsec, 0.2\arcsec, and 0.4\arcsec, and corresponding Gaussian tapers of 0.3\arcsec, 0.6\arcsec, and 1.2\arcsec. For efficient deep cleaning and to better recover extended diffuse emission, we apply `auto' masking, multi-scale deconvolution, and an RMS box equal to 50 times the synthesized beam size \citep{cornwell2008, offringa2017}. \texttt{WSClean} ends by applying a final full primary beam correction to correctly account for the attenuation of the primary beam.

\subsection{Facets and mosaics}\label{sec:mosaics}

\begin{figure*}[htbp]
 \centering
    \includegraphics[width=0.49\linewidth]{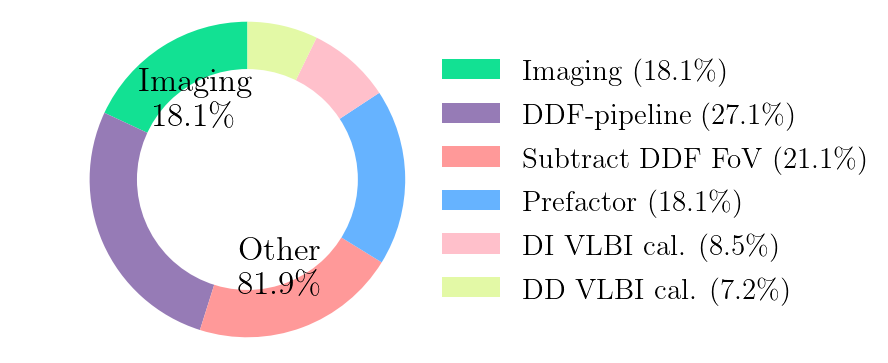}
    \includegraphics[width=0.49\linewidth]{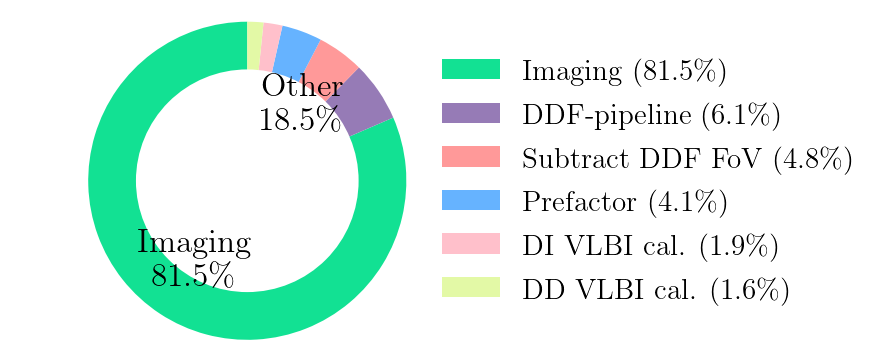}
  \caption{Pie plots depicting the fraction of CPU~hrs of each of the major calibration and imaging steps for 1.2\arcsec-imaging (\textit{left panel}) and 0.3\arcsec-imaging (\textit{right panel}). The 1.2\arcsec~processing was done for one observation with the facet-mode from \texttt{WSClean} in about 7,000 CPU~hrs, and the 0.3\arcsec-imaging was done for one observation using the predict-subtract method for one observation in about 139,000 CPU~hrs. These numbers scale roughly linearly with the number of observations. The `Subtract DDF FoV' includes the subtraction of sources outside the 2.5\degree$\times$2.5\degree~field of view from the last paragraph of Section \ref{sec:initdutchcalibration}. The `DI VLBI calibration' includes the DI calibration from Section \ref{sec:infieldcal}. The `DD VLBI calibration' includes the DD calibration selection and self-calibration from Section \ref{sec:ddcal}. To highlight the computational costs for imaging compared to the other data reduction steps combined, we indicate in the figure 'Imaging' and 'Other' in the pie plots.}
\label{fig:pieplot}
\end{figure*}

Table \ref{table:facetinformation} gives the resolutions, RMS noise, and source density of each individual facet.
We reach a best RMS noise value of 14~$\mu$Jy~beam$^{-1}$ for the 0.3\arcsec~facets, 21~$\mu$Jy~beam$^{-1}$ for the 0.6\arcsec~facets, and 39~$\mu$Jy~beam$^{-1}$ for the 1.2\arcsec~facets near the pointing centre (see Section \ref{sec:rmsanalysis} for a more detailed RMS noise analysis). 
This is about twice as deep as the sensitivities reported by \cite{sweijen2022} at 0.3\arcsec~and \cite{ye2023} at 1.2\arcsec, who utilised data with four times less integration time. This aligns with the expected behaviour from the radiometer equation, which states that sensitivity improves as the square root of the integration time \citep{kraus1966}. 
We find our best resolutions near the pointing centre, with resolutions of 0.33\arcsec$\times$0.38\arcsec~for the 0.3\arcsec~facets, 0.54\arcsec$\times$0.59\arcsec~for the 0.6\arcsec~facets, 1.00\arcsec$\times$1.40\arcsec~for the 1.2\arcsec~facets. The stronger elongation of the synthesized beam for the 1.2\arcsec~target resolution is due to the sparsity of LOFAR stations between 80 and 180~km (see Figure \ref{fig:uvcoverage}).

To make wide-field images, we convert the individual resolutions from the facets to one common resolution. Due to issues with one of our computing nodes, we lost 5 of our residual and model images of the 0.3\arcsec~facets. We were therefore only able to convolve our 0.3\arcsec~map to a common resolution equal to the facet with the lowest resolution, using CASA \texttt{imsmooth} \citep{casa2022}. This gives us a resolution of 0.36\arcsec$\times$0.45\arcsec. Having all model and residual images available for the other resolutions, we were able to restore these maps with \texttt{WSClean} to a common resolution of 0.58\arcsec$\times$0.62\arcsec~and 1.0\arcsec$\times$1.5\arcsec.

After mosaicing the individual facets, we have our wide-field images for all three resolutions with image sizes of 90,000$\times$90,000, 45,000$\times$45,000, and 22,500$\times$22,500 pixels for the 0.36\arcsec$\times$0.45\arcsec, 0.58\arcsec$\times$0.62\arcsec~and 1.0\arcsec$\times$1.5\arcsec~respectively. In Figure \ref{fig:widefield}, we present one of our wide-field images with a few cutouts of areas at different resolutions. In order to assess visually the quality and level of detail across various resolutions, we showcase selected cutouts from our radio maps at resolutions of 0.36\arcsec$\times$0.45\arcsec, 0.58\arcsec$\times$0.62\arcsec~and 1.0\arcsec$\times$1.5\arcsec, and 6\arcsec~in Figure \ref{fig:compare_res}, where for the 6\arcsec~counterparts we utilised the deep wide-field image recently created by Shimwell et al. (in prep.), who used more than 500~hrs of LOFAR data. These selected cutouts reveal the structural details at the higher resolutions, notably evident in the lobes of radio galaxies, while the lower resolutions highlight the diffuse emission from these same sources. during the rest of this paper, we use our individual facet images, as these have the best fitted resolutions and depths. We therefore continue to refer to these images by 0.3\arcsec, 0.6\arcsec, and 1.2\arcsec~resolutions.

\subsection{Computing costs}\label{sec:computingcosts}

Making wide-field images in the order of $10^{9}$ to $10^{10}$ pixels requires significant computing resources. For processing our data we utilised AMD~EPYC~7551 and AMD~EPYC~7702P processor nodes with each 60 cores and with 0.5~TB and 1~TB RAM. Although the predict-subtract method before imaging of the individual facets (see Section~\ref{sec:imaging_method}) incurs a large computational cost, accounting for approximately 76\% of the total imaging costs, we managed to reduce the wall-clock time by a factor 16 through additional parallelisation by splitting our total frequency bandwidth in smaller blocks and performing the prediction and subtraction step for each block separately. The final imaging costs for four observations total 550,000 CPU~hrs, which brings us to a total of about 680,000 CPU~hrs for full data processing including calibration. With the large data volumes, we found a linear relationship between data volume and computing costs. Using this linear relationship, we find an improvement of about a factor 2 compared to \cite{sweijen2022}, who worked with only a sixth of the data size we processed (taking into account that they pre-averaged their 8~hrs dataset by a factor 2 in time). This speedup is primarily due to a combination of software enhancements in \texttt{WSClean} and the optimisation of our software containers for the appropriate hardware. While we observe an improvement in processing speed, it is notable that when mapping the sky at the highest resolution the imaging expenses account for 81.5\% of our overall processing costs, as depicted in the right panel of Figure \ref{fig:pieplot}, which is slightly higher compared to the 76\% reported by \cite{sweijen2022}. This difference could be due to a combination of different numbers of facets and software improvements that have affected parts of the pipeline differently than other parts. That computational demands for 0.3\arcsec-imaging are predominantly driven by imaging, highlights that full data reduction speedups need development for this step. For creating the 0.6\arcsec~resolution wide-field image, we averaged our data to 2~sec and 24.42~kHz, before imaging and changed the final imaging parameters. We could similarly for the facets at 1.2\arcsec~resolution average again by a factor two (4~sec and 48.84~kHz) compared to the 0.6\arcsec~resolution. The averaging makes only the final imaging after the prediction and subtraction 4 (0.6\arcsec) or 16 (1.2\arcsec) times faster, compared to 0.3\arcsec-imaging. 

As a part of the prediction and subtraction step for the 0.3\arcsec~and 0.6\arcsec~resolution imaging, we made wide-field model images at 1.2\arcsec~for each of our four observations. The computing costs for this step, using the \texttt{WSClean} facet-based imaging mode, required 7,000~CPU~hrs per observation. This is almost two times faster than what \cite{ye2023} reported for wide-field imaging. The improvement is again due to a combination of recent software improvements and the different computing nodes they used for imaging. In the left panel of Figure \ref{fig:pieplot} we display the imaging costs for 1.2\arcsec-imaging. Comparing this with the plot corresponding to 0.3\arcsec-imaging from the right panel, it is evident that 1.2\arcsec-imaging with the facet-mode from \texttt{WSClean} significantly reduces the weight of the imaging step on the complete data reduction workflow. The reduced computational costs, relative to sub-arcsecond imaging, make imaging at 1.2\arcsec~resolution an interesting intermediate resolution for specific science goals or surveys, as was also highlighted by \cite{ye2023}. The scientific benefits of the different resolutions are discussed further in Section \ref{sec:res_sens}.

\section{Cataloguing}\label{sec:catalogue}

We constructed catalogues with radio sources for all three of our image resolutions (0.3\arcsec, 0.6\arcsec, and 1.2\arcsec) by employing \texttt{PyBDSF}\footnote{\url{https://pybdsf.readthedocs.io}} on our individual facets \citep{mohan2015}. All parameter settings that we modified from the default settings are displayed in Table \ref{table:pybdsf}. The \texttt{rms\_box} sets the sliding box parameters for calculating the RMS and mean flux density per beam over the entire image. With the \texttt{rms\_box\_bright} parameter we enable \texttt{PyBDSF} to more effectively increase the noise in regions of artefacts, by using a smaller box around brighter sources. The \texttt{group\_tol} parameter groups Gaussian components fitted by \texttt{PyBDSF}. We opted to use the value 10 for this parameter, as this value has been often adopted for source detections at the same or similar frequencies \citep[e.g.][]{williams2019, ocran2020, sabater2021, ye2023}. \texttt{PyBDSF} detected with these settings 24251 objects for our 0.3\arcsec~resolution radio map, 14099 objects for our 0.6\arcsec~resolution radio map, and 10229 objects for our 1.2\arcsec~resolution radio map.

\begin{table}
\caption{\texttt{PyBDSF} settings modified from the default values.}
\centering
\begin{tabular}{ll} \toprule
    \textbf{Parameter} & \textbf{Value} \\ \midrule
    \texttt{rms\_box} & (120, 15) \\
    \texttt{rms\_box\_bright} & (40, 10) \\
    \texttt{adaptive\_rms\_box} & True \\
    \texttt{atrous\_do} & True \\
    \texttt{group\_tol} & 10.0 \\
   \bottomrule
\end{tabular}
\tablefoot{We refer to the documentation for a full description of these parameters.}
\label{table:pybdsf}
\end{table}

Although we set the \texttt{group\_tol} higher than the default value to enhance the association of components, we find by eye unassociated detections by \texttt{PyBDSF} that are part of the same physical object. We therefore decided to apply additional source association. Automated source association approaches at 6\arcsec, using for instance a convolutional neural network, have shown to provide a similar accuracy to what visual inspection by astronomers would obtain \citep{mostert2022}. Since we have radio maps at higher resolutions, these models require extra training and testing, introducing additional challenges that are beyond the scope of this paper. Fortunately, our field of view is confined to 2.5\degree$\times$2.5\degree, limiting the number of large resolved sources. We therefore decided to visually inspect all clusters of sources within a distance of 50\arcsec~from each other and associate components that are likely part of the same source. For guidance, we used images at each of the four different resolutions, which allowed us, in case of doubt, to make comparisons during the association of extended objects. After this visual inspection, we were left with 22804, 13119, and 9577 sources at 0.3\arcsec, 0.6\arcsec, and 1.2\arcsec~respectively.

\begin{figure}[htbp]
 \centering
     \includegraphics[width=0.48\linewidth]{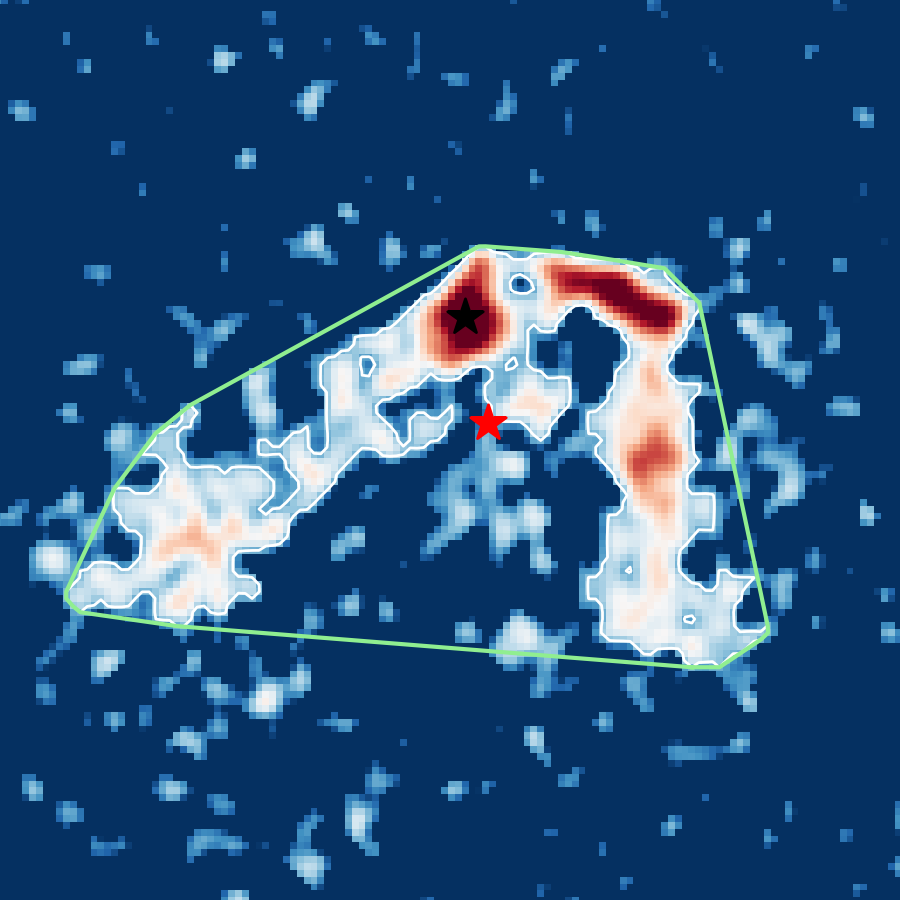}
      \includegraphics[width=0.48\linewidth]{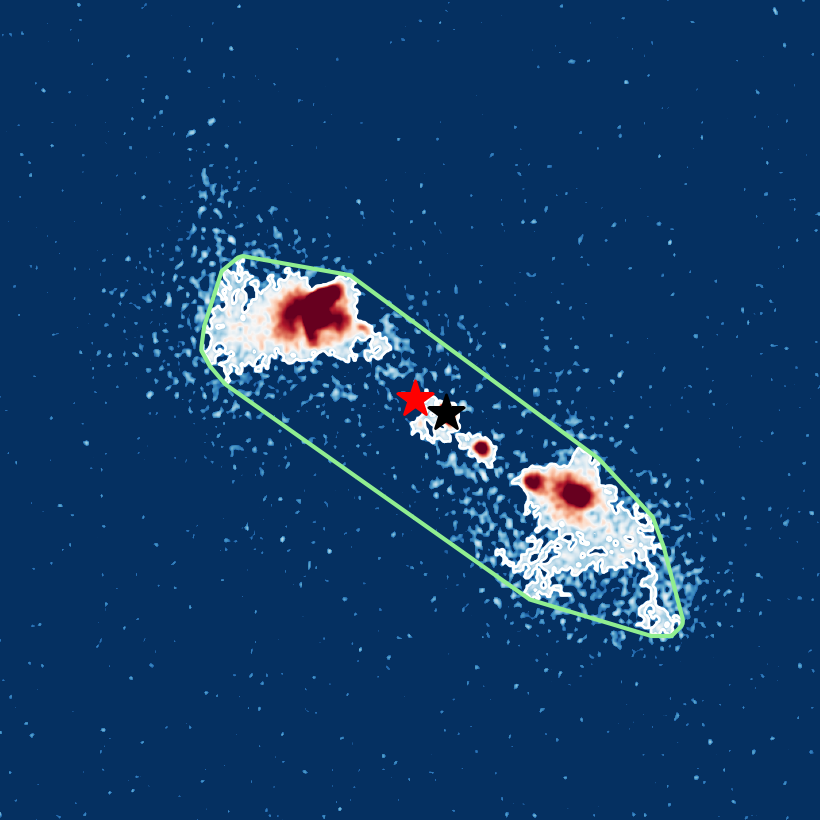}
  \caption{Examples of resolved sources where the integrated flux density and source position were obtained by using an enveloping polygon (light green) circumventing all source islands. Similar to \texttt{PyBDSF}, we require each polygon to have a peak intensity exceeding 5 times the local RMS. The final source position, determined using moment analysis, is marked by the red star. The black star is at the position of the peak intensity of the source within the polygon boundary.}
\label{fig:resolved}
\end{figure}

Following component associations, we employed the \texttt{shapely}\footnote{\url{https://shapely.readthedocs.io}} Python package to automatically find the integrated flux densities of the visually associated components. With this package, we automatically drew polygons enveloping all islands from each source and summed the pixels within to obtain the integrated flux density. For the final central source position, we applied, similar to \texttt{PyBDSF}, moment analysis \citep{hu1962}, which calculates the weighted mean of the brightness distribution, thus providing a robust measure of the centroid of extended sources. We carried out an additional visual inspection of the sources that we identified by eye as having a bad polygon fitting. Consequently, this led us to manually calculate the integrated flux density of 67 sources. To illustrate our method, we show in Figure \ref{fig:resolved} two examples of extended sources that have been subjected to our fitting procedure.

To ensure the reliability of the sources in our final catalogue, we decided to use a S/N threshold at 5$\sigma$, which implies that we reject all sources with a peak intensity below 5 times the local RMS. This is reported by \texttt{PyBDSF} in the \texttt{Isl\_rms} column. After removing these sources we had 13058, 10241, 6997 sources at respectively 0.3\arcsec, 0.6\arcsec, and 1.2\arcsec. The remaining sources were cross-matched with the catalogue from the 6\arcsec~ELAIS-N1 LOFAR HBA map by Shimwell et al. (in prep.), where we reject sources from our catalogue that do not have a cross-match within 6\arcsec. Their map has a minimal sensitivity of 11~$\mu$Jy~beam$^{-1}$ and is therefore deeper than our images. This additional selection step ensured the reliability of our catalogue content and left us with final source counts of 9203, 8567, 5872 sources at 0.3\arcsec, 0.6\arcsec, and 1.2\arcsec~respectively. 

  \begin{figure}[htbp]
 \centering
    \includegraphics[width=1\linewidth]{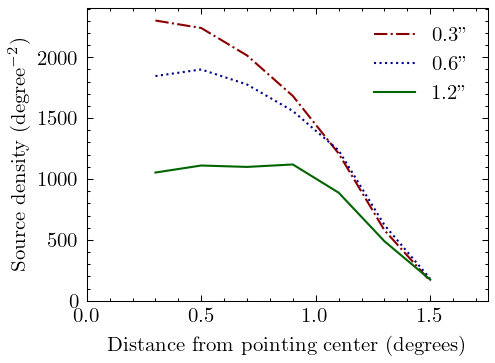}
  \caption{Source density as a function of distance from the pointing centre in degrees for all three resolutions. This figure was constructed by evaluating the median source density in bins of 0.2\degree.}
\label{fig:density}
\end{figure}

Going about two times deeper than \cite{sweijen2022} and \cite{ye2023}, we find respectively 4 and 2.5 times more objects at the same resolutions. To illustrate the source distribution across our different facets and resolutions, we give in Table \ref{table:facetinformation} the source densities across our 30 facets and plot in Figure \ref{fig:density} the source density as a function of distance from the pointing centre. In Section \ref{sec:res_sens} we further discuss the different sources detected at different resolutions.

\section{Discussion}\label{sec:discussion}

We have created the deepest (sub-)arcsecond wide-field images of ELAIS-N1 at 140MHz, accomplished by processing together four 8~hrs observations including all available LOFAR stations. In this section, we do additional analysis of the image and source detection quality.

\subsection{RMS noise}\label{sec:rmsanalysis}

\begin{figure*}
\centering
\includegraphics[width=0.33\linewidth]{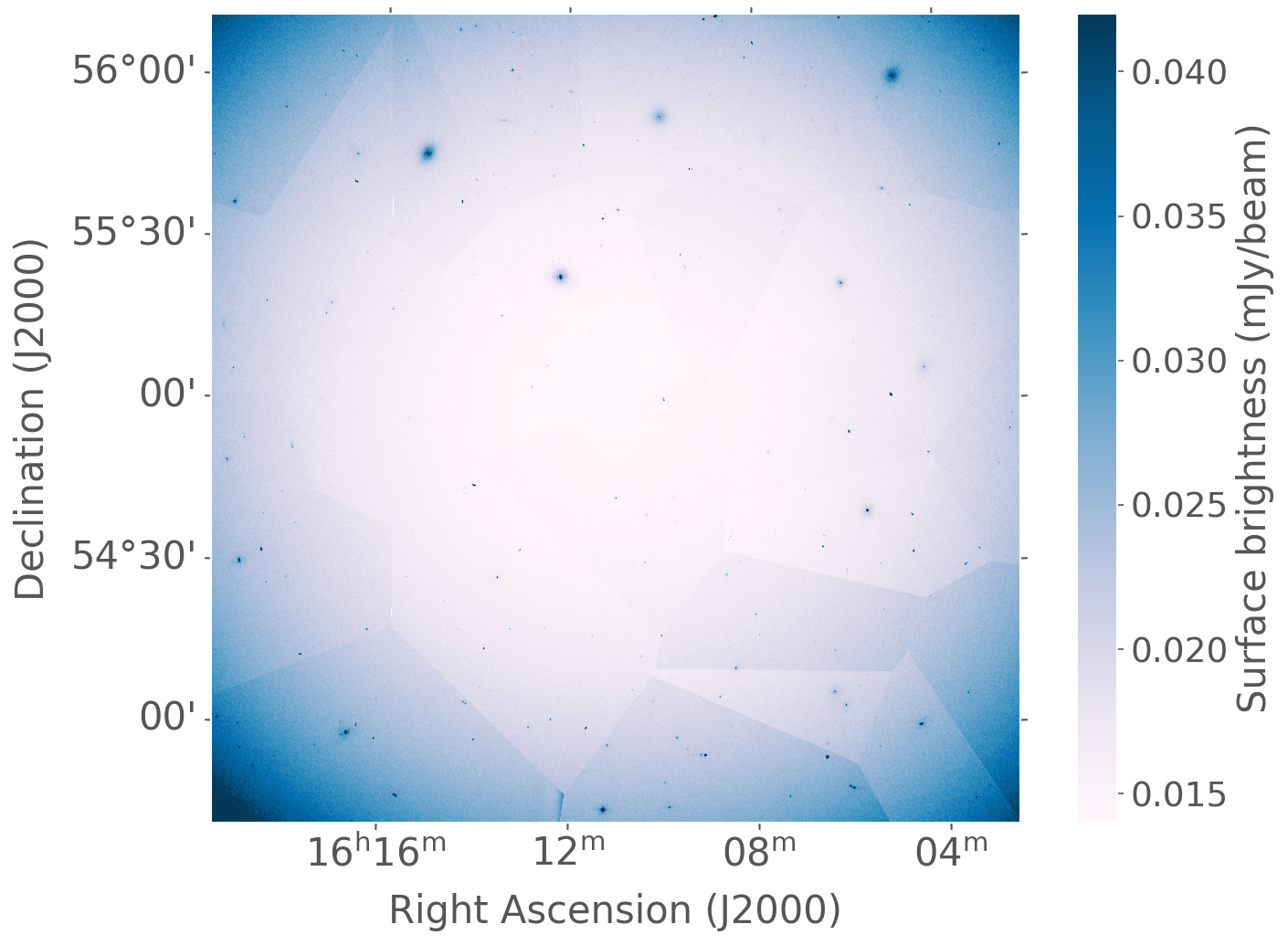}
\includegraphics[width=0.33\linewidth]{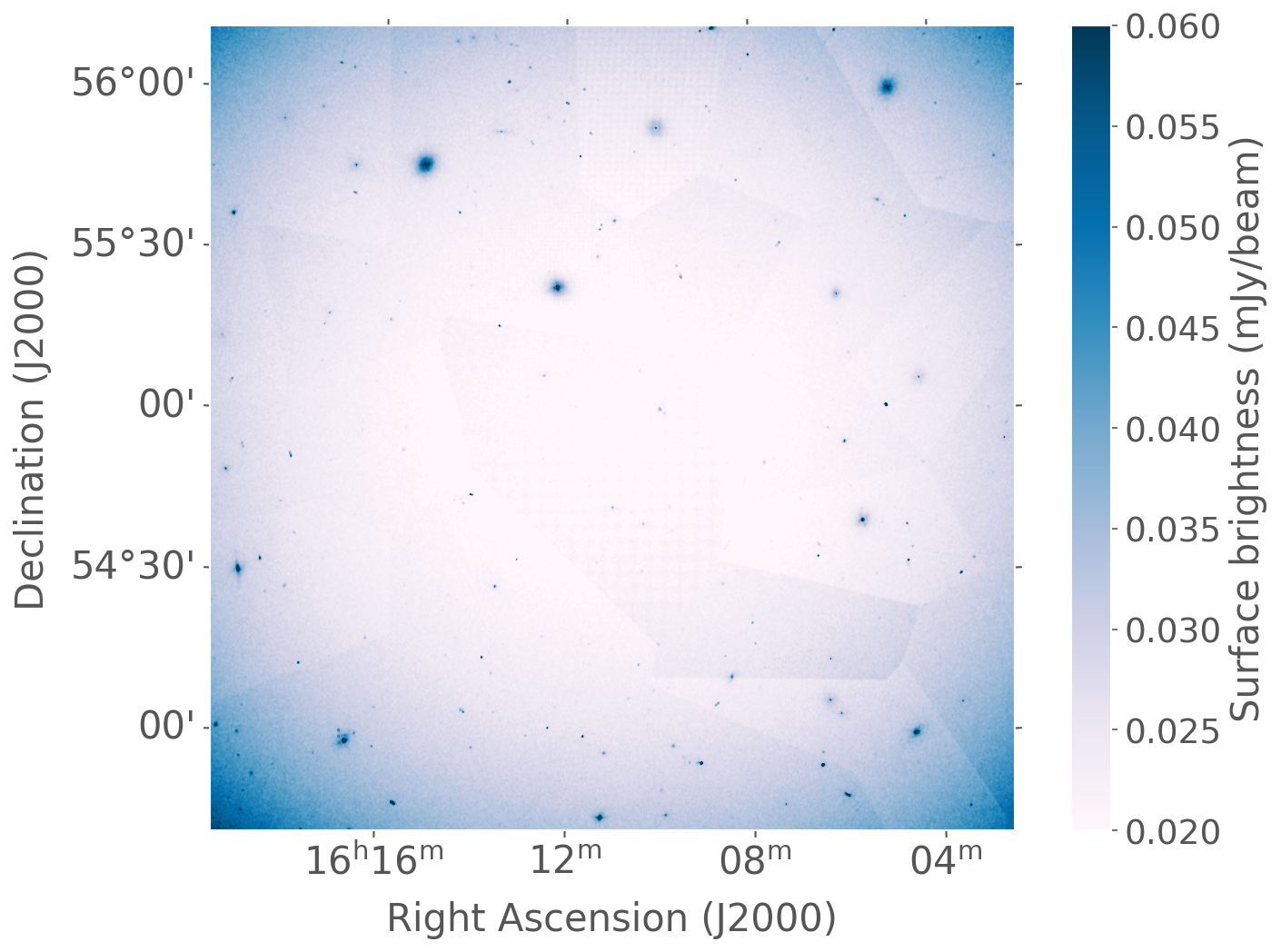}
\includegraphics[width=0.33\linewidth]{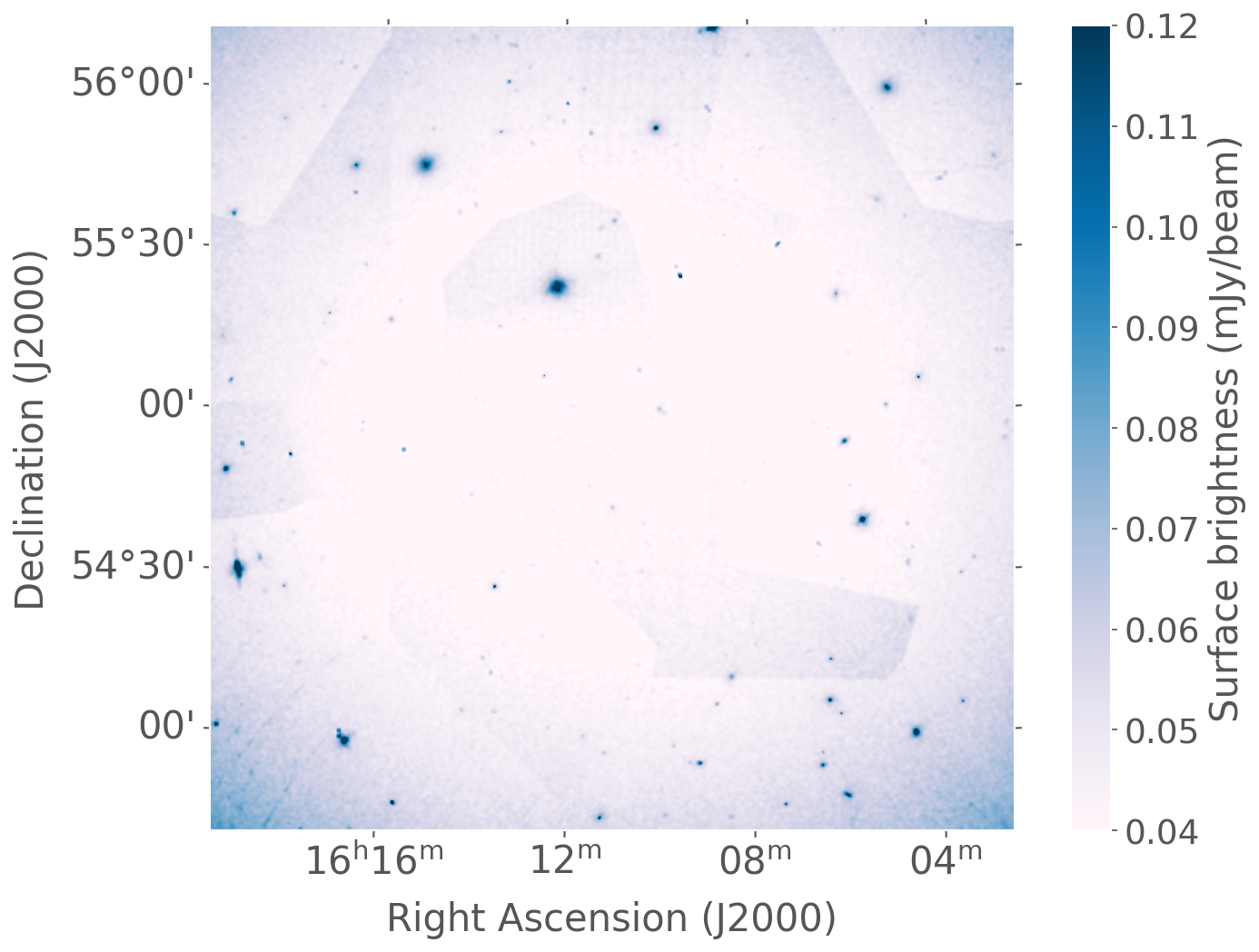}
\caption{RMS maps for different resolutions. \textit{Left:} 0.3\arcsec. \textit{Middle:} 0.6\arcsec. \textit{Right:} 1.2\arcsec. These maps are made with \texttt{PyBDSF} and scaled between one and three times the best RMS noise in the image (see Figure \ref{fig:rms_relative}).}
\label{fig:rmsmaps}
\end{figure*}

\begin{figure*}
\centering
\includegraphics[width=0.48\linewidth]{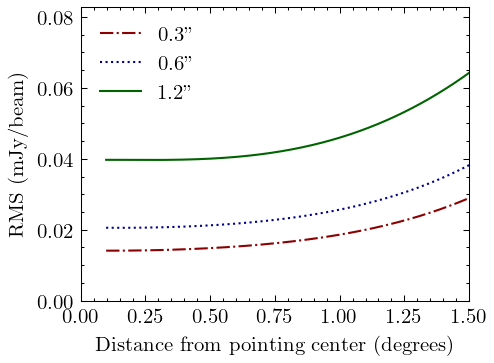}
        \includegraphics[width=0.48\linewidth]{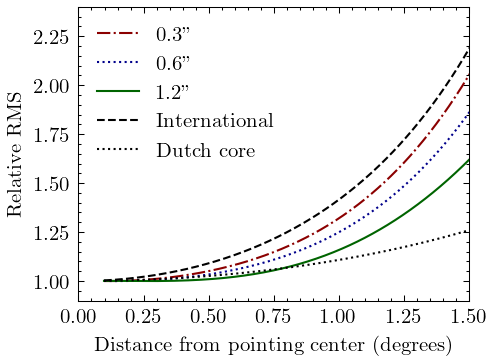}
  \caption{Median RMS noise as a function of distance from the pointing centre in degrees for all three of our resolutions. The median RMS noise values are calculated in bins of 0.06\degree~and fitted by a second-order polynomial. \textit{Left panel:} Absolute RMS. \textit{Right panel:} RMS divided by the RMS at the pointing centre, where we added the inverse primary beam intensity for an international and Dutch core station.}
\label{fig:rms_relative}
\end{figure*}

In Section \ref{sec:mosaics} we touched upon the RMS noise across different facets. This variation is also notable in the RMS noise map from \texttt{PyBDSF}, as shown in Figure \ref{fig:rmsmaps}. The higher RMS noise values at lower resolution are due to a combination of tapering and Briggs weighting, which we set for all three resolutions at $-1.5$ (see Section \ref{sec:imaging_method}). 
Given that we have not optimized the Briggs weighting for all resolutions, this may particularly negatively effect the RMS noise at 1.2\arcsec~resolution, as this resolution is most susceptible to the $uv$ sampling gaps between 80~and~180~km (see Figure \ref{fig:uvcoverage}). However, optimising the Briggs weighting parameter is with the current computing costs too expensive (see Section \ref{sec:computingcosts}). The RMS noise offsets  between facets in Figure \ref{fig:rmsmaps} are attributed to solution quality differences across the DD calibrators, with each facet having its own set of DD solutions. The peaks in the local RMS noise at the positions of bright sources are due to remaining DDEs that are not completely removed around the brightest sources. 
The sensitivity variation due to the attenuation from the primary beam is also a key contributor to the RMS noise increases towards the edge of the field. This is demonstrated with the median RMS noise as a function of distance in the left panel of Figure \ref{fig:rms_relative}. 

In the right panel of Figure \ref{fig:rms_relative} we compare the shape of the RMS curves as a function of distance from the pointing centre for the 0.3\arcsec, 0.6\arcsec, and 1.2\arcsec~resolution. The larger steepness of the relative RMS noise for higher resolution reflects the primary beams of the stations used at that particular resolution. This explains why the source densities converge for the three resolutions at the edges of the field, as shown in Figure \ref{fig:density} and Table \ref{table:facetinformation}. In the right panel of Figure \ref{fig:rms_relative} we plot for comparison also the inverse primary beam intensity ($I_{\text{P}}$) for an international and a Dutch core station as a function of distance ($\theta$) from the pointing centre, using a simple Gaussian model given by
$$I_{\text{P}}=\text{exp}\left(-4\ln{2}\frac{\theta^{2}}{\text{FWHM}^{2}}\right),$$
where the FWHM is the full width at half maximum of the synthesized beam and determined by
$$\text{FWHM}=\alpha\cdot\frac{\lambda}{D}.$$
In this formula $\lambda$ represents the wavelength (corresponding to 140~MHz) and $D$ signifies the diameter of the stations, where we used a diameter of 30.75 meters for a Dutch core station, and 56.5 for the international station \citep{haarlem2013}. The value for $\alpha$ varies based on the station layout and additional tapering, being 1.02 for a perfect circular aperture \citep{napier1999}, and 1.3 for LOFAR.\footnote{\url{https://science.astron.nl/telescopes/lofar/lofar-system-overview/observing-modes/lofar-imaging-capabilities-and-sensitivity}} As anticipated, we observe that the relative RMS noise across all resolutions falls between the primary beam intensities of the international and Dutch core stations.

\subsection{Smearing}\label{sec:smearing}

\begin{figure*}[htbp]
 \centering
    \includegraphics[width=0.48\linewidth]{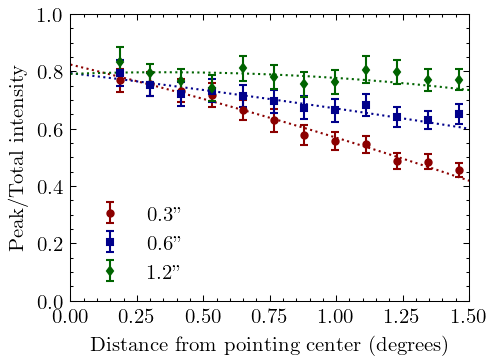}
    \includegraphics[width=0.48\linewidth]{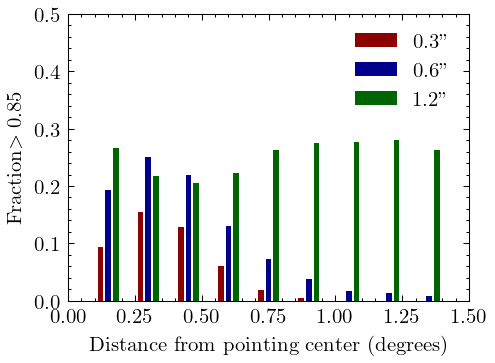}
  \caption{Smearing across distance from the pointing centre. \textit{Left panel:} Ratio of peak over integrated flux density as a function of the distance from the pointing centre for the 3 different resolutions considered in this paper. The plot is made by taking the median ratio for several distance bins. The error bars are based on the errors on the peak and integrated flux densities measured by \texttt{PyBDSF}. 
  \textit{Right panel:} Fraction of sources with peak over integrated flux densities above 0.85 for different distance bins.
  To enhance the reliability of our source sample, we considered for both panels only sources fitted by a single Gaussian by \texttt{PyBDSF} and a peak intensity at least 15 times above the local RMS noise.}
\label{fig:smearing}
\end{figure*}

Bandwidth and time smearing are important effects that degrade the quality of our radio maps and cannot be corrected by calibration. These effects cause sources to appear smeared or elongated in the radial direction in the case of bandwidth smearing and tangentially in the case of time smearing with respect to the pointing centre. This makes accounting for smearing effects essential for accurately measuring source sizes, morphologies, and peak intensities. 
Since smearing becomes more pronounced closer to the edge of the field of view, it contributes to the decrease in source detections in Figure \ref{fig:density}.

An effective measure for assessing smearing levels is to compare the peak intensity with the integrated flux density of unresolved sources across the field. This is because smearing affects the peak intensity much more than the integrated flux density, as the integrated flux density only reduces when flux disappears below the detection threshold, while the peak intensity always decreases. For an ideal, unsmeared point source, the ratio of the peak intensity over the integrated flux density should be one. In the left panel of Figure \ref{fig:smearing}, we illustrate the variation of this ratio with distance from the pointing centre, by selecting sources with peak intensities at least 15 times above their local RMS and fitted by a single Gaussian, across our three resolution images. The figure demonstrates a noticeable reduction in this ratio towards the edge of our field of view, most pronounced at the highest resolution. These declining trends are due to the inevitable effects of smearing, which in our case are intensified by the fact that half of our observations were stored in the LTA with a factor of two extra time averaging. Although we find the peak over integrated flux densities for some sources to be close to one near the pointing centre, the sources with much lower ratios push the median peak over integrated flux density down to below 0.8 in our distance bins. Part of the reason why our ratios are not at unity is due to our source selection where we are plotting all sources that are fitted by \texttt{PyBDSF} by a single Gaussian, while a source might still be a partly resolved source. Therefore, to quantify the number of sources with a peak over integrated flux density closer to 1, we show in the right panel of Figure \ref{fig:smearing} the fraction of sources with a peak over integrated flux density above 0.85. Similar to the left panel, we find only the 0.3\arcsec~and 0.6\arcsec~to have a declining trend from the pointing centre towards the edge of the field, which implies that smearing is negligible at 1.2\arcsec.

\subsection{Astrometry}\label{sec:astrometry}

\begin{figure*}[htbp]
 \centering
    \includegraphics[width=0.44\linewidth]{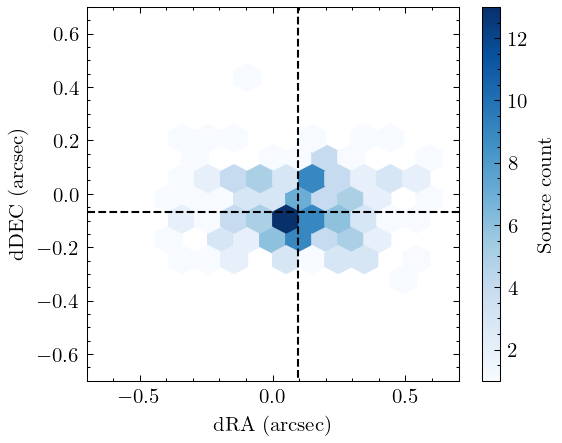}
    \includegraphics[width=0.48\linewidth]{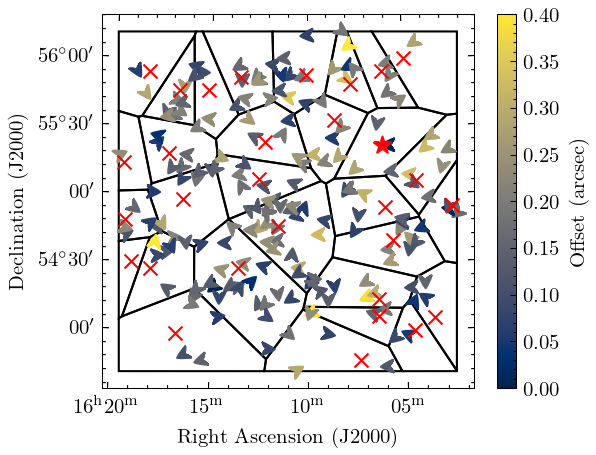}
  \caption{Position offsets between our 0.3\arcsec~radio source detections and optical counterparts according to the 6\arcsec~catalogue from \cite{kondapally2021}, where we used selection filters based on the accuracy of the associations at 6\arcsec~as described in the text. \textit{Left panel:} Two-dimensional hex-bin histogram with the RA/DEC offsets (dRA and dDEC) between the 0.3\arcsec~catalogue and the optical counterparts. The median offsets are given by the black dashed line at dRA=$0.094\pm0.093$\arcsec~and dDEC=$-0.067\pm0.064$\arcsec. \textit{Right panel:} Position offsets of the selected 192 sources at 0.3\arcsec~resolution, where the colourbar corresponds to their absolute offsets and the arrow direction to their directional offset. We added the positions of the DD calibrator with red crosses and the in-field calibrator with a red star. Facets are given by black contours.}
\label{fig:dRA_dDEC}
\end{figure*}

We used an in-field calibrator source in our sky model of which the position was known with milliarcsecond-level accuracy (see Section \ref{sec:skymodel}). 
To evaluate the quality of the final astrometry from our radio maps, we conduct a comparison between our catalogue and the 6\arcsec~catalogue from the LOFAR deep field DR1 \citep{kondapally2021, sabater2021}. The wide-field image behind the 6\arcsec~catalogue has a sensitivity up to 20$\mu$Jy~beam$^{-1}$. The source detections have been associated with sources across multiple wavelengths, enabling an astrometric reference frame. The multi-wavelength data includes the Hyper-Suprime-Cam Subaru Strategic Program (HSC-SSP) survey \citep{aihara2018}, Panoramic Survey Telescope and Rapid Response System \citep[Pan-STARRS;][]{kaiser2010}, the UK Infrared Deep Sky Survey Deep Extra-galactic Survey \citep[UKIDSS-DXS][]{lawrence2007}, the Spitzer Adaptation of the Red-sequence Cluster Survey \citep[SpARCS;][]{wilson2009, muzzin2009}, the Space Infrared Telescope Facility (SIRTF) Wide-Area Infrared Extragalactic Survey \citep[SWIRE;][]{lonsdale2003} and the Spitzer Extragalactic Representative Volume Survey \citep[SERVS;][]{mauduit2012}. The multi-wavelength catalogues were generated by combining imaging datasets from all of these surveys across optical-infrared. The astrometry for the optical datasets used was calibrated to Two-Micron All Sky Survey \citep[2MASS][]{skrutskie2006}. \cite{kondapally2021} compared the astrometry between their generated catalogues to publicly available catalogues, finding typical offsets of around 0.1\arcsec~to 0.2\arcsec.

We first associate for each of our sources at 0.3\arcsec~a nearest radio counterpart from the 6\arcsec~catalogues from \cite{kondapally2021}, where we allowed a maximum distance of 6\arcsec. Since point sources are most accurately cross-matched, we only select sources fitted by \texttt{PyBDSF} with a single Gaussian with major and minor axes less than 1.25 the size of the synthesized beam. To ensure that we cross-match sources that are with high certainty detected in both radio observations, we select from both catalogues sources with peak intensities 15 times larger than the local RMS. As our goal is to determine the astrometry using the optical cross-matches from \cite{kondapally2021}, we use flags in their catalogue to exclude sources in regions around bright stars, as the positions of the optical counterparts of our radio galaxies might be affected due to close proximity to these bright stars. We also use a threshold on the likelihood ratio (LR), as determined by \cite{kondapally2021}. This value indicates the ratio of the probability that a galaxy is a correct cross-match against being an incorrect cross-match. The LR is an often used statistical value to assess the quality of the counterpart cross-matching \citep[e.g.][]{deruiter1977, sutherland1992, smith2011, mcalpine2012}. 
We opt for selecting the top 30\% of sources with the highest LR scores, as this leaves us with a sufficient number of sources (192) with the best optical-radio cross-matches. With this sample, we find median offsets of dRA=$0.094\pm0.093$\arcsec~and dDEC=$-0.067\pm0.064$\arcsec~between our 0.3\arcsec~radio positions and the optical positions selected from \cite{kondapally2021} catalogue. These offsets are within the astrometric uncertainty of the optical positions reported by \cite{kondapally2021}. 

The scatter on the positional offsets is illustrated in the left panel of Figure \ref{fig:dRA_dDEC}. 
A positional `random' scatter around the median value is expected because, due to the possible complex morphological nature of radio sources, there is no guarantee that the brightest point of a radio source aligns exactly with the position of an optical host. However, calibration errors can lead to systematic offsets. Since each facet has its own calibration solutions (see Section \ref{sec:ddcal}), we assess systematic offsets in the right panel of Figure \ref{fig:dRA_dDEC}, by plotting the distribution of the selected radio-optical associations with absolute offsets and offset directions across our 30 facets. While most facets do not show any preferred positional offset direction, we only find for facets 17, 22, and 23 hints of a positional offset direction in the +RA direction (see Figure \ref{fig:facets} for the corresponding facet numbers). However, noting that the absolute offsets for these facets are not significantly larger than what we find for other facets, we do not apply astrometric corrections.
To conclude, our small astrometric offsets confirm the high accuracy of our astrometry as a result of calibrating the in-field calibrator against an accurate sky model. Despite the presence of hints of minor systematic offsets in a few facets, the accuracy is largely maintained during DD calibration.

Additionally, we conducted a similar astrometric analysis with our 0.6\arcsec~and 1.2\arcsec~data and found within the uncertainties the same accurate results. This consistency is expected because the same calibration solutions were applied for all resolutions, with the exception of some extra facets at 1.2\arcsec~that received additional calibration for the Dutch core and remote stations (see Section \ref{sec:extradutch}). Nevertheless, we do not observe any impact from this additional Dutch calibration on the positional offsets because the calibration solutions for the international stations, which primarily determine the smallest angular scales, have remained unchanged.

\subsection{Flux scale}\label{sec:fluxscale}

Similar to the astrometry evaluation, we verified our flux density scale by using the catalogue from \cite{kondapally2021} (see Section \ref{sec:astrometry}). For this purpose, we utilised only the radio source information from their catalogue. We selected in both catalogues sources that were fitted by \texttt{PyBDSF} with a single Gaussian, have a maximum position offset of 6\arcsec, and which exhibited a peak intensity at least 25 times greater than the local RMS noise. This brightness threshold ensures we select sources without any loss of flux density, considering the resolution difference. 
With the remaining 368 sources, we find a median flux density ratio of $\frac{S_{6}}{S_{0.3}}=0.97\pm0.14$ between the cross-matches of the 6\arcsec~catalogue and the 0.3\arcsec~catalogue, demonstrating the consistency of our flux scale. Similar to the astrometry, the accurate flux density scale stems from the sky model that we used for calibrating the primary in-field calibrator (see Section \ref{sec:skymodel}).

We also compared the flux scales across our three resolutions. After selecting again sources identified by \texttt{PyBDSF} by a single Gaussian and with peak intensities 25 times above the local RMS noise, we cross-matched the three catalogues using a maximum position offset of 1\arcsec. This yielded 215 sources. For these sources, we find flux density ratios of $\frac{S_{0.3}}{S_{0.6}}=1.00\pm0.04$, $\frac{S_{0.6}}{S_{1.2}}=0.99\pm0.08$, $\frac{S_{0.3}}{S_{1.2}}=0.98\pm0.09$. The consistency of these ratios supports the robustness of our flux density scale across different resolutions. When we remove the brightness constraint but keep the 1\arcsec~position offset for sources fitted by single Gaussians, we find with the 3607 remaining sources the following flux density ratios $\frac{S_{0.3}}{S_{0.6}}=0.97\pm0.17$, $\frac{S_{0.6}}{S_{1.2}}=0.92\pm0.24$, $\frac{S_{0.3}}{S_{1.2}}=0.89\pm0.28$. Despite the large uncertainties, these results align with our expectations that lower resolutions tend to capture more diffuse emission compared to higher resolutions. The resolution differences are discussed further in the next section.

\subsection{Sensitivity versus resolution}\label{sec:res_sens}

Higher resolutions allow for accurate astrometry and detailed characterisation of compact and extended structures whilst also allowing for more precise optical and near-infrared identification of host galaxies. The enhanced ability to discriminate between sources is demonstrated in the lower panel of Figure \ref{fig:compare_res}, where at 6\arcsec~we initially identify 3 Gaussians that appear to belong to the same source, while at 0.3\arcsec~resolution, we find this to actually be multiple sources with complex components. However, the detection of objects with low surface brightness is more challenging in high-resolution images and requires further image processing (e.g. smoothing), causing some extended sources, that are visible at lower resolutions, to not be trivially detectable at higher resolutions. For example, we observe that 22\% of the sources in our 1.2\arcsec~catalogue do not have a counterpart at 0.3\arcsec, even though the 0.3\arcsec~map is deeper. This demonstrates the importance of considering a trade-off between resolution, sensitivity, and scientific objective.

\begin{figure*}[htbp]
 \centering
    \includegraphics[width=0.48\linewidth]{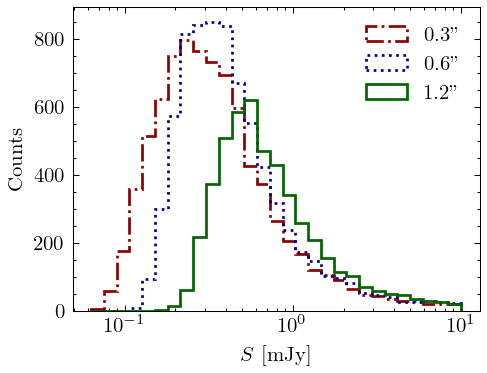}
    \includegraphics[width=0.48\linewidth]{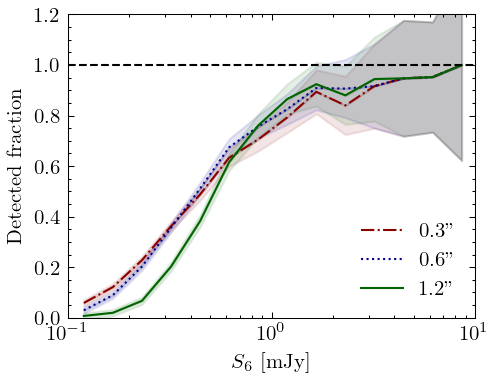}
  \caption{Source detection across resolution. \textit{Left panel:} Source counts as a function of the integrated flux density for our 3 resolutions below 10 mJy. \textit{Right panel:} Detected fraction between sources from our radio maps and the compact sources detected by Shimwell et al. (in prep.) at 6\arcsec. For the errors, we propagate Poisson uncertainties.}
\label{fig:detectability}
\end{figure*}

To illustrate the above, we plot in the left panel of Figure \ref{fig:detectability} the number of sources as a function of integrated flux density across resolution. This figure demonstrates the effect of enhanced sensitivity at higher resolutions on the number of sources detected at lower flux densities. As most of our sources are below 1~mJy, we expect a large fraction of our detections to be part of the star-forming galaxy population \citep{best2023}. Above \textasciitilde0.25~mJy, we find the number of detections at 0.6\arcsec~resolution to be higher than the 0.3\arcsec~resolution detections. This is likely due to the fact that at these flux densities the S/N at 0.6\arcsec~is large enough to detect many sources and because the source population here contains many distant star-forming galaxies (with redshifts of approximately $z\sim 0.5-1.0$) that have typical sizes of a few tenths of an arcsecond. These sources are therefore more likely to be resolved (out) at 0.3\arcsec~compared to the 0.6\arcsec. Similarly, above \textasciitilde0.5~mJy the 1.2\arcsec~resolution detects the most sources, which is because fewer sources are resolved at this resolution.

The 1.2\arcsec~resolution has the best surface brightness sensitivity of our three resolutions and is therefore the best to identify extended sources. This includes the population of LERGs, which are the most dominant radio source population in the LOFAR Deep Fields DR1 above \textasciitilde1.5~mJy \citep{best2023}. However, at higher redshifts, LERGs will be fainter and more compact and more often hosted by star-forming galaxies \citep{kondapally2022}, leading to selection biases when being unable to separate radio jets from star formation \citep{mingo2022, dejong2024}.
This underscores again the value of making wide-field images at multiple resolutions with the same calibrated data.

We assess in the right panel of Figure \ref{fig:detectability} the fraction of sources detected as a function of resolution by cross-matching our catalogues with the compact sources (fitted by a single Gaussian) from a deeper 6\arcsec~catalogue (Shimwell et al., in prep.). This figure shows that the fractions detected at 0.3\arcsec~and 0.6\arcsec~resolutions yield similar results, despite the lower sensitivity of the 0.6\arcsec~resolution map. The 1.2\arcsec~resolution has a low fraction of sources detected below 0.4~mJy, while above this flux density, all resolution images have about the same detection fraction and approach completeness between 1 and 10~mJy. The decline in the detected fraction towards lower flux densities is explained by a combination of being less sensitive to detect low surface brightness sources at higher resolutions and the fact that the 6\arcsec~resolution map based on 500~hrs of LOFAR data is slightly deeper than our maps.

\section{Summary and conclusion}\label{sec:conclusion}

We have presented the currently deepest wide-field image of ELAIS-N1 at about 0.3\arcsec~resolution and 140~MHz. This image has a field of view of 2.5\degree$\times$2.5\degree~with a sensitivity down to 14~$\mu$Jy~beam$^{-1}$ at the pointing centre. This was achieved by implementing an improved DI and DD calibration strategy built upon the existing VLBI calibration and imaging strategies \citep{morabito2022, sweijen2022, ye2023} and applying it to four 8-hr LOFAR HBA observations. As additional products, we produced wide-field images at 0.6\arcsec~and 1.2\arcsec~resolution with sensitivities of respectively 21~$\mu$Jy~beam$^{-1}$ and 39~$\mu$Jy~beam$^{-1}$ near the pointing centre. In these radio maps, we report the detection of 9203 sources at 0.3\arcsec, 8567 sources at 0.6\arcsec, and 5872 sources at 1.2\arcsec.

For accurate calibration, we generated a sky model for our primary in-field calibrator by fitting the spectral index using different surveys and by imaging ELAIS-N1 at 54~MHz with LOFAR LBA data. This approach, along with refinements in the calibration steps for calibrating the in-field calibrator, improved our DI calibration of the international LOFAR stations. We adopted a quantitative approach to assess the selection of the DD calibrators. This enabled us to quickly and robustly select the best secondary calibrators to correct for the varying ionosphere across our field of view. Although we improved the DD calibration, we identified existing inaccuracies in the calibration solutions for the Dutch core and remote stations. This likely stems from the fact that in this work, due to computational cost, we initially ignored bright sources in our field beyond the facet image boundary during self-calibration. To rectify this, we introduced an extra calibration round specifically for the Dutch stations after subtracting the sky outside each facet. After imaging each individual facet separately in parallel, we mosaiced everything back together into our final wide-field images for each resolution. We find our complete data processing from calibration to imaging to be about two times faster compared to the previous work by \cite{sweijen2022} and \cite{ye2023}. This is due to software and hardware improvements.

We find the smearing in our images to be the most severe at the highest resolution, which is intensified by the fact that half of our observations from the LTA were pre-averaged to 2~sec. As a result of our primary in-field calibration strategy, using an accurate sky model, we achieved precise astrometry with median offsets of dRA=$0.094\pm0.093$\arcsec~and dDEC=$-0.067\pm0.064$\arcsec~after comparing with optical counterparts selected from the catalogues by \cite{kondapally2021}. We also found accurate flux density scales for the wide-field images. 

By comparing the three resolutions, we find 1.2\arcsec~to be a good intermediate resolution to detect sources with extended low surface brightness emission, while the depth and detail in our 0.3\arcsec~resolution map are expected to be great for separating source components or detecting compact objects at higher redshifts. The 0.6\arcsec~resolution map complements these two resolutions for objects that are resolved out at 0.3\arcsec~and unresolved at 1.2\arcsec, such as star-forming galaxies at low flux densities. We also find the detected fraction across the three resolutions to reach completeness between 1 and 10 mJy.

Our work demonstrates the feasibility of making deep wide-field images at sub-arcsecond resolutions with LOFAR. Near the pointing centre, we reached RMS noise values close to what has been recently achieved with the Dutch LOFAR stations at 6\arcsec, with about 16 times less observation time.
Currently, computational costs are the primary obstacle in processing all 500~hrs of LOFAR observations of ELAIS-N1 stored in the LTA (Shimwell et al. in prep.). Addressing the computational challenges will enable the creation of the deepest LOFAR wide-field image, which can be used to uncover objects at the smallest angular scales in the low-frequency radio sky.

 \begin{acknowledgements} 

This publication is part of the project CORTEX (NWA.1160.18.316) of the research programme NWA-ORC which is (partly) financed by the Dutch Research Council (NWO). This work made use of the Dutch national e-infrastructure with the support of the SURF Cooperative using grant no. EINF-6218. This work is co-funded by the EGI-ACE project (Horizon 2020) under Grant number 101017567.

This work was sponsored by NWO Domain Science for the use of the national computer facilities.

LKM and EE are grateful for support from a UKRI Future Leaders Fellowship [MR/T042842/1].
 
RJvW acknowledges support from the ERC Starting Grant ClusterWeb 804208. 

MB acknowledges support from INAF under the Large Grant 2022 funding scheme (project “MeerKAT and LOFAR Team up: a Unique Radio Window on Galaxy/AGN co-Evolution”.

RK and PNB are grateful for support from the UK STFC via grant ST/V000594/1.

JP acknowledges support for their PhD studentship from grants ST/T506047/1 and ST/V506643/1.

The Pan-STARRS1 Surveys (PS1) and the PS1 public science archive have been made possible through contributions by the Institute for Astronomy, the University of Hawaii, the Pan-STARRS Project Office, the Max-Planck Society and its participating institutes, the Max Planck Institute for Astronomy, Heidelberg and the Max Planck Institute for Extraterrestrial Physics, Garching, The Johns Hopkins University, Durham University, the University of Edinburgh, the Queen’s University Belfast, the Harvard-Smithsonian Center for Astrophysics, the Las Cumbres Observatory Global Telescope Network Incorporated, the National Central University of Taiwan, the Space Telescope Science Institute, the National Aeronautics and Space Administration under Grant No. NNX08AR22G issued through the Planetary Science Division of the NASA Science Mission Directorate, the National Science Foundation Grant No. AST-1238877, the University of Maryland, Eotvos Lorand University (ELTE), the Los Alamos National Laboratory, and the Gordon and Betty Moore Foundation.

LOFAR data products were provided by the LOFAR Surveys Key Science project (LSKSP; \url{https://lofar-surveys.org/}) and were derived from observations with the International LOFAR Telescope (ILT). LOFAR \citep{haarlem2013} is the Low Frequency Array designed and constructed by ASTRON. It has observing, data processing, and data storage facilities in several countries, which are owned by various parties (each with their own funding sources), and which are collectively operated by the ILT foundation under a joint scientific policy. The efforts of the LSKSP have benefited from funding from the European Research Council, NOVA, NWO, CNRS-INSU, the SURF Co-operative, the UK Science and Technology Funding Council and the Jülich Supercomputing Centre.

\end{acknowledgements} 

\bibliographystyle{bst}
\bibliography{bib}

\appendix

\section{Towards an automated VLBI pipeline}\label{sec:automation}

Given the large data volumes, processing LOFAR data with international stations incurs significant computing costs (see also Section \ref{sec:computingcosts}). This makes it essential to carefully optimize each of the steps in the current calibration strategy. In addition, removing the manual data processing and visual inspection from the current strategy could lead to the possibility of conducting more automated processing, allowing for a larger high-resolution survey of the northern hemisphere survey. We explore in this appendix section parts of the pipeline that could be replaced by automated approaches with strategies to test and implement these.

The initial steps for obtaining the DI calibration for Dutch stations with \texttt{Prefactor} pipeline\footnote{Note that this step has been recently replaced by the LOFAR Initial Calibration (LINC) pipeline: \url{https://linc.readthedocs.io}.} and running the \texttt{DDF} pipeline are already automated. However, human intervention is still required to inspect the output from the \texttt{Prefactor} and \texttt{DDF} pipelines, as it is important to validate the quality of the observations before proceeding with the data reduction of the long-baselines. A preliminary examination of the data flagging percentages catches the severeness of RFI and can help eliminate corrupted observations. Nonetheless, the flagging fraction does not catch all instances of corrupted data. One therefore typically manually inspects calibration solution plots from the calibrator and target solutions generated by \texttt{LoSoTo}. The wide-field image at 6\arcsec, after running the \texttt{DDF} pipeline, also allowed us to assess the image quality with only Dutch stations of the observation. This provides information about the effects of the ionosphere on the final wide-field image quality and is therefore a direct indicator of the calibratability of our data. The manual interactions with the data in these steps could be replaced by adopting an automated inspection process such that the future pipeline could determine which observations require modifications to the input data, adjust calibration parameters in the pipeline, or perhaps decide to reject a particular observation.

The next important manual input starts when selecting suitable calibrators in both the in-field DI calibrator and DD calibrators. These steps are highlighted in red in Figures \ref{fig:pipeline1} and \ref{fig:pipeline2}. 
For the primary in-field calibrator selection (see Section \ref{sec:infieldselection}), one typically selects first the calibrator candidates from the LBCS catalogue \citep{jackson2016, jackson2022}. If no sources are available or the quality of the solutions and calibration proves after visual inspection to be inadequate, one tries other bright unresolved sources from the LoTSS catalogue \citep{shimwell2017, shimwell2019, shimwell2022}. The standard method for selecting DD calibrators is similar, but begins by directly identifying candidates from the LoTSS catalogue above a specified flux density threshold. Currently, users inspect visually the calibration solutions and image qualities and perhaps change the parameters or decide to exclude the candidate calibrator entirely.
In both steps, we suggest applying the selection procedure described in Section \ref{sec:ddsourceselection}, where in the selection we evaluate the circular standard deviation on the phase RR and LL phase difference. This is computationally cheap, as it requires just about 1~CPU~hr for each source, and eliminates candidates with an insufficient S/N at the longest baselines. Following this up by an additional selection after self-calibration, will eliminate falsely selected sources (as discussed in \ref{sec:ddsourceinspection}). 

It is important to stress that we can currently not guarantee that our empirically selected circular standard deviation score of 2.3~rad from the \texttt{scalarphasediff} calibration will be generally good enough for selecting DD calibrators of other fields. For instance, we found the scores to differ up to \textasciitilde0.5~rad between the individual nights, which is a significant difference on a score that has a maximum around $\pi$. One could for example expect that sky areas at lower declinations would have on average much higher scores. Hence, for implementing these steps for automation, additional tests on different observations and fields are required. Moreover, the process of manually adjusting calibration parameters for self-calibration could be automated by linking it to specific circular standard deviation scores from the \texttt{scalarphasediff} calibration and by incorporating additional metrics. We have not yet explored this in detail, but as the circular standard deviation links to the brightness of a source, we expect this to link to the solution interval as well.

After obtaining the image output products, it requires -- despite having tools such as \texttt{PyBDSF} -- a vast amount of work to catalogue source detections. The primary challenge involves source association. Work has been done to automate this through machine learning \citep[e.g.][]{mostert2022}. However, work needs to be done to re-train the models and improve them for our high-resolution data. 

To summarize, we suggest replacing the following manual steps in the calibration strategy with automated approaches:
\begin{itemize}[leftmargin=*,labelsep=5mm]
  \item[$\square$] \textbf{Solution inspection:} Inspection of for instance \texttt{Prefactor}/\texttt{LINC} calibration solution output to validate data quality before moving to the calibration for the international LOFAR stations. Solution inspection can be similarly done after every calibration step, such as the DI and DD calibration (see Sections \ref{sec:dical_strategy} and \ref{sec:selfcal}).
    \item[$\square$] \textbf{DDF-image inspection:} Inspection of the DD-corrected wide-field image quality at 6\arcsec, which is produced by the \texttt{DDF}-pipeline. This indicates the severeness of the ionosphere and therefore the calibratability of our data.
    \item[$\square$] \textbf{Quantity inspection:} An inspection of the output data after every main step in the pipeline can ensure that no data gets `lost', which involves monitoring for excessive flagging or tracking job failures on portions of the data.
  \item[$\square$] \textbf{Calibrator selection:} Select the best in-field calibrator source and the best DD calibrator sources by using computational cheap metrics, such as the phase noise metric discussed in Section \ref{sec:ddsourceselection}.
  \item[$\square$] \textbf{Calibration parameters:} The metrics for the calibrator selection could be linked to optimising the calibrator parameters, such as the solution interval or smoothness constraints.
  \item[$\square$] \textbf{Source association:} To prepare our output images for scientific analysis, it would be advantageous to automatically and accurately perform source association for our high-resolution data.
\end{itemize}

\section{Potential self-calibration issues}\label{appendix:selfcalibrationissues}

In Section \ref{sec:selfcal}, we discuss the self-calibration of our DD calibrators. While our source selection performed well (see Section \ref{sec:ddsourceinspection}), we ran during testing also self-calibration on a few sources that were not selected by our selection metric. We highlight below two examples of sources that were above our selection threshold from Section \ref{sec:ddsourceselection} (so were not selected) but diverged strongly due to various issues.

\begin{figure*}
    \begin{minipage}{0.73\textwidth}
        \centering
        \includegraphics[width=0.73\textwidth]{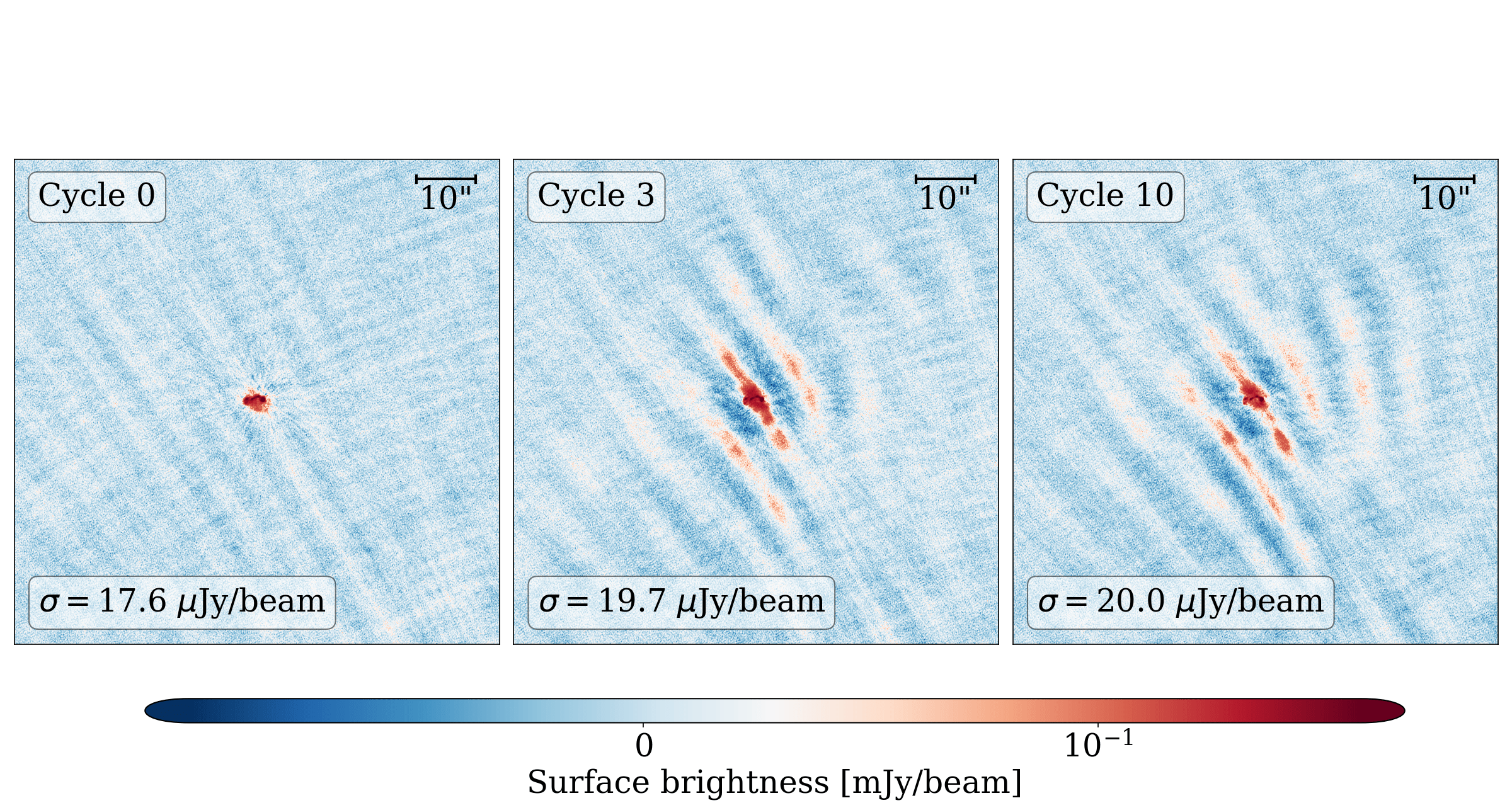}\\
        \includegraphics[width=0.73\textwidth]{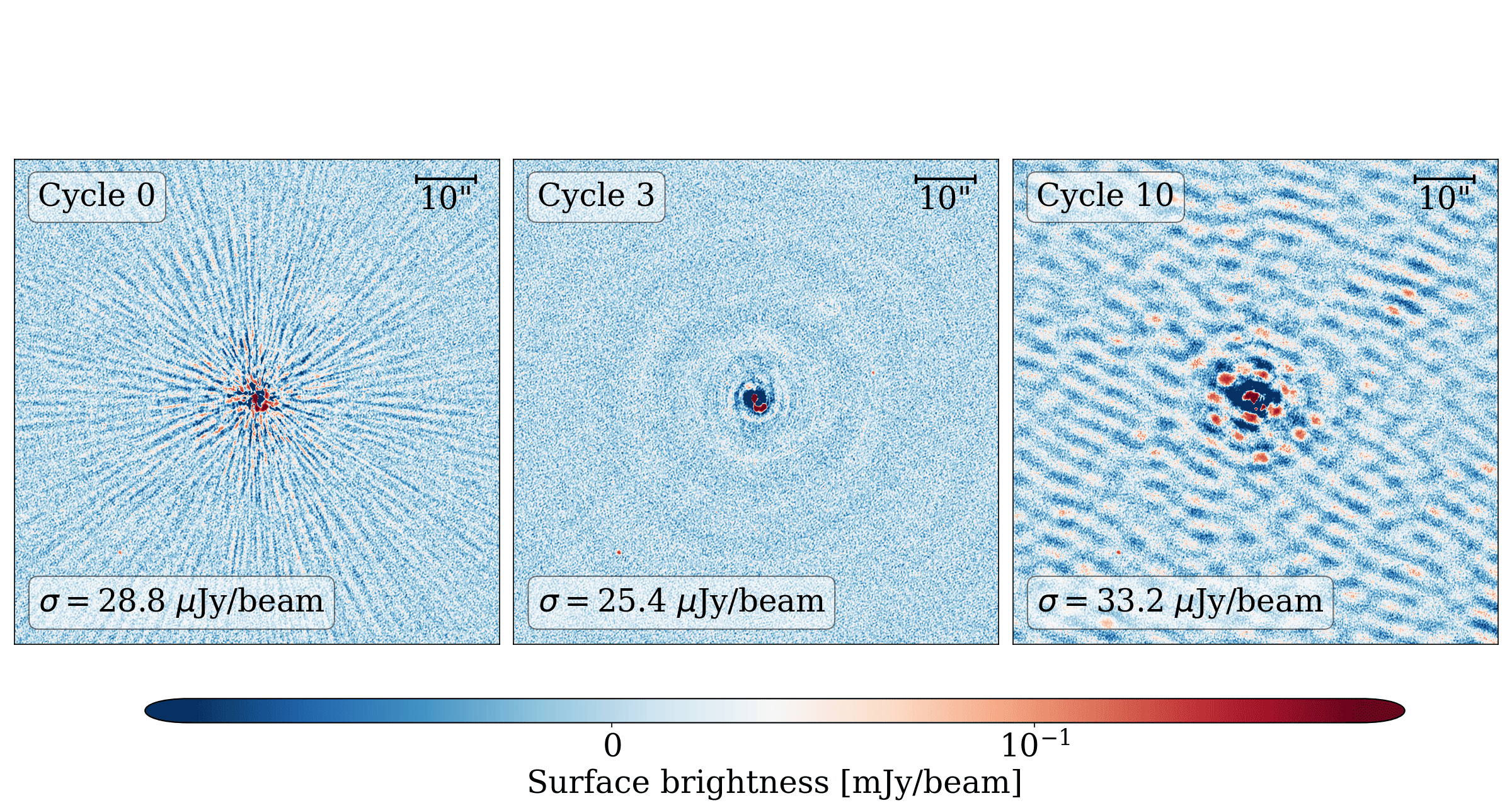}
    \end{minipage}\hfill
    \begin{minipage}{0.24\textwidth}
    \vspace{0.25\textheight}
        \caption{Two examples of self-calibration issues. Cycle 0 is the first image with only DI solutions applied from the in-field calibrator (see Section \ref{sec:infieldcal}). Cycle 3 corresponds to the self-calibration image after 3 rounds of \texttt{scalarphase} calibration. After this cycle, \texttt{scalarcomplexgain} calibration is added. This also calibrates for amplitude errors. Cycle 10 shows the result after the 10th self-calibration round.}
    \label{fig:selfcal_issues}
    \end{minipage}
\end{figure*}

\begin{figure*}
\centering
\includegraphics[width=0.48\linewidth]{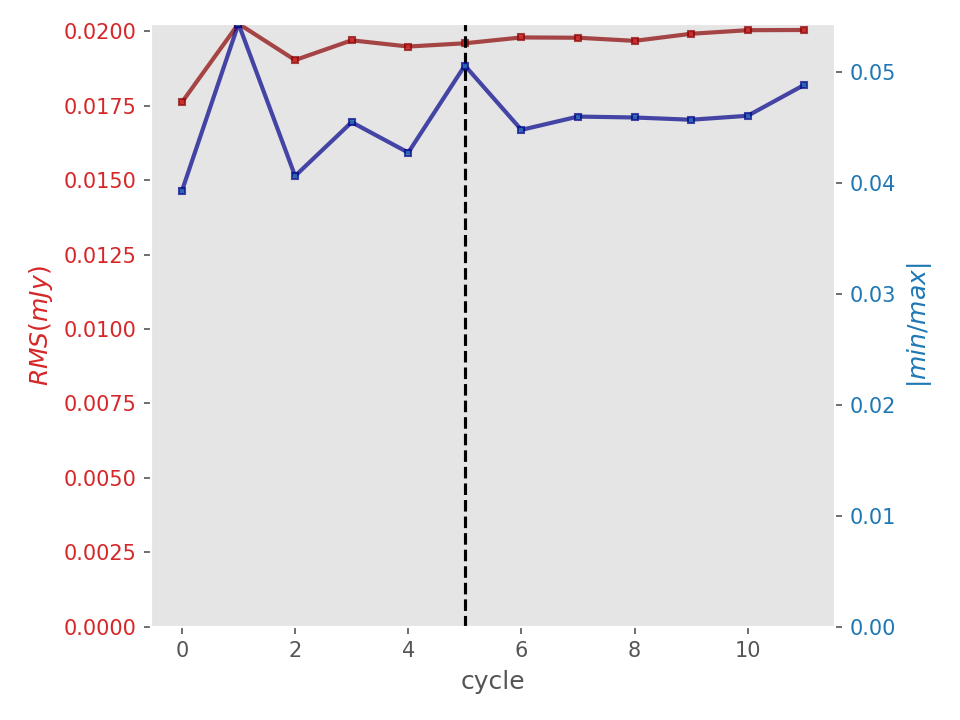}
\includegraphics[width=0.48\linewidth]{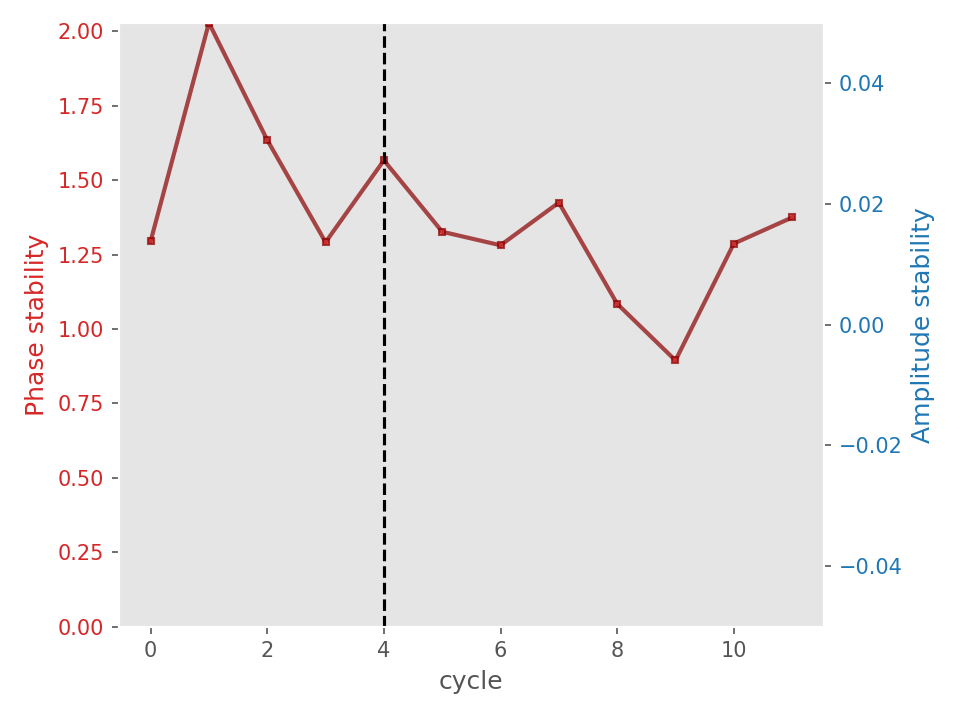}
\caption{Self-calibration and solution stabilities for the source in the upper panel of Figure \ref{fig:selfcal_issues}. The interpretation of these figures is discussed in the captions of Figures \ref{fig:imagestability} and \ref{fig:solutionsstability}. It is important to note that this source did not have amplitude corrections, as according to the \texttt{auto} settings from \texttt{facetselfcal}, this source was not sufficiently bright enough to trigger \texttt{scalarcomplexgain} corrections.}
\label{fig:P34825stability}
\end{figure*}

\begin{figure*}
\centering
\includegraphics[width=0.48\linewidth]{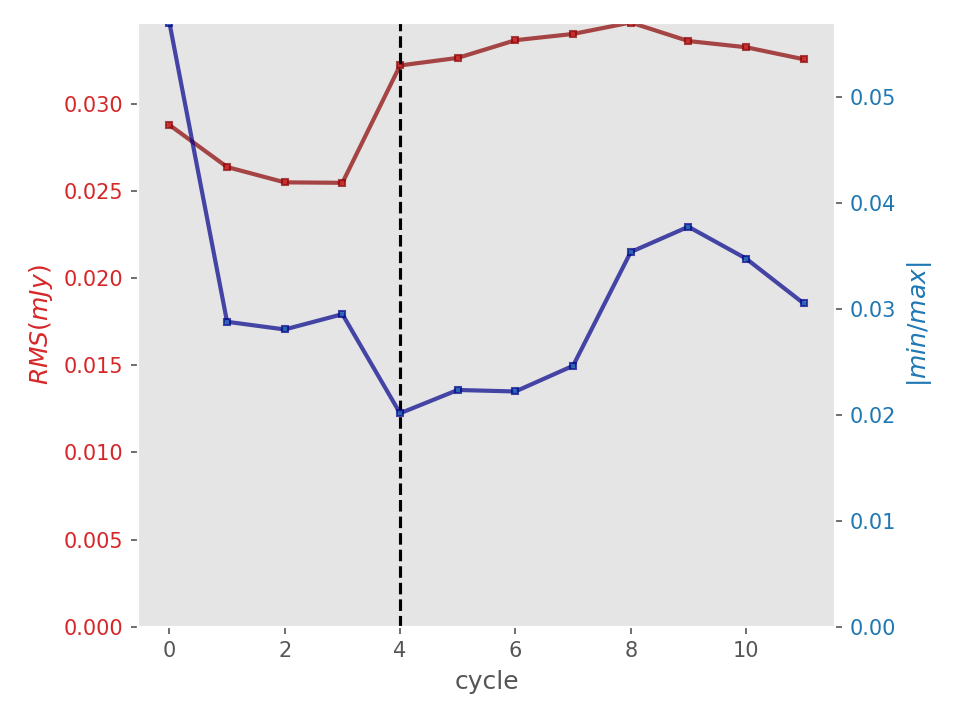}
\includegraphics[width=0.48\linewidth]{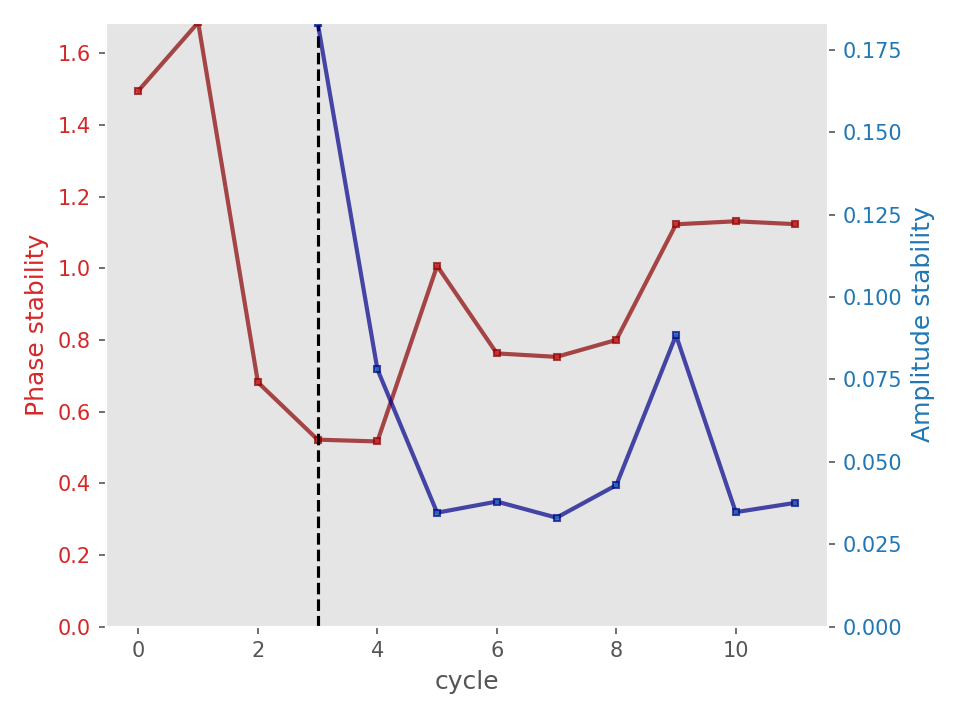}
\caption{Self-calibration and solution stabilities for the source in the lower panel of Figure \ref{fig:selfcal_issues}. The interpretation of these figures is discussed in the captions Figures \ref{fig:imagestability} and \ref{fig:solutionsstability}.}
\label{fig:P45240stability}
\end{figure*}

In Figure \ref{fig:selfcal_issues} we find in the upper panel a source that has a bright calibrator nearby, which introduces during phase (\texttt{scalarphase}) calibration strong artefacts. This results in bad calibration solutions and no improvements of their images, as is evident from the graphs in the left panel of Figure \ref{fig:P34825stability}, where neither the RMS nor the dynamic range shows any improvement. The phase solutions remain also unstable in the right panel.
In the lower panel of Figure \ref{fig:selfcal_issues} we display self-calibration cycles from a source that was partly subtracted on the edge of our 2.5\degree$\times$2.5\degree~field of view. This introduced strong artefacts after amplitude calibration, starting from cycle 3. In Figure \ref{fig:P45240stability}, we find in the left panel the RMS goes up after cycle 4, which corresponds in the right panel to the instability of the phase solutions after this same cycle.
Although these sources were not selected by our algorithm, they do demonstrate the effectiveness of performing a pre-selection for bright compact sources and for self-calibration inspection in case of similar or other issues that are not guaranteed to be captured by the phase noise metric selection in Section \ref{sec:ddsourceselection}.

\end{document}